\documentclass[onecolumn,fleqn,usenatbib]{mnras}
\usepackage{txfonts}
\usepackage[T1]{fontenc}
\usepackage{ae,aecompl}
\usepackage{graphicx}

\newcommand{\overbar}[1]{\mkern 1.5mu\overline{\mkern-1.5mu#1\mkern-1.5mu}\mkern 1.5mu}  
\newcommand       \be           {\begin{equation}}
\newcommand       \ee           {\end{equation}}

\title[Alignment of grains by mechanical torques]
{Alignment of interstellar grains by mechanical torques: 
suprathermally rotating Gaussian random spheres}
\author[I. Das \&  J. C. Weingartner]
{Indrajit Das$^{1}$\thanks{E-mail: idas@masonlive.gmu.edu; jweinga1@gmu.edu}
and  Joseph C. Weingartner$^{1}$\\
$^{1}$Department of Physics and Astronomy, George Mason University, 
4400 University Drive, Fairfax, VA 22030, USA}

\date{Accepted XXX. Received YYY; in original form ZZZ}

\pubyear{2015}

\begin{document}
\label{firstpage}
\pagerange{\pageref{firstpage}--\pageref{lastpage}}
\maketitle

\begin{abstract}

Collisions of gas particles with a drifting grain give rise to a 
mechanical torque on the grain.  Recent work by Lazarian \& Hoang showed that
mechanical torques might play a significant role in aligning helical grains
along the interstellar magnetic field direction, even in the case of subsonic 
drift.  We compute the mechanical torques on 13 different irregular grains and 
examine their resulting rotational dynamics, assuming steady rotation about
the principal axis of greatest moment of inertia.  We find that the alignment 
efficiency in the subsonic drift regime depends sensitively on the grain 
shape, with more efficient alignment for shapes with a substantial 
mechanical torque even in the case of no drift.  The alignment is typically
more efficient for supersonic drift.  A more rigorous analysis of the 
dynamics is required to definitively appraise the role of mechanical 
torques in grain alignment.    

\end{abstract}


\section{Introduction}

Observations of starlight polarization and polarized thermal emission from 
dust indicate that interstellar grains are nonspherical and aligned.  
Despite over 60 years of effort, the theory of grain alignment is not
yet complete; see \citet{Lazarian07} and \citet{Andersson15} for reviews.

Among the early proposals for alignment mechanisms, \citet{Gold52a, Gold52b}
considered ``mechanical torques'' arising from collisions of gas particles
with an elongated grain moving through the gas supersonically.  Numerous
authors have further elaborated and extended this model; see references in 
\citet{LH07mech}.  While the alignment described by Gold is a stochastic 
process, \citet{Lazarian07} and \citet{LH07rad} noted that irregularly 
shaped grains could experience systematic mechanical torques associated with
their helicity.  \citet{LH07mech} examined the torque on a highly idealized
helical grain.  They concluded that the resulting alignment can be efficient
even for grains moving subsonically, likely dominates over Gold-type 
alignment, and aligns grains with their long axes perpendicular to the magnetic
field.  

\citet{LH07mech} noted that detailed studies of the mechanical torques on
irregular grains are needed to clarify the efficiency of helicity-related 
mechanical torques, since the helicities of realistic grain shapes are 
unknown.  That is our aim in this work.  We examine the mechanical torques,
for a variety of gas-grain drift speeds, on 13 irregular grains, whose shapes
are described in \S \ref{sec:grs}.  We describe the theoretical and 
computational aspects of the torque calculations in \S \S 
\ref{sec:torque-theory} and \ref{sec:torque-comp}, respectively.  The results 
of these calculations are presented in \S \ref{sec:torque-results0}.  
In \S \ref{sec:dynamics}, we examine the grain rotational dynamics under
the influence of the mechanical, drag, and magnetic torques, assuming that
the grain rotates about its principal axis of greatest moment of inertia,
$\bmath{\hat{a}}_1$.  We discuss the implications for the efficiency of grain 
alignment by helicity-induced mechanical torques, but defer a detailed
examination to an upcoming study, where the assumption of 
rotation about $\bmath{\hat{a}}_1$ will be relaxed.  
Conclusions and future work are summarized in \S \ref{sec:conclusions}.  

\section{Grain Shapes} \label{sec:grs}

We examine Gaussian random spheres (GRSs) using a slightly
modified version of the prescription of \citet{Muinonen96}. 
Consider a coordinate system $(x, y, z)$ fixed with respect to the grain with
the origin located inside the grain.   
In spherical coordinates, the distance from 
the origin to the surface of the GRS, 
as a function of the polar angle $\theta$ (with $\bmath{\hat{z}}$ 
as the reference
axis) and azimuthal angle $\phi$ (with $\bmath{\hat{x}}$ as the reference 
axis), is
\be
r_{\mathrm{surf}}(\theta, \phi) = a (1 + \sigma^2)^{-1/2} \exp[w_1(\theta, 
\phi)]
\ee
where
\be
w_1(\theta, \phi) = \sum_{l=1}^{l_{\mathrm{max}}} \sum_{m=0}^l P_l^m(\cos 
\theta) \, (a_{lm} \cos m \phi + b_{lm} \sin m \phi)
\ee
and $P_l^m(u)$ denotes the associated Legendre functions.
The expansion coefficients $a_{lm}$ and $b_{lm}$ are taken as independent
Gaussian random variables with zero means and equal variances
$\beta_{lm}^2$ given by
\be
\beta_{lm}^2 = (2 - \delta_{m0}) \frac{(l-m)!}{(l+m)!} c_l \ln 
(1 + \sigma^2) 
\ee
with 
\be
c_l = l^{- \alpha} \ \left( \sum_{l=1}^{l_{\mathrm{max}}} l^{- \alpha} \right)^{
-1}.
\ee
For a given direction $(\theta, \phi)$, the mean and variance of the distance
$r$ to the surface, over an ensemble of realizations of the grain
geometry, are given by $a$ and $a^2 \sigma^2$, respectively (in the 
limit $l_{\rm max} \rightarrow \infty$).  Thus, the parameter $\sigma$ controls
the amplitude of deviations from sphericity, while $\alpha$  controls the 
angular scale of the deviations.

\begin{figure}
\begin{minipage}{8.5cm}
\includegraphics[width=70mm]{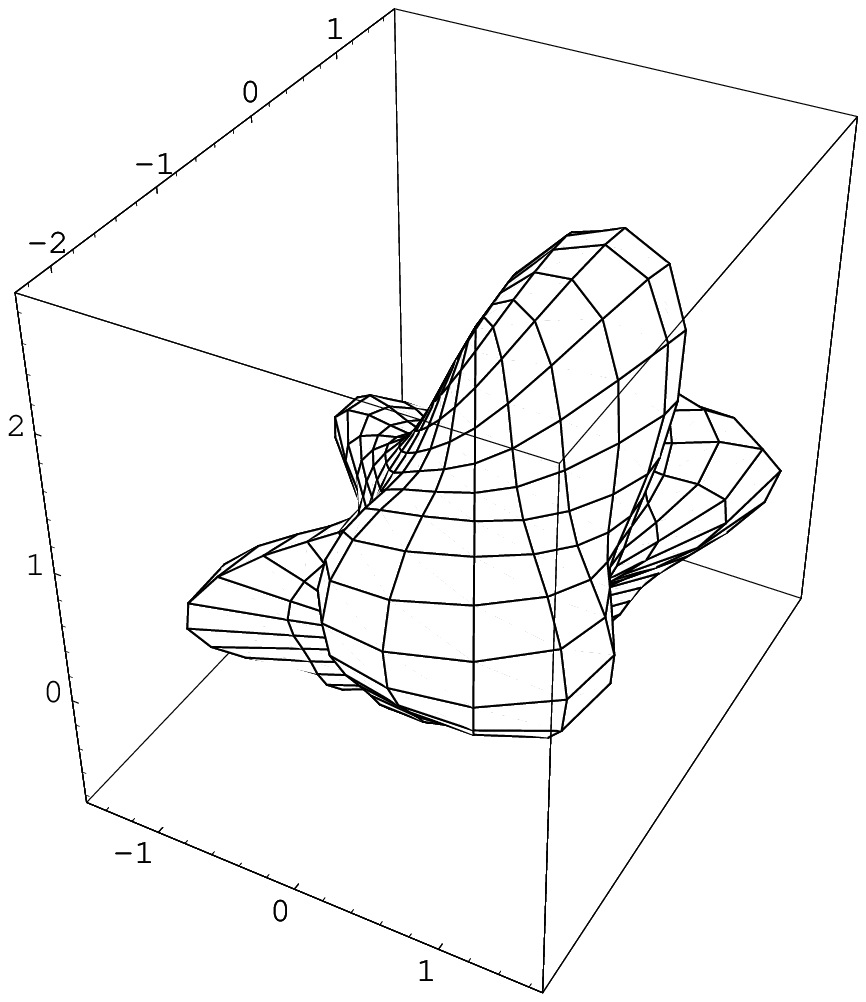}
\end{minipage}
\begin{minipage}{8.5cm}
\includegraphics[width=70mm]{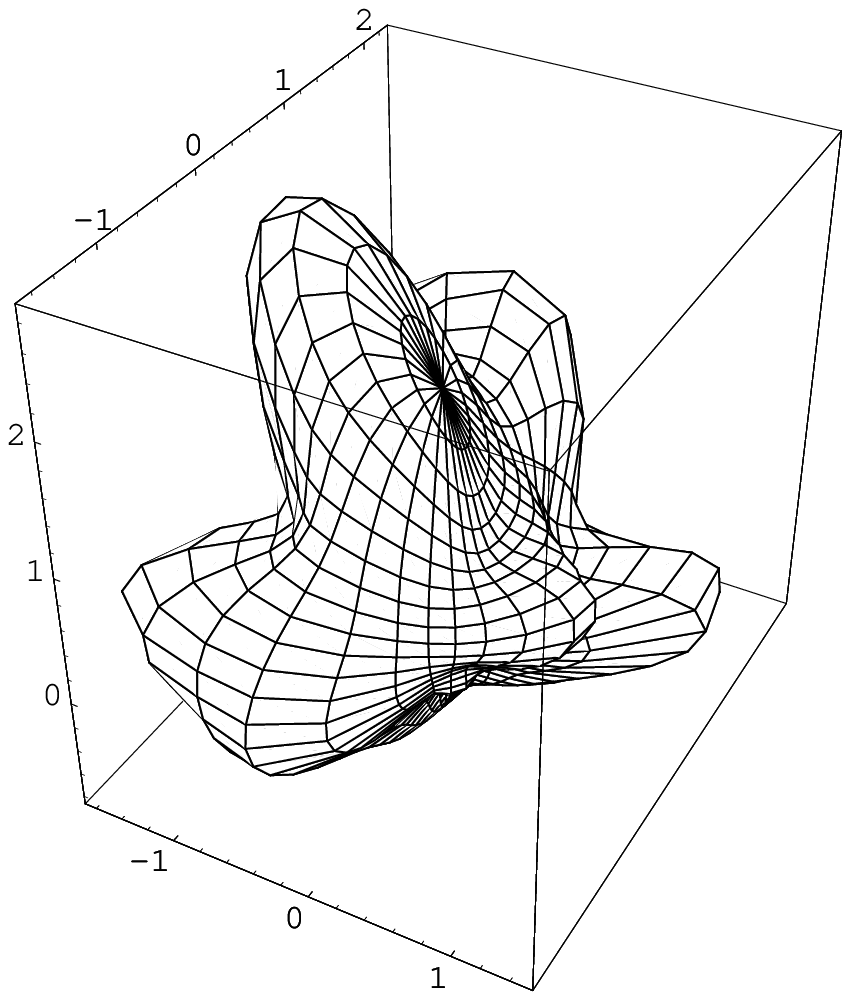}
\end{minipage}
\caption{
Views of grain 1 from two opposite directions.
        }
\label{fig:grs1}
\end{figure}

We generated 20 different grains, each with $\sigma = 0.5$ and half with 
$\alpha = 2$ and the other half with $\alpha = 3$.  
In each case, we took $l_{\mathrm{max}} = 8$ and used a slightly 
modified version of the Gaussian deviate routine {\sc gasdev} from 
\citet{Press92} to select values for the coefficients $a_{lm}$ and $b_{lm}$.  
If the ratio of the maximum to minimum principal moments of inertia of the
resulting grain was
less than 1.5 or greater than 3, then the shape was discarded as too symmetric
or too extreme.  (Preliminary scattering calculations indicate that these
grains can produce polarization consistent with that observed in the ISM.
This will be examined in detail in a study of radiative torques on these
grains.)    
Also, we required that the centre of mass lies within the grain.
Of the 13 grains that satisfied these criteria, grains 1--7
have $\alpha = 2$ and grains 8--13 have $\alpha = 3$. 
The values of $a_{lm}$ and $b_{lm}$ for these grains are given in 
Table \ref{tab:grs-coeffs}.  The resulting shape for
grain 1 is displayed in Fig. \ref{fig:grs1}.  

\begin{table}
\caption{GRS expansion coefficients.  The full table is available online.} 
\label{tab:grs-coeffs}
\begin{tabular}{lllll}
\hline
Grain & $l$ & $m$ & $a_{lm}$ & $b_{lm}$\\
\hline
1 & 1 & 0 &  0.566669585    & -0.0851941355 \\
1 & 1 & 1 & -0.36704101     & 0.213834444 \\
1 & 2 & 0 & -0.0535803218   & -0.0895122548 \\
1 & 2 & 1 & -0.00556386957  & -0.00870338194 \\
1 & 2 & 2 & -0.0668200056   & 0.0882865188 \\
\hline
\end{tabular}
\end{table}

The volume $V$, coordinates of the centre of
mass $x_{i, \mathrm{cm}}$, and components of
the inertia tensor $I_{ij}$ are given by
\be
V = \frac{1}{3} \int_0^{2 \pi} \mathrm{d}\phi \int_{-1}^1 \mathrm{d}(\cos \theta) 
[r_{\mathrm{surf}}(\theta, \phi)]^3 ,
\ee
\be
x_{i, \mathrm{cm}} = \frac{1}{4V} \int_0^{2\pi} \mathrm{d}\phi \int_{-1}^1 
\mathrm{d}(\cos \theta) 
[r_{\mathrm{surf}}(\theta, \phi)]^3 x_{i, \mathrm{surf}}(\theta, \phi) ,
\ee
\be
\label{eq:inertia-tensor1}
I_{ii} = \int_0^{2\pi} \mathrm{d}\phi \int_{-1}^1 \mathrm{d}(\cos \theta) 
\int_0^{r_{\mathrm{surf}}(\theta, \phi)} r^2 \mathrm{d}r [(x_j - x_{j, \mathrm{cm}})^2 + 
(x_k - x_{k, \mathrm{cm}})^2] , 
\ee
and
\be
I_{ij} = - \int_0^{2\pi} \mathrm{d}\phi \int_{-1}^1 \mathrm{d}(\cos \theta) 
\int_0^{r_{\mathrm{surf}}(\theta, \phi)} r^2 \mathrm{d}r \, x_i x_j ,
\ee
where $x_1 = x$, $x_2 = y$, and $x_3 = z$.  
In equation (\ref{eq:inertia-tensor1}), $j$ and $k$ stand for the two 
index values that are not equal to the value of index $i$.

For a given $(\theta, \phi)$, the direction normal to the grain 
surface is found by taking the cross product of the tangent vectors 
$\mathbfit{T}_{\theta}$ and $\mathbfit{T}_{\phi}$ along 
$\bmath{\hat{\theta}}$ and $\bmath{\hat{\phi}}$, respectively.  For a GRS,
\begin{eqnarray}
\mathbfit{T}_{\theta} \bmath{\times} \mathbfit{T}_{\phi} & = &
[r_{\mathrm{surf}}(\theta, \phi)]^2 \sin \theta \biggl\{ \left[ \sin \theta \cos 
\phi +  w_2(\theta, \phi) \sin \theta \cos \theta \cos \phi + 
w_3 (\theta, \phi) \csc \theta \sin \phi \right] \bmath{\hat{x}} \nonumber \\
& & + \left[ \sin \theta \sin \phi + w_2(\theta, \phi) \sin \theta \cos 
\theta \sin \phi - w_3(\theta, \phi) \csc \theta \cos \phi \right]
\bmath{\hat{y}} + \left[ \cos \theta - w_2(\theta, \phi) \sin^2 \theta
\right] \bmath{\hat{z}} \biggr\}
\label{eq:normal}
\end{eqnarray}
where
\be
w_2(\theta, \phi) = \sum_{l=1}^{l_{\mathrm{max}}} \sum_{m=0}^l \frac{
\mathrm{d}P_l^m(\cos \theta)}{\mathrm{d}(\cos \theta)} \,
(a_{lm} \cos m \phi + b_{lm} \sin m \phi) 
\ee
and 
\be
w_3(\theta, \phi) = \sum_{l=1}^{l_{\mathrm{max}}} \sum_{m=0}^l m P_l^m(\cos 
\theta) \, (- a_{lm} \sin m \phi + b_{lm} \cos m \phi) .
\ee
The surface area of the grain is given by
\be
S = \int_0^{\pi} \mathrm{d}\theta \int_0^{2\pi} \mathrm{d}\phi \, 
|\mathbfit{T}_{\theta} \bmath{\times} \mathbfit{T}_{\phi}| 
= \int_{-1}^1 \mathrm{d}(\cos \theta) \int_0^{2 \pi} \mathrm{d}\phi \, 
\eta_S(\theta, \phi)
\ee
where
\be
\eta_S(\theta, \phi) = [r_{\mathrm{surf}}(\theta, \phi)]^2
\left\{1 + [w_2(\theta, \phi) \sin \theta]^2 + [w_3(\theta, \phi) \csc 
\theta]^2 \right\}^{1/2} .
\ee
The outward-pointing unit normal to the grain surface is given by
\be
\label{eq:n-hat}
\bmath{\hat{N}} = \frac{(\mathbfit{T}_{\theta} \bmath{\times} 
\mathbfit{T}_{\phi}) \, \nu_N}{\sin \theta \, \eta_S}
\ee
where $\nu_N = 1$ if 
$\bmath{\hat{r}} \bmath{\cdot} (\mathbfit{T}_{\theta} \bmath{\times} 
\mathbfit{T}_{\phi}) > 0$
and $\nu_N = -1$ if 
$\bmath{\hat{r}} \bmath{\cdot} (\mathbfit{T}_{\theta} \bmath{\times} 
\mathbfit{T}_{\phi}) < 0$.
  
The effective radius $a_{\mathrm{eff}}$ is defined as the radius of a sphere
with volume equal to that of the grain:
\be
a_{\mathrm{eff}} = \left( \frac{3V}{4 \pi} \right)^{1/3} .
\ee
The grain principal axes are denoted $\bmath{\hat{a}}_i$ such that the 
associated moments of inertia satisfy $I_1 \ge I_2 \ge I_3$.  We define shape
parameters $\alpha_i$ by
\be
I_i = \frac{2}{5} \alpha_i \rho V a_{\mathrm{eff}}^2
\ee
where $\rho$ is the density of the grain material, assumed to be constant
throughout the grain volume.  

We take 8000 values (each) of $\cos \theta$ and $\phi$ in performing
the integrals for $V$, $S$, and $x_{i, \mathrm{cm}}$ and 4000 values (each) of
$\cos \theta$, $\phi$, and $r$ in performing the integrals for the
inertia tensor.  We use the recurrence relation
\be
(l-m) P_l^m(x) = (2l-1) x P_{l-1}^m(x) - (l+m-1) P_{l-2}^m(x) 
\ee
and the expressions
\be
P_m^m(x) = (-1)^m (2m-1)!! (1-x^2)^{m/2} 
\ee
and
\be
P_{m+1}^m(x) = (2m+1) x P_m^m(x)
\label{eq:rec-rel2}
\ee
to efficiently compute $P_l^m(\cos \theta)$ and 
$\mathrm{d}P_l^m(\cos \theta)/\mathrm{d}(\cos \theta)$ for the 44 combinations 
$(l,m)$ with $l=1$ through $l=8$.  During the volume integration, we keep 
track of the largest value of $r_{\mathrm{surf}}(\theta, \phi)$, which we denote 
$r_{\mathrm{max}}$.

The derived quantities that characterize the 13 grains examined in this study  
are given in Tables \ref{tab:grs-derived-quantities} and 
\ref{tab:grs-principal-axes}.  

\begin{table}
\caption{GRS derived quantities.}
\label{tab:grs-derived-quantities}
\begin{tabular}{llllllll}
\hline
Grain & $V a^{-3} (1+\sigma^2)^{3/2}$ & $r_{\mathrm{max}}/a_{\mathrm{eff}}$ &
$S/(4 \pi a^2_{\mathrm{eff}})$ & $\mathbfit{r}_{\mathrm{cm}}/a_{\mathrm{eff}}$ &
$\alpha_1$ & $\alpha_2$ & $\alpha_3$ \\
\hline
1 & 10.568 & 2.2349 & 1.3430 & (0.28992, -0.19559, 0.50464) & 1.7130 &
1.2900 & 1.0919 \\
2 & 7.3329 & 2.4285 & 1.4011 & (0.02461, 0.19339, -0.44991) & 2.1253 & 
1.9392 & 0.81156 \\
3  & 8.7756 &  1.9046 &  1.2935 & (0.12517, -0.31383, 0.20120) & 1.8185 &  
1.5170 & 0.82255 \\
4 & 16.336 & 1.9635 & 1.2528 & (-0.59578, -0.05017, 0.19873) & 1.9565 &  
1.9084 & 0.69020 \\
5 & 6.8640 & 1.9173 & 1.2684 & (0.21208, -0.32522, -0.30808) & 1.4298 &  
1.3524 & 0.81991 \\
6 & 6.6200 & 2.1619 & 1.2632 & (-0.12706, 0.46337, 0.18607) & 1.4923 &  
1.3382 & 0.88205 \\
7 & 7.4806 & 2.2211 & 1.3596 & (0.08203, -0.44095, -0.12676) & 1.8843 &  
1.4960 & 0.89730 \\
8 & 7.4378 &  1.6693 & 1.1266 & (0.33362, 0.18462, -0.28530) & 1.5227 &  
1.1675 & 0.85452 \\
9 & 10.468 & 2.1160 & 1.1180 & (0.45635, 0.09777, -0.47634) & 1.4005 &  
1.2106 & 0.87044 \\
10 & 7.0687 & 1.8321 & 1.0893 & (-0.26718, 0.17087, 0.36227) & 1.3109 &  
1.1339 & 0.86201 \\
11 & 8.6542 & 1.9723 & 1.1240 & (0.13915, 0.28140, 0.48741) & 1.4962 &  
1.1479 & 0.87868 \\
12 & 12.710 & 2.1524 & 1.0968 & (-0.74042, 0.25973, 0.17926) & 1.4447 &  
1.3281 & 0.73723 \\
13 & 5.6324 & 1.5987 & 1.1207 & (0.02609, -0.19623, 0.08627) & 1.5328 &  
1.4072 & 0.72268 \\
\hline
\end{tabular}
\end{table}

\begin{table}
\caption{GRS principal axes.}
\label{tab:grs-principal-axes}
\begin{tabular}{llll}
\hline
Grain & $\bmath{\hat{a}}_1$ & $\bmath{\hat{a}}_2$ & $\bmath{\hat{a}}_3$ \\
\hline
1 & (0.86599, -0.10272, -0.48939) & (0.49815, 0.26264, 0.82636) &
(0.04366, -0.95941, 0.27861) \\
2 & (0.47329, 0.81876, 0.32501) & (0.83196, -0.53673, 0.14058) & 
(0.28954, 0.20386, -0.93520) \\
3 & (0.66304, -0.08554, 0.74368) & (0.74508, 0.17139, -0.64457) & 
(-0.07232, 0.98148, 0.17737) \\
4 & (0.22458, -0.27127, -0.93593) & (0.76316, 0.64620, -0.00417) & 
(0.60593, -0.71333, 0.35215) \\
5 & (0.16410, 0.69832, -0.69673) & (0.66826, 0.44084, 0.59924) & 
(0.72561, -0.56393, -0.39431) \\
6 & (0.66350, -0.35368, 0.65930) & (0.74549, 0.23798, -0.62258) & 
(0.06330, 0.90458, 0.42157) \\
7 & (0.41190, 0.05621, -0.90949) & (0.85298, -0.37491, 0.36314) & 
(-0.32056, -0.92536, -0.20237) \\
8 & (0.29777, 0.86430, 0.40535) & (0.47572, 0.23379, -0.84796) & 
(-0.82766, 0.44533, -0.34155) \\
9 & (0.54570, 0.83219, -0.09833) & (0.71475, -0.40099, 0.57301) & 
(0.43742, -0.38297, -0.81363) \\
10 & (0.69300, -0.72083, 0.01197) & (0.61914, 0.60358, 0.50235) & 
(-0.36933, -0.34072, 0.86458) \\
11 & (0.94710, 0.31792, 0.04384) & (0.01855, -0.19059, 0.98149) & 
(0.32039, -0.92876, -0.18641) \\
12 & (0.49703, -0.30507, 0.81234) & (0.60056, 0.79666, -0.06827) & 
(-0.62633, 0.52179, 0.57918) \\
13 & (0.47866, -0.35395, -0.80349) & (0.33235, 0.92009, -0.20733) & 
(0.81267, -0.16780, 0.55804) \\
\hline
\end{tabular}
\end{table}

\section{Torque Calculations:  Theory} \label{sec:torque-theory}

\subsection{Collisions of gas particles with the grain}

The first step in evaluating the effects of gas-particle collisions with 
a grain is to determine which gas-particle trajectories, as observed in 
the rest frame of the grain, hit the grain.  
To this end, consider an enclosing sphere that is at rest relative to the GRS, 
is centred on the origin used in constructing the GRS, and has a radius 
$r_{\mathrm{sph}}$ that exceeds the maximum value of 
$r_{\mathrm{surf}}(\theta, \phi)$.
When a gas particle strikes the enclosing sphere, its position 
$\mathbfit{r}_0 = r_{\mathrm{sph}} \bmath{\hat{r}}$ is represented
by spherical coordinates $(r_{\mathrm{sph}}, \theta_{\mathrm{sph}}, \phi_{\mathrm{sph}})$
and its velocity
$\mathbfit{v} = v_{\mathrm{th}} s \, \bmath{\hat{s}}$ 
is represented by spherical coordinates 
$(v_{\mathrm{th}} s, \theta_{\mathrm{in}}, \phi_{\mathrm{in}})$, where
the radial vector 
$\bmath{\hat{r}}$ is the reference axis for the polar angle 
$\theta_{\mathrm{in}}$, the vector $\bmath{\hat{\theta}}$ is the reference axis 
for the azimuthal
angle $\phi_{\mathrm{in}}$, and the ``reduced speed'' $s$ is the particle's 
speed divided by the gas thermal speed, 
\be
\label{eq:v-th}
v_{\mathrm{th}} = \left( \frac{2 k T_{\mathrm{gas}}}{m} \right)^{1/2};
\ee
$k$ is Boltzmann's constant, $T_{\mathrm{gas}}$ is the gas temperature, and
$m$ is the mass of the gas particle.  Explicitly,
\be
\label{eq:r-hat}
\bmath{\hat{r}} = \sin \theta_{\mathrm{sph}} \cos \phi_{\mathrm{sph}} \, 
\bmath{\hat{x}} + \sin \theta_{\mathrm{sph}} \sin \phi_{\mathrm{sph}} \, 
\bmath{\hat{y}} + \cos \theta_{\mathrm{sph}} \, \bmath{\hat{z}} , 
\ee
\be
\label{eq:s-hat}
\bmath{\hat{s}}  =  -
(\sin \theta_{\mathrm{in}} \cos \phi_{\mathrm{in}} \, \bmath{\hat{\theta}} +
\sin \theta_{\mathrm{in}} \sin \phi_{\mathrm{in}} \, \bmath{\hat{\phi}} + \cos
\theta_{\mathrm{in}} \, \bmath{\hat{r}}) .
\ee

Assuming $T_{\mathrm{gas}} \ga 20 \,$K, 
$v_{\mathrm{th}} \ga 6 \times 10^4 \, \mathrm{cm} \, \mathrm{s}^{-1}$
for incoming H atoms.  Even for highly suprathermal rotation, we expect
grain angular rotational speeds 
$\omega \la 10^7 \, \mathrm{rad} \, \mathrm{s}^{-1}$ \citep{DW97}, 
corresponding to linear speeds of the grain surface  
$v_{\mathrm{surf}} \sim \omega a_{\mathrm{eff}} \la 2 \times 10^2 
\, \mathrm{cm} \, \mathrm{s}^{-1}$ for 
$a_{\mathrm{eff}} \approx 0.2 \, \mu \mathrm{m}$.  Since 
$v_{\mathrm{surf}} \ll v_{\mathrm{th}}$, the grain rotation can be neglected 
during the time that the gas particle traverses the enclosing sphere.
In other words, we neglect the 
curvature of the gas-particle trajectories as observed in the rest frame
of the grain.  

A gas particle that approaches the grain and enclosing sphere along
a radial path has $\theta_{\mathrm{in}} = 0$ and will hit the grain.  A 
gas particle that approaches with $\theta_{\mathrm{in}} = \pi/2$ will 
not hit the grain.  By construction, there is a unique distance from the 
origin to the grain surface for each direction $(\theta, \phi)$.  Thus,
for each set of angles $(\theta_{\mathrm{sph}}, \phi_{\mathrm{sph}}, 
\phi_{\mathrm{in}})$, there is a 
critical value $u_c$ of $\cos \theta_{\mathrm{in}}$
such that a gas particle hits the grain when 
$\cos \theta_{\mathrm{in}} \ge u_c$ and does not hit when 
$\cos \theta_{\mathrm{in}} < u_c$.  Our computational approach for determining
$u_c$ as a function of $(\theta_{\mathrm{sph}}, \phi_{\mathrm{sph}}, \phi_{\mathrm{in}})$
is described in \S \ref{sec:incoming-trajectories}.

\subsection{Torque due to incoming and reflected atoms}

In this section, we calculate the torque due to gas particles (hereafter 
referred to as ``atoms'', though the analysis is equally valid for 
molecules) that
strike the grain, assuming that they stick to the grain or reflect
specularly.  In the next section we will examine the torque associated
with atoms or molecules that depart the grain after sticking to the
surface.  

Consider atoms with mass $m$ and number density $n$ in a gas with 
temperature $T_{\mathrm{gas}}$.  An atom's velocity in the rest frame of 
the gas is $\mathbfit{v}_g = v_{\mathrm{th}} \, \mathbfit{s}_g$; the thermal speed
$v_{\mathrm{th}}$ was defined in equation (\ref{eq:v-th}).  
The Maxwell velocity distribution is
\be
P_g(\mathbfit{s}_g) s_g^2 \, \mathrm{d}s_g \, \mathrm{d}\Omega_g 
= \pi^{-3/2} \exp(-s_g^2) s_g^2 \, \mathrm{d}s_g \, \mathrm{d}\Omega_g
\ee
where $\mathrm{d}\Omega_g$ is the solid angle element.  The subscript ``$g$''
indicates that the quantities are evaluated in the rest frame of the gas.

Now suppose the grain moves through the gas with velocity 
$v_{\mathrm{th}} \mathbfit{s}_d$ where
\be
\label{eq:sd}
\mathbfit{s}_d = s_d (\sin \theta_{\mathrm{gr}} \cos \phi_{\mathrm{gr}} \, 
\bmath{\hat{x}} + \sin \theta_{\mathrm{gr}} \sin \phi_{\mathrm{gr}} \, 
\bmath{\hat{y}} + \cos \theta_{\mathrm{gr}} \, \bmath{\hat{z}}).
\ee
The reduced velocity of the gas atom 
as observed in the rest frame of the grain is 
$\mathbfit{s} = \mathbfit{s}_g - \mathbfit{s}_d$.  Since $\mathbfit{s}_d$ is 
constant, 
\be
s^2 \mathrm{d}s \, \mathrm{d}\Omega = s_g^2 \mathrm{d}s_g \, \mathrm{d}\Omega_g
\ee
and the distribution of atom velocities as observed in the rest frame
of the grain is
\be
P(\mathbfit{s}) s^2 \mathrm{d}s \, \mathrm{d}\Omega = \pi^{-3/2} \exp(-
|\mathbfit{s} + \mathbfit{s}_d|^2) s^2 \mathrm{d}s \, \mathrm{d}\Omega
\ee
with
\be
|\mathbfit{s} + \mathbfit{s}_d|^2 = s^2 + s_d^2 - 2 \beta s s_d 
\ee
and
\be
\beta = \beta_1 \cos \theta_{\mathrm{in}} + \beta_2 \sin \theta_{\mathrm{in}}
\ee
with
\be
\beta_1 = \sin \theta_{\mathrm{sph}} \sin \theta_{\mathrm{gr}} \cos (\phi_{\mathrm{sph}}
- \phi_{\mathrm{gr}}) + \cos \theta_{\mathrm{sph}} \cos \theta_{\mathrm{gr}} 
\ee
and 
\be
\beta_2 = 
[\cos \theta_{\mathrm{sph}} \sin \theta_{\mathrm{gr}} \cos (\phi_{\mathrm{sph}} - 
\phi_{\mathrm{gr}}) - \sin \theta_{\mathrm{sph}} \cos \theta_{\mathrm{gr}}] \cos 
\phi_{\mathrm{in}} - \sin \theta_{\mathrm{gr}} \sin (\phi_{\mathrm{sph}} - 
\phi_{\mathrm{gr}}) \sin \phi_{\mathrm{in}}.
\ee
Note that $-1 \le \beta \le 1$.  It is convenient to define the functions
\be
\label{eq:i-s}
I_s(p, s_d, \beta) = \int_0^{\infty} \mathrm{d}s \, s^p \exp[-(s^2 + s_d^2 - 2 
\beta s s_d)] .
\ee

The rate at which gas atoms arrive at a surface element on the 
enclosing sphere with area 
$r_{\mathrm{sph}}^2 \mathrm{d}(\cos \theta_{\mathrm{sph}}) \, \mathrm{d}
\phi_{\mathrm{sph}}$,
with reduced speeds between $s$ and $s+\mathrm{d}s$ and from within a solid 
angle element $\mathrm{d}(\cos \theta_{\mathrm{in}}) \, \mathrm{d}\phi_{\mathrm{in}}$
around direction $(\theta_{\mathrm{in}}, \phi_{\mathrm{in}})$, is
\be
\label{eq:dR}
\mathrm{d}R_{\mathrm{arr}} = n v_{\mathrm{th}} [\mathbfit{s} \bmath{\cdot} 
(-\bmath{\hat{r}})] \pi^{-3/2} s^2 
\exp[-(s^2 + s_d^2 - 2 \beta s s_d)] \, \mathrm{d}s \, \mathrm{d}(\cos 
\theta_{\mathrm{in}}) \, \mathrm{d}\phi_{\mathrm{in}} \, r_{\mathrm{sph}}^2 \, 
\mathrm{d}(\cos \theta_{\mathrm{sph}}) \, \mathrm{d}\phi_{\mathrm{sph}};
\ee
\be
\mathbfit{s} \bmath{\cdot} (- \bmath{\hat{r}}) = s \cos \theta_{\mathrm{in}}.
\ee
The total rate at which gas atoms strike the grain (to be used
in \S \ref{sec:torque-outgoing}) is thus
\be
R_{\mathrm{arr}} = n v_{\mathrm{th}} a_{\mathrm{eff}}^2 Q_{\mathrm{arr}}
\ee
with
\be
Q_{\mathrm{arr}} =  \pi^{-3/2} \left( \frac{r_{\mathrm{sph}}}{a_{\mathrm{eff}}} 
\right)^2
\int_{-1}^1 \mathrm{d}(\cos \theta_{\mathrm{sph}}) \int_0^{2\pi} \mathrm{d}
\phi_{\mathrm{sph}} \int_0^{2\pi} \mathrm{d}\phi_{\mathrm{in}} \int_{u_c}^1 \mathrm{d}
(\cos \theta_{\mathrm{in}}) \cos \theta_{\mathrm{in}} I_s(3, s_d, \beta) .
\ee

Each atom that strikes and sticks to the grain transfers angular momentum
(relative to the grain's centre of mass) 
$\Delta \mathbfit{J}_{\mathrm{arr}} = 
m (\mathbfit{r}_0 - \mathbfit{r}_{\mathrm{cm}}) \bmath{\times} v_{\mathrm{th}} 
\mathbfit{s}$.
The mean torque due to arriving atoms is
$\bmath{\Gamma}_{\mathrm{arr}} = 
\int \mathrm{d}R_{\mathrm{arr}} \, \Delta \mathbfit{J}_{\mathrm{arr}}$.  Thus,
\be
\bmath{\Gamma}_{\mathrm{arr}} = \pi^{-3/2} m n v_{\mathrm{th}}^2 r_{\mathrm{sph}}^3
\int_{-1}^1 \mathrm{d}(\cos \theta_{\mathrm{sph}}) \int_0^{2\pi} \mathrm{d}
\phi_{\mathrm{sph}} \int_0^{2\pi} \mathrm{d}\phi_{\mathrm{in}} \int_{u_c}^1 \mathrm{d}
(\cos \theta_{\mathrm{in}}) \int_0^{\infty} \mathrm{d}s \, s^2 
\exp[-(s^2 + s_d^2 - 2 \beta s s_d)] \, s \cos \theta_{\mathrm{in}}
\left( \bmath{\hat{r}} - \frac{\mathbfit{r}_{\mathrm{cm}}}{r_{\mathrm{sph}}} \right) 
\bmath{\times} \mathbfit{s} .
\ee
Expressing the mean torque in terms of an efficiency factor
$\mathbfit{Q}_{\Gamma, \mathrm{arr}}$,
\be
\bmath{\Gamma}_{\mathrm{arr}} = m n v_{\mathrm{th}}^2 a_{\mathrm{eff}}^3
\mathbfit{Q}_{\Gamma, \mathrm{arr}}
\ee
with 
\be
\mathbfit{Q}_{\Gamma, \mathrm{arr}} = \pi^{-3/2} \left( \frac{r_{\mathrm{sph}}}
{a_{\mathrm{eff}}} \right)^3 \int_{-1}^1 \mathrm{d}(\cos \theta_{\mathrm{sph}}) 
\int_0^{2\pi} \mathrm{d}\phi_{\mathrm{sph}} \int_0^{2\pi} \mathrm{d}\phi_{\mathrm{in}} 
\int_{u_c}^1 \mathrm{d}(\cos \theta_{\mathrm{in}}) \cos \theta_{\mathrm{in}} 
\left( \bmath{\hat{r}} - \frac{\mathbfit{r}_{\mathrm{cm}}}{r_{\mathrm{sph}}} 
\right) \bmath{\times} \bmath{\hat{s}} \, I_s(4, s_d, \beta) .
\label{eq:Q-Gamma-arr}
\ee
See Appendix \ref{sec:calc-details-incoming} for explicit expressions for
$[(\bmath{\hat{r}} - \mathbfit{r}_{\mathrm{cm}}/r_{\mathrm{sph}}) \bmath{\times} 
\bmath{\hat{s}}]_i$.

We calculate the components of the mean 
torque along the $\bmath{\hat{x}}$, $\bmath{\hat{y}}$, and
$\bmath{\hat{z}}$ directions that are fixed relative to the grain body.  
Of course, these are identical to the components in an inertial frame with
basis vectors that are instantaneously aligned with those of the grain frame.
When these quantities are used to examine the grain rotational
dynamics, they will be transformed to a single inertial frame and averaged
over the grain rotation.  

Now consider the case that atoms reflect specularly from the grain 
surface.  Following a 
reflection, the atom may escape the grain or strike the grain surface at
another location.  In the latter case, the atom undergoes another 
specular reflection; this continues until the atom ultimately escapes the
grain.

Since the speed of the atom does not change upon reflection,
the recoil angular momentum delivered to the grain is
\be
\Delta \mathbfit{J}_{\mathrm{spec}} = - m v_{\mathrm{th}} s r_{\mathrm{sph}} \left(
\bmath{\hat{r}}_f - \frac{\mathbfit{r}_{\mathrm{cm}}}{r_{\mathrm{sph}}} \right) 
\bmath{\times} \bmath{\hat{s}}_f
\ee
where $r_{\mathrm{sph}} \, \bmath{\hat{r}}_f$ and $s \bmath{\hat{s}}_f$ are the 
final position and
reduced velocity of the reflected atom as it leaves the enclosing sphere.
Thus, the mean recoil torque associated with 
specular reflection is
\be
\bmath{\Gamma}_{\mathrm{spec}} = m n v_{\mathrm{th}}^2 a_{\mathrm{eff}}^3
\mathbfit{Q}_{\Gamma, \mathrm{spec}} 
\ee
with 
\be
\mathbfit{Q}_{\Gamma, \mathrm{spec}} = - \pi^{-3/2} \left( \frac{r_{\mathrm{sph}}}
{a_{\mathrm{eff}}} \right)^3 \int_{-1}^1 \mathrm{d}(\cos \theta_{\mathrm{sph}}) 
\int_0^{2\pi} \mathrm{d}\phi_{\mathrm{sph}} \int_0^{2\pi}
\mathrm{d}\phi_{\mathrm{in}} \int_{u_c}^1 \mathrm{d}(\cos \theta_{\mathrm{in}})
\cos \theta_{\mathrm{in}} 
\left( \bmath{\hat{r}}_f - \frac{\mathbfit{r}_{\mathrm{cm}}}{r_{\mathrm{sph}}} 
\right) \bmath{\times} \bmath{\hat{s}}_f \, I_s(4, s_d, \beta) .
\label{eq:Q-Gamma-spec}
\ee

\subsection{Mechanical torque due to outgoing atoms or molecules
\label{sec:torque-outgoing}}

We assume that the rate at which H atoms depart the grain (either
in atomic form or as part of an H$_2$ molecule) equals the rate at which
they arrive at the grain.  In this section, we consider only particles that
stick to the grain surface upon arrival (as opposed to those that reflect
specularly).  We further assume that these outgoing particles
depart along the direction $\bmath{\hat{N}}(\theta, \phi)$
normal to the grain surface (see equation \ref{eq:n-hat}).  In order to keep
the computational time manageable, we do not consider 
a distribution of outgoing directions for atoms/molecules that have been 
accommodated on the grain surface.  We consider the following scenarios
for the departing particles.

(1)  Atoms or molecules depart from an arbitrary location on the grain surface.
The rate of departure from a surface element is proportional to its area.

(2)  Atoms or molecules depart from approximately the same location where
they arrived on the grain surface.  

In future work we will also examine the case that 
molecules depart from a set of special sites of molecule formation on
the grain surface.  

The angular momentum imparted to the grain when an atom or molecule departs
is
\be
\Delta \mathbfit{J}_{\mathrm{out}} = - m_{\mathrm{out}}
 v_{\mathrm{out}} a_{\mathrm{eff}} \left(
\frac{\mathbfit{r}_{\mathrm{surf}} - \mathbfit{r}_{\mathrm{cm}}}{a_{\mathrm{eff}}} 
\right) \bmath{\times} \bmath{\hat{N}} 
\ee
where $ m_{\mathrm{out}}$ and $v_{\mathrm{out}}$ are the mass and speed of the
outgoing particle, respectively.  
Next, we introduce a function $\kappa_{\mathrm{esc}}(\theta, \phi)$ such that
$\kappa_{\mathrm{esc}} = 1$ if an atom or molecule that departs the surface 
along $\bmath{\hat{N}}$ at $(\theta, \phi)$ escapes to infinity and 
$\kappa_{\mathrm{esc}} = 0$ if the departing particle instead strikes the 
grain at another location on the surface.  

For scenario (1), the rate at which particles depart a surface element is
\be
\mathrm{d}R_{\mathrm{out}, (1)} = g R_{\mathrm{arr}} S_{\mathrm{esc}}^{-1} 
\kappa_{\mathrm{esc}}(\theta, \phi) \mathrm{d}(\cos \theta) \mathrm{d}\phi \, 
\eta_S(\theta, \phi) ,
\ee
where $g = 1$ if the departing species is an H atom and $g=1/2$ if it is
an H$_2$ molecule;
\be
S_{\mathrm{esc}} = \int_{-1}^1 \mathrm{d}(\cos \theta) \int_0^{2\pi} \mathrm{d}\phi 
\, \eta_S(\theta, \phi) \kappa_{\mathrm{esc}}(\theta, \phi) .
\ee
Thus, the mean torque is
\be
\bmath{\Gamma}_{\mathrm{out}, (1)} = g m_{\mathrm{out}} n v_{\mathrm{th}} v_{\mathrm{out}} 
a_{\mathrm{eff}}^3 \mathbfit{Q}_{\Gamma, \mathrm{out}, (1)} = m  n v_{\mathrm{th}} 
v_{\mathrm{out}} a_{\mathrm{eff}}^3 \mathbfit{Q}_{\Gamma, \mathrm{out}, (1)} 
\ee
with 
\be
\label{eq:Q-Gamma-out-1}
\mathbfit{Q}_{\Gamma, \mathrm{out}, (1)} = - Q_{\mathrm{arr}} S_{\mathrm{esc}}^{-1} 
\int_{-1}^1 \mathrm{d}(\cos \theta) \int_0^{2\pi} \mathrm{d}\phi \, \eta_S(\theta,
\phi) \kappa_{\mathrm{esc}}(\theta, \phi) \left(
\frac{\mathbfit{r}_{\mathrm{surf}} - \mathbfit{r}_{\mathrm{cm}}}{a_{\mathrm{eff}}} 
\right) \bmath{\times} \bmath{\hat{N}} .
\ee
The relation $g m_{\mathrm{out}} = m$ follows from the assumption that H atoms
depart the grain at the same rate at which they arrive.  

For scenario (2), consider a gas-phase atom arriving at the enclosing 
sphere at $(\theta_{\mathrm{sph}}, \phi_{\mathrm{sph}}, \phi_{\mathrm{in}}, 
\theta_{\mathrm{in}})$.  
After arriving at the grain surface, it departs along the surface normal.
Its path either takes it away from the grain (beyond $r_{\mathrm{sph}}$)
or intersects the grain at another surface location, from which it then
departs along the local normal.  After some number
of surface intersections, the departing particle hits the surface at a
location $(\theta^{\prime}, \phi^{\prime})$ such that its path along 
$\bmath{\hat{N}}(\theta^{\prime}, \phi^{\prime})$ takes it away from the grain.  
Thus, the mean torque due to outgoing atoms or molecules in scenario (2) is
\be
\bmath{\Gamma}_{\mathrm{out}, (2)} = g \int \mathrm{d}R_{\mathrm{arr}}
(\theta_{\mathrm{sph}}, \phi_{\mathrm{sph}}, \phi_{\mathrm{in}}, \theta_{\mathrm{in}}) 
\Delta \mathbfit{J}_{\mathrm{out}}( \theta^{\prime}, \phi^{\prime}) 
= m n v_{\mathrm{th}} v_{\mathrm{out}} a_{\mathrm{eff}}^3 
\mathbfit{Q}_{\Gamma, \mathrm{out}, (2)}  
\ee
with 
\be
\mathbfit{Q}_{\Gamma, \mathrm{out}, (2)} = - \pi^{-3/2} \left( \frac{r_{\mathrm{sph}}}
{a_{\mathrm{eff}}} \right)^2 \int_{-1}^1 \mathrm{d}(\cos \theta_{\mathrm{sph}}) 
\int_0^{2\pi} \mathrm{d}\phi_{\mathrm{sph}} \int_0^{2\pi} \mathrm{d}\phi_{\mathrm{in}} 
\int_{u_c}^1 \mathrm{d}(\cos \theta_{\mathrm{in}}) \cos \theta_{\mathrm{in}} 
\left[ \frac{\mathbfit{r}_{\mathrm{surf}}(\theta^{\prime}, \phi^{\prime}) - 
\mathbfit{r}_{\mathrm{cm}}}{a_{\mathrm{eff}}} \right] \bmath{\times} 
\bmath{\hat{N}}(\theta^{\prime}, \phi^{\prime}) \, I_s(3, s_d, \beta) .
\ee

\subsection{Total mechanical torque}

If a fraction $f_{\mathrm{spec}}$ of the gas-phase atoms that strike the grain
surface reflect specularly, then the total mechanical torque is
\be
\label{eq:total-torque}
\bmath{\Gamma}_{\mathrm{mech}} = m n v^2_{\mathrm{th}} a_{\mathrm{eff}}^3 
\mathbfit{Q}_{\Gamma, \mathrm{mech}} 
\ee
with 
\be
\label{eq:q-total-torque}
\mathbfit{Q}_{\Gamma, \mathrm{mech}} = \mathbfit{Q}_{\Gamma, \mathrm{arr}} + 
f_{\mathrm{spec}}
\mathbfit{Q}_{\Gamma, \mathrm{spec}}  + (1 - f_{\mathrm{spec}}) \frac{v_{\mathrm{out}}}
{v_{\mathrm{th}}} \, \mathbfit{Q}_{\Gamma, \mathrm{out}}
\ee
where $\mathbfit{Q}_{\Gamma, \mathrm{out}}$ is the efficiency factor for one of the
scenarios (1 or 2) for outgoing particles.  

\subsection{Rotational averaging \label{sec:suprathermal}}

We assume that the grain rotates steadily about $\bmath{\hat{a}}_1$, as is 
appropriate for suprathermal rotation, and average the torque efficiency 
factors over this rotation.  

Consider a coordinate system $(x_v, y_v, z_v)$ fixed in space with
$\bmath{\hat{z}}_v$ along the direction of the grain's velocity and with 
$\bmath{\hat{a}}_1$ lying in the $x_v$-$z_v$ plane.  From equation 
(\ref{eq:sd}),
\be
\label{eq:exp1}
\sin \theta_{\mathrm{gr}} \cos \phi_{\mathrm{gr}} = \bmath{\hat{z}}_v 
\bmath{\cdot} \bmath{\hat{x}} ,
\ee
\be
\label{eq:exp2}
\sin \theta_{\mathrm{gr}} \sin \phi_{\mathrm{gr}} = \bmath{\hat{z}}_v 
\bmath{\cdot} \bmath{\hat{y}} ,
\ee
\be
\label{eq:exp3}
\cos \theta_{\mathrm{gr}} = \bmath{\hat{z}}_v \bmath{\cdot} \bmath{\hat{z}} .
\ee

Take the angle between $\bmath{\hat{v}}_{\mathrm{gr}}$ and $\bmath{\hat{a}}_1$ 
to be $\theta_{va}$.  Since $\bmath{\hat{a}}_1$ lies in the $x_v$-$z_v$ plane,
\be
\label{eq:a1a}
\bmath{\hat{a}}_1 = \sin \theta_{va} \, \bmath{\hat{x}}_v + \cos \theta_{va} \, 
\bmath{\hat{z}}_v .
\ee
Next introduce angle $\Phi_2$ to describe the rotation of $\bmath{\hat{a}}_2$ 
about $\bmath{\hat{a}}_1$.  Define it such that $\bmath{\hat{a}}_2$ lies along 
the $y_v$-axis when $\Phi_2 = 0$ and in the $x_v$-$z_v$ plane when 
$\Phi_2 = \pi/2$.  Specifically, $\bmath{\hat{a}}_2 = \bmath{\hat{y}}_v$ when 
$\Phi_2 = 0$ and $\bmath{\hat{a}}_2 = - \cos \theta_{va} \, \bmath{\hat{x}}_v 
+ \sin \theta_{va} \, \bmath{\hat{z}}_v$ when $\Phi_2 = \pi/2$.  Thus, 
\be
\label{eq:a2a}
\bmath{\hat{a}}_2 = \cos \Phi_2 \, \bmath{\hat{y}}_v  + \sin \Phi_2 (- \cos 
\theta_{va} \, \bmath{\hat{x}}_v + \sin \theta_{va} \, \bmath{\hat{z}}_v) 
\ee
and 
\be
\label{eq:a3a}
\bmath{\hat{a}}_3 = \bmath{\hat{a}}_1 \bmath{\times} \bmath{\hat{a}}_2 = 
\cos \Phi_2 (- \cos \theta_{va} \,
\bmath{\hat{x}}_v + \sin \theta_{va} \, \bmath{\hat{z}}_v) - \sin \Phi_2 \, 
\bmath{\hat{y}}_v .
\ee
Expressing the principal axes in equations (\ref{eq:a1a})--(\ref{eq:a3a}) in 
terms of their components in the $(x, y, z)$ system,
\be
\label{eq:a1b}
a_{1x} \, \bmath{\hat{x}} + a_{1y} \, \bmath{\hat{y}} + a_{1z} \, 
\bmath{\hat{z}}
= \sin \theta_{va} \, \bmath{\hat{x}}_v + \cos \theta_{va} \, 
\bmath{\hat{z}}_v , 
\ee
\be
\label{eq:a2b}
a_{2x} \, \bmath{\hat{x}} + a_{2y} \, \bmath{\hat{y}} + a_{2z} \, \bmath{\hat{z}}
= \cos \Phi_2 \, \bmath{\hat{y}}_v  + \sin \Phi_2 (- \cos \theta_{va} \, 
\bmath{\hat{x}}_v + \sin \theta_{va} \, \bmath{\hat{z}}_v) ,
\ee
\be
\label{eq:a3b}
a_{3x} \, \bmath{\hat{x}} + a_{3y} \, \bmath{\hat{y}} + a_{3z} \, \bmath{\hat{z}}
= \cos \Phi_2 (- \cos \theta_{va} \,
\bmath{\hat{x}}_v + \sin \theta_{va} \, \bmath{\hat{z}}_v) - \sin \Phi_2 \, 
\bmath{\hat{y}}_v .
\ee
Taking the dot product of equations (\ref{eq:a1b})--(\ref{eq:a3b})
with $\bmath{\hat{x}}$ yields 
\be
\label{eq:a1x}
a_{1x} = \sin \theta_{va} (\bmath{\hat{x}}_v \bmath{\cdot} \bmath{\hat{x}}) 
+ \cos \theta_{va} (\bmath{\hat{z}}_v \bmath{\cdot} \bmath{\hat{x}}) ,
\ee
\be
a_{2x} = - \sin \Phi_2 \cos \theta_{va} (\bmath{\hat{x}}_v \bmath{\cdot} 
\bmath{\hat{x}}) + \cos \Phi_2 (\bmath{\hat{y}}_v \bmath{\cdot} 
\bmath{\hat{x}}) + \sin \Phi_2 \sin \theta_{va} (\bmath{\hat{z}}_v \bmath{\cdot}
\bmath{\hat{x}}) ,
\ee
\be
\label{eq:a3x}
a_{3x} = - \cos \Phi_2 \cos \theta_{va} (\bmath{\hat{x}}_v \bmath{\cdot} 
\bmath{\hat{x}}) - \sin \Phi_2 (\bmath{\hat{y}}_v \bmath{\cdot} 
\bmath{\hat{x}}) + \cos \Phi_2 \sin \theta_{va} (\bmath{\hat{z}}_v 
\bmath{\cdot} \bmath{\hat{x}}) .
\ee
Equations of identical structure result when taking the dot product with 
$\bmath{\hat{y}}$ or $\bmath{\hat{z}}$.  Solving for the dot products,
\be
\label{eq:dot1}
\bmath{\hat{x}}_v \bmath{\cdot} \bmath{\hat{x}}_i = a_{1i} \sin \theta_{va} - 
(a_{2i} \sin \Phi_2 + a_{3i} \cos \Phi_2) \cos \theta_{va} ,
\ee
\be
\bmath{\hat{y}}_v \bmath{\cdot} \bmath{\hat{x}}_i = a_{2i} \cos \Phi_2 - 
a_{3i} \sin \Phi_2 ,
\ee
\be
\label{eq:dot3}
\bmath{\hat{z}}_v \bmath{\cdot} \bmath{\hat{x}}_i = a_{1i} \cos \theta_{va} + 
(a_{2i} \sin \Phi_2 + a_{3i} \cos \Phi_2) \sin \theta_{va} .
\ee
In equations (\ref{eq:dot1})--(\ref{eq:dot3}), the subscript $i = 1$--3
denotes coordinates $x$, $y$, $z$ in the original coordinate 
system used to define the GRS.  

Substituting the expressions in equation (\ref{eq:dot3}) into 
equations (\ref{eq:exp1})--(\ref{eq:exp3}) yields 
$(\theta_{\mathrm{gr}}, \phi_{\mathrm{gr}})$ as functions of 
$(\theta_{va}, \Phi_2)$:
\be
\label{eq:rot-avg-transf1}
\sin \theta_{\mathrm{gr}} \cos \phi_{\mathrm{gr}} = 
a_{1x} \cos \theta_{va} + (a_{2x} \sin \Phi_2 + 
a_{3x} \cos \Phi_2) \sin \theta_{va} ,
\ee
\be
\sin \theta_{\mathrm{gr}} \sin \phi_{\mathrm{gr}} =
a_{1y} \cos \theta_{va} + (a_{2y} \sin \Phi_2 + 
a_{3y} \cos \Phi_2) \sin \theta_{va} ,
\ee
\be
\label{eq:rot-avg-transf3}
\cos \theta_{\mathrm{gr}} = 
a_{1z} \cos \theta_{va} + (a_{2z} \sin \Phi_2 + 
a_{3z} \cos \Phi_2) \sin \theta_{va} .
\ee

The rotationally averaged value of the scalar efficiency factor 
$Q_{\mathrm{arr}}$ is
\be
\overbar{Q}_{\mathrm{arr}}(\theta_{va}) = \frac{1}{2 \pi} \int_0^{2\pi} \mathrm{d}
\Phi_2 \, Q_{\mathrm{arr}}(\theta_{va}, \Phi_2) .
\ee

The rotationally averaged value of the $x_v$-component of a vector efficiency 
factor
$\mathbfit{Q}_i$ ($i$ equals, e.g., `$\Gamma, \mathrm{arr}$') is
\be
\overbar{Q}_{i, x_v}(\theta_{va}) = \frac{1}{2 \pi} \int_0^{2\pi} \mathrm{d}\Phi_2 
\, Q_{i, x_v}(\theta_{va}, \Phi_2) , 
\ee
\be
\label{eq:Q_i-xv}
Q_{i, x_v}(\theta_{va}, \Phi_2) = \sum_{j=1}^3 
Q_{i, j}(\theta_{\mathrm{gr}}, \phi_{\mathrm{gr}}) \, 
(\bmath{\hat{x}}_v \bmath{\cdot} \bmath{\hat{x}}_j) .
\ee
Similarly for the $y_v$- and $z_v$-components.  

It is convenient to express the averaged torque components in terms of 
spherical unit vectors $\bmath{\hat{a}}_1$, 
$\bmath{\hat{\theta}}_v = \bmath{\hat{x}}_v \cos \theta_{va} - \bmath{\hat{z}}_v
\sin \theta_{va}$, and $\bmath{\hat{\phi}}_v = \bmath{\hat{y}}_v$.  

\subsection{Drag torque}

A rotating grain experiences a drag torque.  Only the outgoing particles
(reflected or otherwise) contribute since the angular momenta of the
incoming atoms (as observed in an inertial frame) are not affected by the
grain rotation.  In scenarios (1) and (2) the outgoing particle departs
along the local surface normal $\bmath{\hat{N}}$.  After some number of times
intersecting the grain surface (possibly zero), the particle's path along the
local $\bmath{\hat{N}}$ leads it to escape the grain.  The outgoing particle's
velocity in the torque expressions is $v_{\mathrm{out}} \bmath{\hat{N}}$.  For
a rotating grain, this velocity is replaced with
$v_{\mathrm{out}} \bmath{\hat{N}} + \mathbfit{v}_{\mathrm{surf}}$, where 
$\mathbfit{v}_{\mathrm{surf}} = \bomega \bmath{\times} 
(\mathbfit{r}_{\mathrm{surf}} - \mathbfit{r}_{\mathrm{cm}})$
is the velocity of the surface element due to the grain rotation.  Thus,
the expressions for the drag torque efficiency factors are identical to
those for the mechanical torque except that $v_{\mathrm{out}} \bmath{\hat{N}}$ is
replaced with $\mathbfit{v}_{\mathrm{surf}}$.

Since the orientation $(\theta_{\mathrm{gr}}, \phi_{\mathrm{gr}})$ of a rotating 
grain relative to the direction of the drift velocity is not constant, the 
drag torque efficiency factors must be averaged over the rotation.  For
steady rotation about $\bmath{\hat{a}}_1$, this is done as described in \S
\ref{sec:suprathermal}.  

Since the motion of the grain can be neglected during the time interval
that an outgoing particle is in the grain vicinity and the outgoing
particle is always assumed to travel along $\bmath{\hat{N}}$ in scenarios (1)
and (2), the details of whether and where an outgoing particle strikes
the grain surface are unaffected by the grain rotation.  However, the
velocity vector of the reflected particle does depend on rotation in the
case of specular reflection, since the law of reflection applies in the
rest frame of the surface element.  This would introduce a major
computational burden, since the particle paths would have to be traced
anew for each value of the angular velocity.  Thus, we do not compute the
drag torque for the case of specular reflection.  

For steady rotation about $\bmath{\hat{a}}_1$,
\be
\bmath{\Gamma}_{\mathrm{drag, \, out}, (i)} = m n v_{\mathrm{th}} a_{\mathrm{eff}}^4
\omega \mathbfit{Q}_{\Gamma, \mathrm{drag, \, out}, (i)} 
\ee
with
\be
\mathbfit{Q}_{\Gamma, \mathrm{drag, \, out}, (1)} = - Q_{\mathrm{arr}} S_{\mathrm{esc}}^{-1} 
\int_{-1}^1 \mathrm{d}(\cos \theta) \int_0^{2\pi} \mathrm{d}\phi \, \eta_S(\theta,
\phi) \kappa_{\mathrm{esc}}(\theta, \phi) \left[
\frac{r^2_{\mathrm{surf}} + r^2_{\mathrm{cm}} - 2 \mathbfit{r}_{\mathrm{cm}} 
\bmath{\cdot} 
\mathbfit{r}_{\mathrm{surf}}}{a_{\mathrm{eff}}^2} \ \bmath{\hat{a}}_1 - 
\bmath{\hat{a}}_1 \bmath{\cdot} \left( 
\frac{\mathbfit{r}_{\mathrm{surf}} - \mathbfit{r}_{\mathrm{cm}}}{a_{\mathrm{eff}}} 
\right) \frac{(\mathbfit{r}_{\mathrm{surf}} - \mathbfit{r}_{\mathrm{cm}})}
{a_{\mathrm{eff}}} \right] 
\ee
and 
\begin{eqnarray}
\mathbfit{Q}_{\Gamma, \mathrm{drag, \, out}, (2)} & = & - \pi^{-3/2} \left( 
\frac{r_{\mathrm{sph}}}{a_{\mathrm{eff}}} \right)^2 \int_{-1}^1 \mathrm{d}
(\cos \theta_{\mathrm{sph}}) \int_0^{2\pi} \mathrm{d}\phi_{\mathrm{sph}} \int_0^{2\pi} 
\mathrm{d}\phi_{\mathrm{in}} \int_{u_c}^1 \mathrm{d}(\cos 
\theta_{\mathrm{in}}) \cos \theta_{\mathrm{in}} \, I_s(3, s_d, \beta) \nonumber \\
& & \times \left\{ \frac{r^2_{\mathrm{surf}}(\theta^{\prime}, \phi^{\prime}) 
+ r^2_{\mathrm{cm}}
- 2 \mathbfit{r}_{\mathrm{cm}} \bmath{\cdot} 
\mathbfit{r}_{\mathrm{surf}}(\theta^{\prime}, 
\phi^{\prime})}{a^2_{\mathrm{eff}}} \ \bmath{\hat{a}}_1 
- \bmath{\hat{a}}_1 \bmath{\cdot} \left[ \frac{
\mathbfit{r}_{\mathrm{surf}}(\theta^{\prime}, \phi^{\prime}) - 
\mathbfit{r}_{\mathrm{cm}}}
{a_{\mathrm{eff}}} \right] \frac{[\mathbfit{r}_{\mathrm{surf}}(\theta^{\prime}, 
\phi^{\prime}) - \mathbfit{r}_{\mathrm{cm}}]}{a_{\mathrm{eff}}} \right\} .
\label{eq:q-drag-out-2}
\end{eqnarray}
The above expressions must be averaged over the rotation about 
$\bmath{\hat{a}}_1$ as described in \S \ref{sec:suprathermal}.

In the case of a spherical grain at rest relative to the gas, 
$u_c = 0$, $r_{\mathrm{sph}} = a_{\mathrm{eff}}$, $\mathbfit{r}_{\mathrm{cm}} = 0$, 
$s_d = 0$, and equation (\ref{eq:q-drag-out-2}) simply evaluates to 
$\mathbfit{Q}_{\Gamma, \mathrm{drag, \, out}, (2)} = - 4 \pi^{1/2} \bmath{\hat{a}}_1 /3$,
which is a well known result \citep[see, e.g.,][]{DW96}. 

\subsection{Extreme subsonic limit \label{sec:subsonic}}

When the grain's motion through the gas is highly subsonic ($s_d \ll 1$),
simple approximations for the integrals over $s$ are available:
\be
I_s(3, s_d, \beta) \approx \frac{1}{2} + \frac{3 \sqrt{\pi}}{4} \, \beta
s_d , 
\label{eq:is3}
\ee
\be
I_s(4, s_d, \beta) \approx \frac{3 \sqrt{\pi}}{8} + 2 \beta s_d .
\label{eq:is4}
\ee

The efficiency factors associated with arriving atoms simplify to
\be
Q_{\mathrm{arr}} \approx Q_{\mathrm{arr}}(s_d = 0) + 
Q^{\prime}_{\mathrm{arr}} \ s_d , 
\ee
\be
\label{eq:q_arr_sd0}
Q_{\mathrm{arr}}(s_d = 0) =  \frac{\pi^{-3/2}}{4} \left( \frac{r_{\mathrm{sph}}}
{a_{\mathrm{eff}}} \right)^2 \int_{-1}^1 \mathrm{d}(\cos \theta_{\mathrm{sph}}) 
\int_0^{2 \pi} \mathrm{d}\phi_{\mathrm{sph}} \int_0^{2 \pi} \mathrm{d}\phi_{\mathrm{in}}
(1-u_c^2) ,
\ee
\be
\label{eq:q-prime-subsonic}
Q^{\prime}_{\mathrm{arr}} = \frac{1}{4 \pi} \left( \frac{r_{\mathrm{sph}}}
{a_{\mathrm{eff}}} \right)^2 \int_{-1}^1 \mathrm{d}(\cos \theta_{\mathrm{sph}}) 
\int_0^{2 \pi} \mathrm{d}\phi_{\mathrm{sph}} \int_0^{2 \pi} \mathrm{d}\phi_{\mathrm{in}}
\left[ \beta_1 (1 - u_c^3) + \beta_2 (1-u_c^2)^{3/2} \right] , 
\ee
\be
\mathbfit{Q}_{\Gamma, \mathrm{arr}} \approx \mathbfit{Q}_{\Gamma, \mathrm{arr}}(s_d = 0) 
+ \mathbfit{Q}^{\prime}_{\Gamma, \mathrm{arr}} \ s_d , 
\ee
\be
\mathbfit{Q}_{\Gamma, \mathrm{arr}}(s_d = 0) = 
\frac{3}{8 \pi} \left( \frac{r_{\mathrm{sph}}}{a_{\mathrm{eff}}} \right)^3
\int_{-1}^1 \mathrm{d}(\cos \theta_{\mathrm{sph}}) \int_0^{2\pi} \mathrm{d}
\phi_{\mathrm{sph}} \int_0^{2\pi} \mathrm{d}\phi_{\mathrm{in}} \int_{u_c}^1 \mathrm{d}
(\cos \theta_{\mathrm{in}}) \cos \theta_{\mathrm{in}} 
\left( \bmath{\hat{r}} - \frac{\mathbfit{r}_{\mathrm{cm}}}{r_{\mathrm{sph}}} 
\right) \bmath{\times} \bmath{\hat{s}} , 
\label{eq:q-gamma-arr-sd0}
\ee
\be
\mathbfit{Q}^{\prime}_{\Gamma, \mathrm{arr}} = 2 \pi^{-3/2} \left( \frac{r_{\mathrm{sph}}}
{a_{\mathrm{eff}}} \right)^3 \int_{-1}^1 \mathrm{d}(\cos \theta_{\mathrm{sph}}) 
\int_0^{2\pi} \mathrm{d}\phi_{\mathrm{sph}} \int_0^{2\pi} \mathrm{d}\phi_{\mathrm{in}} 
\int_{u_c}^1 \mathrm{d}(\cos \theta_{\mathrm{in}}) \cos \theta_{\mathrm{in}} 
\left( \bmath{\hat{r}} - \frac{\mathbfit{r}_{\mathrm{cm}}}{r_{\mathrm{sph}}} 
\right) \bmath{\times} \bmath{\hat{s}} \, \beta .
\label{eq:q-gamma-prime-arr}
\ee
See Appendix \ref{sec:calc-details-subsonic} for explicit expressions that 
simplify the calculation of $\mathbfit{Q}_{\Gamma, \mathrm{arr}}(s_d = 0)$ and
$\mathbfit{Q}^{\prime}_{\Gamma, \mathrm{arr}}$.
Although it is not evident from equations (\ref{eq:q-prime-subsonic}) and
(\ref{eq:q-gamma-arr-sd0}) or (\ref{eq:q-gamma-arr-sd0-2}), both 
$Q^{\prime}_{\mathrm{arr}}$ and $\mathbfit{Q}_{\Gamma, \mathrm{arr}}(s_d = 0)$ vanish,
as shown in Appendix \ref{sec:arrival-special-results}.

For specular reflection,
\be
\mathbfit{Q}_{\Gamma, \mathrm{spec}} \approx \mathbfit{Q}_{\Gamma, \mathrm{spec}}
(s_d = 0) + \mathbfit{Q}^{\prime}_{\Gamma, \mathrm{spec}} \ s_d 
\ee
with 
\be
\mathbfit{Q}_{\Gamma, \mathrm{spec}}(s_d = 0) = - \frac{3}{8 \pi} \left( 
\frac{r_{\mathrm{sph}}}{a_{\mathrm{eff}}} \right)^3 \int_{-1}^1 \mathrm{d}(\cos 
\theta_{\mathrm{sph}}) \int_0^{2\pi} \mathrm{d}\phi_{\mathrm{sph}} \int_0^{2\pi} 
\mathrm{d}\phi_{\mathrm{in}} \int_{u_c}^1 \mathrm{d}(\cos \theta_{\mathrm{in}})
\cos \theta_{\mathrm{in}} 
\left( \bmath{\hat{r}}_f - \frac{\mathbfit{r}_{\mathrm{cm}}}{r_{\mathrm{sph}}} 
\right) \bmath{\times} \bmath{\hat{s}}_f , 
\label{eq:Q-Gamma-spec-sd=0}
\ee
\be
\mathbfit{Q}^{\prime}_{\Gamma, \mathrm{spec}} = - 2 \pi^{-3/2} \left( 
\frac{r_{\mathrm{sph}}}{a_{\mathrm{eff}}} \right)^3 \int_{-1}^1 \mathrm{d}(\cos 
\theta_{\mathrm{sph}}) \int_0^{2\pi} d\phi_{\mathrm{sph}} \int_0^{2\pi}
\mathrm{d}\phi_{\mathrm{in}} \int_{u_c}^1 \mathrm{d}(\cos \theta_{\mathrm{in}})
\cos \theta_{\mathrm{in}}
\left( \bmath{\hat{r}}_f - \frac{\mathbfit{r}_{\mathrm{cm}}}{r_{\mathrm{sph}}} 
\right) \bmath{\times} \bmath{\hat{s}}_f \, \beta .
\ee
When $s_d=0$, and in the limit that the grain rotation can be neglected
during the collision, for every specularly reflected atom, there is an 
arriving atom whose velocity has the same magnitude but opposite sign.
Thus, 
$\mathbfit{Q}_{\Gamma, \mathrm{spec}}(s_d = 0) = \mathbfit{Q}_{\Gamma, \mathrm{arr}}
(s_d = 0) = 0$.

As a check of our computer codes, we implement equations 
(\ref{eq:q-prime-subsonic}), (\ref{eq:q-gamma-arr-sd0-2}), and
(\ref{eq:Q-Gamma-spec-sd=0}) to compute
$Q^{\prime}_{\mathrm{arr}}$, $\mathbfit{Q}_{\Gamma, \mathrm{arr}}(s_d = 0)$, and 
$\mathbfit{Q}_{\Gamma, \mathrm{spec}}(s_d = 0)$ and verify 
that they tend to zero as the numerical resolution improves.

The efficiency factors for outgoing scenario (1) depend on 
$s_d$ only through $Q_{\mathrm{arr}}$.  For scenario (2), 
\be
\mathbfit{Q}_{\Gamma, \mathrm{out}, (2)} \approx 
\mathbfit{Q}_{\Gamma, \mathrm{out}, (2)}(s_d = 0) + 
\mathbfit{Q}^{\prime}_{\Gamma, \mathrm{out}, (2)} \ s_d 
\ee
with 
\be
\mathbfit{Q}_{\Gamma, \mathrm{out}, (2)}(s_d = 0) = - \frac{\pi^{-3/2}}{2} \left( 
\frac{r_{\mathrm{sph}}}{a_{\mathrm{eff}}} \right)^2 \int_{-1}^1 \mathrm{d}(\cos 
\theta_{\mathrm{sph}}) \int_0^{2\pi} \mathrm{d}\phi_{\mathrm{sph}} \int_0^{2\pi} 
\mathrm{d}\phi_{\mathrm{in}} \int_{u_c}^1 \mathrm{d}(\cos 
\theta_{\mathrm{in}}) \cos \theta_{\mathrm{in}}
\left[ \frac{\mathbfit{r}_{\mathrm{surf}}(\theta^{\prime}, \phi^{\prime}) - 
\mathbfit{r}_{\mathrm{cm}}}{a_{\mathrm{eff}}} \right] \bmath{\times} 
\bmath{\hat{N}}(\theta^{\prime}, \phi^{\prime}) ,
\ee
\be
\mathbfit{Q}^{\prime}_{\Gamma, \mathrm{out}, (2)} = - \frac{3}{4 \pi} \left( 
\frac{r_{\mathrm{sph}}}{a_{\mathrm{eff}}} \right)^2 \int_{-1}^1 \mathrm{d}(\cos 
\theta_{\mathrm{sph}}) \int_0^{2\pi} \mathrm{d}\phi_{\mathrm{sph}} \int_0^{2\pi} 
\mathrm{d}\phi_{\mathrm{in}} \int_{u_c}^1 \mathrm{d}(\cos 
\theta_{\mathrm{in}}) \cos \theta_{\mathrm{in}} 
\left[ \frac{\mathbfit{r}_{\mathrm{surf}}(\theta^{\prime}, \phi^{\prime}) - 
\mathbfit{r}_{\mathrm{cm}}}{a_{\mathrm{eff}}} \right] \bmath{\times} 
\bmath{\hat{N}}(\theta^{\prime}, \phi^{\prime}) \, \beta 
\ee
and 
\be
\mathbfit{Q}_{\Gamma, \mathrm{drag, \, out}, (2)} \approx 
\mathbfit{Q}_{\Gamma, \mathrm{drag, \, out}, (2)}(s_d = 0) + 
\mathbfit{Q}^{\prime}_{\Gamma, \mathrm{out}, (2)} \ s_d 
\ee
with 
\begin{eqnarray}
\mathbfit{Q}_{\Gamma, \mathrm{drag, \, out}, (2)}(s_d = 0) & = & - \frac{\pi^{-3/2}}{2} 
\left( \frac{r_{\mathrm{sph}}}{a_{\mathrm{eff}}} \right)^2 \int_{-1}^1 \mathrm{d}
(\cos \theta_{\mathrm{sph}}) \int_0^{2\pi} \mathrm{d}\phi_{\mathrm{sph}} \int_0^{2\pi} 
\mathrm{d}\phi_{\mathrm{in}} \int_{u_c}^1 \mathrm{d}(\cos 
\theta_{\mathrm{in}}) \cos \theta_{\mathrm{in}} \nonumber \\
& & \times \left\{ \frac{r^2_{\mathrm{surf}}(\theta^{\prime}, \phi^{\prime}) 
+ r^2_{\mathrm{cm}}
- 2 \mathbfit{r}_{\mathrm{cm}} \bmath{\cdot} 
\mathbfit{r}_{\mathrm{surf}}(\theta^{\prime}, 
\phi^{\prime})}{a^2_{\mathrm{eff}}} \ \bmath{\hat{a}}_1 
- \bmath{\hat{a}}_1 \bmath{\cdot} \left[ \frac{
\mathbfit{r}_{\mathrm{surf}}(\theta^{\prime}, \phi^{\prime}) - 
\mathbfit{r}_{\mathrm{cm}}}
{a_{\mathrm{eff}}} \right] \frac{[\mathbfit{r}_{\mathrm{surf}}(\theta^{\prime}, 
\phi^{\prime}) - \mathbfit{r}_{\mathrm{cm}}]}{a_{\mathrm{eff}}} \right\} ,
\end{eqnarray}
\begin{eqnarray}
\mathbfit{Q}^{\prime}_{\Gamma, \mathrm{drag, \, out}, (2)} & = & - \frac{3}{4 \pi} 
\left( \frac{r_{\mathrm{sph}}}{a_{\mathrm{eff}}} \right)^2 \int_{-1}^1 \mathrm{d}
(\cos \theta_{\mathrm{sph}}) \int_0^{2\pi} \mathrm{d}\phi_{\mathrm{sph}} \int_0^{2\pi} 
\mathrm{d}\phi_{\mathrm{in}} \int_{u_c}^1 \mathrm{d}(\cos \theta_{\mathrm{in}}) \cos 
\theta_{\mathrm{in}} \nonumber \\
& & \times \left\{ \frac{r^2_{\mathrm{surf}}(\theta^{\prime}, \phi^{\prime}) 
+ r^2_{\mathrm{cm}}
- 2 \mathbfit{r}_{\mathrm{cm}} \bmath{\cdot} 
\mathbfit{r}_{\mathrm{surf}}(\theta^{\prime}, 
\phi^{\prime})}{a^2_{\mathrm{eff}}} \ \bmath{\hat{a}}_1
- \bmath{\hat{a}}_1 \bmath{\cdot} \left[ \frac{
\mathbfit{r}_{\mathrm{surf}}(\theta^{\prime}, \phi^{\prime}) - 
\mathbfit{r}_{\mathrm{cm}}}
{a_{\mathrm{eff}}} \right] \frac{[\mathbfit{r}_{\mathrm{surf}}(\theta^{\prime}, 
\phi^{\prime}) - \mathbfit{r}_{\mathrm{cm}}]}{a_{\mathrm{eff}}} \right\} \beta .
\end{eqnarray}

\subsection{Extreme supersonic limit \label{sec:supersonic}}

If the grain motion is highly supersonic ($s_d \gg 1$), then we can neglect
the thermal motion of the gas atoms.  In this case, all of the atoms move
along $- \bmath{\hat{s}}_d$.  The simplest and most efficient computational 
approach dispenses with the enclosing sphere.  The rate at which gas-phase 
particles arrive at a patch on the grain surface is
\be
\mathrm{d}R_{\mathrm{arr}} \approx n v_{\mathrm{th}} s_d \, \mathrm{d}(\cos \theta) 
\, \mathrm{d}\phi \, \eta_S(\theta, \phi) [\bmath{\hat{s}}_d \bmath{\cdot} 
\bmath{\hat{N}}(\theta, \phi)] \mu_{\mathrm{hit}} 
\ee
where $\mu_{\mathrm{hit}} = 1$ (0) if gas atoms do (do not) strike the patch.  
Gas atoms do not strike the patch if they are moving in the wrong direction 
($\bmath{\hat{s}}_d \bmath{\cdot} \bmath{\hat{N}} < 0$) or if the patch is
obstructed by another portion of the grain.  Thus,
\be
\label{eq:Q-arr-sup}
Q_{\mathrm{arr}} \approx s_d \int_{-1}^1 \mathrm{d}(\cos \theta) \int_0^{2 \pi} 
\mathrm{d}\phi \,
\frac{\eta_S(\theta, \phi)}{a_{\mathrm{eff}}^2} [\bmath{\hat{s}}_d \bmath{\cdot} 
\bmath{\hat{N}}(\theta, \phi)] \mu_{\mathrm{hit}} .
\ee
The angular momentum of the arriving particle is 
\be
\Delta \mathbfit{J}_{\mathrm{arr}} = m v_{\mathrm{th}} s_d 
( \mathbfit{r}_{\mathrm{surf}} - \mathbfit{r}_{\mathrm{cm}}) 
\bmath{\times} (- \bmath{\hat{s}}_d) .
\ee
Thus,
\be
\label{eq:Q-gamma-arr-sup}
\mathbfit{Q}_{\Gamma, \mathrm{arr}} \approx  s_d^2 \int_{-1}^1 \mathrm{d}(\cos 
\theta) \int_0^{2 \pi} \mathrm{d}\phi \, \frac{\eta_S(\theta, \phi)}
{a_{\mathrm{eff}}^2} 
[\bmath{\hat{s}}_d \bmath{\cdot} \bmath{\hat{N}}(\theta, \phi)] \mu_{\mathrm{hit}}
\frac{(\mathbfit{r}_{\mathrm{surf}} - \mathbfit{r}_{\mathrm{cm}})}{a_{\mathrm{eff}}} 
\bmath{\times} (- \bmath{\hat{s}}_d) .
\ee
Similarly, for specular reflection, 
\be
\mathbfit{Q}_{\Gamma, \mathrm{spec}} \approx  - s_d^2 \int_{-1}^1 \mathrm{d}(\cos 
\theta) \int_0^{2 \pi} \mathrm{d}\phi \, \frac{\eta_S(\theta, \phi)}
{a_{\mathrm{eff}}^2} 
[\bmath{\hat{s}}_d \bmath{\cdot} \bmath{\hat{N}}(\theta, \phi)] \mu_{\mathrm{hit}}
\frac{(\mathbfit{r}_{\mathrm{surf, \, f}} - \mathbfit{r}_{\mathrm{cm}})}
{a_{\mathrm{eff}}} \bmath{\times} \bmath{\hat{s}}_f
\ee
where $\mathbfit{r}_{\mathrm{surf, \, f}}$ is the position on the surface from which
the departing particle escapes to infinity.   

As noted in \S \ref{sec:subsonic}, 
the efficiency factors for outgoing scenario (1) depend on 
$s_d$ only through $Q_{\mathrm{arr}}$. For scenario (2), 
\be
\mathbfit{Q}_{\Gamma, \mathrm{out}, (2)} \approx - s_d \int_{-1}^1 \mathrm{d}(\cos 
\theta) \int_0^{2 \pi} \mathrm{d}\phi \frac{\eta_S(\theta, \phi)}
{a_{\mathrm{eff}}^2} [\bmath{\hat{s}}_d \bmath{\cdot} \bmath{\hat{N}}(\theta, 
\phi)] \mu_{\mathrm{hit}}
\left[ \frac{\mathbfit{r}_{\mathrm{surf}}(\theta^{\prime}, \phi^{\prime}) - 
\mathbfit{r}_{\mathrm{cm}}}{a_{\mathrm{eff}}} \right] \bmath{\times} \bmath{\hat{N}}
(\theta^{\prime}, \phi^{\prime})
\ee
and
\begin{eqnarray}
\mathbfit{Q}_{\Gamma, \mathrm{drag, \, out}, (2)} & \approx & - s_d \int_{-1}^1 
\mathrm{d}(\cos \theta) \int_0^{2 \pi} \mathrm{d}\phi \frac{\eta_S(\theta, \phi)}
{a_{\mathrm{eff}}^2} [\bmath{\hat{s}}_d \bmath{\cdot} 
\bmath{\hat{N}}(\theta, \phi)] \mu_{\mathrm{hit}} \nonumber \\
& & \times \left\{ \frac{r^2_{\mathrm{surf}}(\theta^{\prime}, \phi^{\prime}) 
+ r^2_{\mathrm{cm}} - 2  
\mathbfit{r}_{\mathrm{cm}} \bmath{\cdot} \mathbfit{r}_{\mathrm{surf}}
(\theta^{\prime}, \phi^{\prime})}
{a_{\mathrm{eff}}^2} \ \bmath{\hat{a}}_1 
- \bmath{\hat{a}}_1 \bmath{\cdot} 
\left[ \frac{\mathbfit{r}_{\mathrm{surf}}(\theta^{\prime}, \phi^{\prime}) - 
\mathbfit{r}_{\mathrm{cm}}}{a_{\mathrm{eff}}} \right] \frac{[\mathbfit{r}_{\mathrm{surf}}
(\theta^{\prime}, \phi^{\prime}) - \mathbfit{r}_{\mathrm{cm}}]}{a_{\mathrm{eff}}} 
\right\} .
\label{eq:Q-gamma-drag-out-sup}
\end{eqnarray}

\subsection{Mechanical/drag force \label{sec:force}}

Of course, collisions with gas atoms give rise to a force as well as a
torque on a grain.  Although the grain rotational dynamics
is our primary concern, for completeness and code verification purposes we 
provide expressions for the force in this section.  

An arriving gas particle deposits momentum 
$\Delta \mathbfit{p} = m v_{\mathrm{th}} s \, \bmath{\hat{s}}$ on the grain.  
The force due to arriving atoms is
\be
\mathbfit{F}_{\mathrm{arr}} = \int \mathrm{d}R_{\mathrm{arr}} \, \Delta \mathbfit{p} 
= m n v_{\mathrm{th}}^2 a_{\mathrm{eff}}^2 \mathbfit{Q}_{F, \mathrm{arr}}
\ee
with
\be
\mathbfit{Q}_{F, \mathrm{arr}} = - \pi^{-3/2} \left( \frac{r_{\mathrm{sph}}}
{a_{\mathrm{eff}}} \right)^2 \int_{-1}^1 \mathrm{d}(\cos \theta) \int_0^{2\pi} 
\mathrm{d}\phi \int_0^{2\pi} \mathrm{d}\phi_{\mathrm{in}} \int_{u_c}^1 \mathrm{d}
(\cos \theta_{\mathrm{in}}) \cos \theta_{\mathrm{in}} (- \bmath{\hat{s}}) 
I_s(4, s_d, \beta) .
\label{eq:Q-F-arr}
\ee
See \S \ref{sec:incoming-trajectories} for an explicit expression for 
$- \bmath{\hat{s}}$.  

For specular reflection, there is an additional term
\be
\mathbfit{Q}_{F, \mathrm{spec}} = - \pi^{-3/2} \left( \frac{r_{\mathrm{sph}}}
{a_{\mathrm{eff}}} \right)^2 \int_{-1}^1 \mathrm{d}(\cos \theta) \int_0^{2\pi} 
\mathrm{d}\phi \int_0^{2\pi} \mathrm{d}\phi_{\mathrm{in}} \int_{u_c}^1 \mathrm{d}
(\cos \theta_{\mathrm{in}}) \cos \theta_{\mathrm{in}} \, \bmath{\hat{s}}_f 
I_s(4, s_d, \beta) .
\label{eq:Q-F-spec}
\ee

For scenarios (1) and (2) for the outgoing particles, 
the force is given by 
\be
\mathbfit{F}_{\mathrm{out}, (i)} = m n v_{\mathrm{th}} v_{\mathrm{out}} a_{\mathrm{eff}}^2
\mathbfit{Q}_{F, \mathrm{out}, (i)}
\ee
with
\be
\mathbfit{Q}_{F, \mathrm{out}, (1)} = - Q_{\mathrm{arr}} S_{\mathrm{esc}}^{-1} \int_{-1}^1
\mathrm{d}(\cos \theta) \int_0^{2\pi} \mathrm{d}\phi \, \eta_S(\theta, \phi) 
\kappa_{\mathrm{esc}}(\theta, \phi) \bmath{\hat{N}}(\theta, \phi) , 
\ee
\be
\mathbfit{Q}_{F, \mathrm{out}, (2)} = - \pi^{-3/2} \left( \frac{r_{\mathrm{sph}}}
{a_{\mathrm{eff}}} \right)^2 \int_{-1}^1 \mathrm{d}(\cos \theta) \int_0^{2\pi} 
\mathrm{d}\phi \int_0^{2\pi} \mathrm{d}\phi_{\mathrm{in}} \int_{u_c}^1 \mathrm{d}
(\cos \theta_{\mathrm{in}}) \cos \theta_{\mathrm{in}}
\bmath{\hat{N}}(\theta^{\prime}, \phi^{\prime}) \, I_s(3, s_d, \beta) .
\ee
Refer to \S \ref{sec:torque-outgoing} for the meaning of 
$(\theta^{\prime}, \phi^{\prime})$ for scenario (2).

The total force is 
\be
\mathbfit{F} = m n v^2_{\mathrm{th}} a_{\mathrm{eff}}^2 \left[
\mathbfit{Q}_{F, \mathrm{arr}} + f_{\mathrm{spec}}
\mathbfit{Q}_{F, \mathrm{spec}} + (1 - f_{\mathrm{spec}}) \frac{v_{\mathrm{out}}}
{v_{\mathrm{th}}} \, \mathbfit{Q}_{F, \mathrm{out}} \right]
\ee
where $\mathbfit{Q}_{F, \mathrm{out}}$ is the efficiency factor for one of the
scenarios (1 or 2) for outgoing particles.

In the extreme subsonic limit,
\be
\mathbfit{Q}_{F, \mathrm{arr}} \approx \mathbfit{Q}_{F, \mathrm{arr}}(s_d = 0) +  
\mathbfit{Q}^{\prime}_{F, \mathrm{arr}} \, s_d
\ee
with 
\be
\label{eq:Q-F-arr-sd0}
\mathbfit{Q}_{F, \mathrm{arr}}(s_d = 0) = - \frac{3}{8 \pi} \left( 
\frac{r_{\mathrm{sph}}}{a_{\mathrm{eff}}} \right)^2 \int_{-1}^1 \mathrm{d}(\cos 
\theta) \int_0^{2\pi} \mathrm{d}\phi \int_0^{2\pi} \mathrm{d}\phi_{\mathrm{in}} 
\int_{u_c}^1 \mathrm{d}(\cos \theta_{\mathrm{in}}) 
\cos \theta_{\mathrm{in}} (- \bmath{\hat{s}}) , 
\ee
\be
\label{eq:Q-F-arr-prime}
\mathbfit{Q}^{\prime}_{F, \mathrm{arr}} = - 2 \pi^{-3/2} \left( \frac{r_{\mathrm{sph}}}
{a_{\mathrm{eff}}} \right)^2 \int_{-1}^1 \mathrm{d}(\cos \theta) \int_0^{2\pi} 
\mathrm{d}\phi \int_0^{2\pi} \mathrm{d}\phi_{\mathrm{in}} \int_{u_c}^1 \mathrm{d}
(\cos \theta_{\mathrm{in}}) \cos \theta_{\mathrm{in}} (- \bmath{\hat{s}}) \, \beta .
\ee
See Appendix \ref{sec:calc-details-force-subsonic} for explicit integration
over $\cos \theta_{\mathrm{in}}$.
The expressions for $\mathbfit{Q}_{F, \mathrm{spec}}$ are identical except that
$(- \bmath{\hat{s}}_d)$ is replaced with $\bmath{\hat{s}}_f$.  
From the arguments given in \S \ref{sec:subsonic} (but with momentum in
place of angular momentum), the force associated with arriving and
specularly reflected atoms vanishes when $s_d=0$.  In other words, these 
processes result exclusively in a drag force.  As with the torques, we 
implement the formulas above in our computer code and check that
$\mathbfit{Q}_{F, \mathrm{arr}}(s_d = 0)$ and $\mathbfit{Q}_{F, \mathrm{spec}}(s_d = 0)$
are consistent with zero.  

The efficiency factors for outgoing scenario (1) depend on 
$s_d$ only through $Q_{\mathrm{arr}}$.  For scenario (2),
\be
\mathbfit{Q}_{F, \mathrm{out}, (2)} \approx 
\mathbfit{Q}_{F, \mathrm{out}, (2)}(s_d = 0) + 
\mathbfit{Q}^{\prime}_{F, \mathrm{out}, (2)} \ s_d
\ee
with 
\be
\mathbfit{Q}_{F, \mathrm{out}, (2)}(s_d = 0) = - \frac{\pi^{-3/2}}{2} \left( 
\frac{r_{\mathrm{sph}}}{a_{\mathrm{eff}}} \right)^2 \int_{-1}^1 \mathrm{d}(\cos 
\theta) \int_0^{2\pi} \mathrm{d}\phi \int_0^{2\pi} \mathrm{d}\phi_{\mathrm{in}} 
\int_{u_c}^1 \mathrm{d}(\cos \theta_{\mathrm{in}}) \cos \theta_{\mathrm{in}} 
\bmath{\hat{N}}(\theta^{\prime}, \phi^{\prime}) , 
\ee
\be
\mathbfit{Q}^{\prime}_{F, \mathrm{out}, (2)} = - \frac{3}{4\pi} \left( 
\frac{r_{\mathrm{sph}}}{a_{\mathrm{eff}}} \right)^2 \int_{-1}^1 \mathrm{d}(\cos 
\theta) \int_0^{2\pi} \mathrm{d}\phi \int_0^{2\pi} \mathrm{d}\phi_{\mathrm{in}} 
\int_{u_c}^1 \mathrm{d}(\cos \theta_{\mathrm{in}}) \cos \theta_{\mathrm{in}} 
\bmath{\hat{N}}(\theta^{\prime}, \phi^{\prime}) \, \beta .
\ee

In the extreme supersonic limit,
\be
\mathbfit{Q}_{F, \mathrm{arr}} \approx - s_d Q_{\mathrm{arr}} \, \bmath{\hat{s}}_d ,
\ee
\be
\mathbfit{Q}_{F, \mathrm{spec}} \approx - s_d^2 \int_{-1}^1 \mathrm{d}(\cos \theta) 
\int_0^{2 \pi} \mathrm{d}\phi \, \frac{\eta_S(\theta, \phi)}{a_{\mathrm{eff}}^2} 
[\bmath{\hat{s}}_d \bmath{\cdot} \bmath{\hat{N}}(\theta, \phi)] \mu_{\mathrm{hit}}
\, \bmath{\hat{s}}_f , 
\ee
and 
\be
\mathbfit{Q}_{F, \mathrm{out}, (2)}  \approx - s_d \int_{-1}^1 \mathrm{d}(\cos 
\theta) \int_0^{2 \pi} \mathrm{d}\phi \, \frac{\eta_S(\theta, \phi)}
{a_{\mathrm{eff}}^2} [\bmath{\hat{s}}_d \bmath{\cdot} \bmath{\hat{N}}(\theta, 
\phi)] \mu_{\mathrm{hit}} \, \bmath{\hat{N}}(\theta^{\prime}, \phi^{\prime}) .
\ee

\section{Torque Calculations:  Computational Approach} \label{sec:torque-comp}

\subsection{Incoming trajectories \label{sec:incoming-trajectories}}

We take the radius $r_{\mathrm{sph}}$ of the enclosing 
sphere to be $1.01 \, r_{\mathrm{max}}$.
(See the text following equation \ref{eq:rec-rel2} for the definition of
$r_{\mathrm{max}}$.)

A gas atom incident on the enclosing sphere has initial
position $\mathbfit{r}_0 = r_{\mathrm{sph}} \bmath{\hat{r}}$ (see equation 
\ref{eq:r-hat}) and velocity 
$\mathbfit{v} = v_{\mathrm{th}} s \, \bmath{\hat{s}}$.  From equation
(\ref{eq:s-hat}), 
\begin{eqnarray}
\mathbfit{v} & = &  -v_{\mathrm{th}} s
\left[ (\sin \theta_{\mathrm{in}} \cos \phi_{\mathrm{in}} \cos \theta_{\mathrm{sph}}
\cos \phi_{\mathrm{sph}} - \sin \theta_{\mathrm{in}} \sin \phi_{\mathrm{in}} \sin 
\phi_{\mathrm{sph}} + \cos \theta_{\mathrm{in}} \sin \theta_{\mathrm{sph}} \cos 
\phi_{\mathrm{sph}}) \bmath{\hat{x}} \right. + \nonumber \\
& & \left. (\sin \theta_{\mathrm{in}} \cos \phi_{\mathrm{in}} \cos 
\theta_{\mathrm{sph}} \sin \phi_{\mathrm{sph}} + \sin \theta_{\mathrm{in}} \sin 
\phi_{\mathrm{in}} \cos \phi_{\mathrm{sph}} + \cos \theta_{\mathrm{in}} \sin 
\theta_{\mathrm{sph}} \sin \phi_{\mathrm{sph}}) \bmath{\hat{y}} + 
(- \sin \theta_{\mathrm{in}} \cos \phi_{\mathrm{in}} \sin \theta_{\mathrm{sph}} +
\cos \theta_{\mathrm{in}} \cos \theta_{\mathrm{sph}}) \bmath{\hat{z}} \right] .
\label{eq:v0}
\end{eqnarray}

Given $(\theta_{\mathrm{sph}}, \phi_{\mathrm{sph}}, \phi_{\mathrm{in}})$, the plane 
containing atom
trajectories for arbitrary $\theta_{\mathrm{in}}$ is spanned by the vectors
$\bmath{\hat{r}}$ and $\bmath{\hat{s}}(\theta_{\mathrm{in}} = \pi/2)$.  Now 
consider plane polar coordinates in this plane with $\mu$ the polar angle; 
$\mu = 0$ along $\bmath{\hat{r}}$ and $\mu = \pi/2$ along 
$\bmath{\hat{s}}(\theta_{\mathrm{in}} = \pi/2)$.
The origin remains at the position within the grain originally adopted in
defining the GRS.  Positions along an 
atom trajectory have $0 \le \mu \le \pi$.  The unit 
vector $\bmath{\hat{V}}$ characterized by angle $\mu$ is
\be
\bmath{\hat{V}} = \bmath{\hat{r}} \cos \mu + \bmath{\hat{s}}
(\theta_{\mathrm{in}} = \pi/2) \sin \mu .
\ee
The spherical coordinates $(\theta, \phi)$ of $\bmath{\hat{V}}$ are found by 
equating expressions for the
$x$-, $y$-, and $z$-components in both systems:
\be
\sin \theta \cos \phi = \sin \theta_{\mathrm{sph}} \cos \phi_{\mathrm{sph}} \cos 
\mu + (\sin \phi_{\mathrm{in}} \sin \phi_{\mathrm{sph}} - \cos \phi_{\mathrm{in}} \cos 
\theta_{\mathrm{sph}} \cos \phi_{\mathrm{sph}}) \sin \mu ,
\ee
\be
\sin \theta \sin \phi = \sin \theta_{\mathrm{sph}} \sin \phi_{\mathrm{sph}} \cos 
\mu - (\sin \phi_{\mathrm{in}} \cos \phi_{\mathrm{sph}} + \cos \phi_{\mathrm{in}} \cos 
\theta_{\mathrm{sph}} \sin \phi_{\mathrm{sph}}) \sin \mu ,
\ee
\be
\cos \theta = \cos \theta_{\mathrm{sph}} \cos \mu + \cos \phi_{\mathrm{in}}
\sin \theta_{\mathrm{sph}} \sin \mu .
\ee
For each $\mu$, there is a unique distance $r_{\mathrm{surf}}(\mu)$ from the 
origin to the grain surface.  A straight line that passes through 
$\mathbfit{r}_0$ and the point on the grain surface characterized by $\mu$ has
\be
\cos \theta_{\mathrm{in}}(\mu) = \left\{ 1 + \sin^2 \mu \left[ 
r_{\mathrm{sph}}/r_{\mathrm{surf}}(\mu) - \cos \mu \right]^{-2} \right\}^{-1/2} .
\ee
Wherever $\cos \theta_{\mathrm{in}}(\mu)$ has a local minimum, the line is
tangent to the grain surface.  The critical value $u_c$ of 
$\cos \theta_{\mathrm{in}}$ is, of course, the global minimum.  To find it,
we first isolate the local minima by calculating 
$\cos \theta_{\mathrm{in}}(\mu)$ for 1000 values of $\mu$ (evenly spaced between
0 and $\pi$).  Then, we apply the routine {\sc brent} from \citet{Press92} 
to the lowest local minimum to find $u_c$.  We tabulate $u_c$ and the 
position where the corresponding trajectory strikes the grain surface as
a function of $(\theta_{\mathrm{sph}}, \phi_{\mathrm{sph}}, \phi_{\mathrm{in}})$ for 
$N_1 + 1$ values of 
$\theta_{\mathrm{sph}}$ (spaced evenly in $\cos \theta_{\mathrm{sph}}$) and $N_1$ 
values of $\phi_{\mathrm{sph}}$ and $\phi_{\mathrm{in}}$ (spaced
evenly between 0 and $2 \pi$, but excluding $2 \pi$).  We discuss the adopted
value of $N_1$, as well as the values of other parameters affecting the 
convergence of the results, in \S \ref{sec:torque-results}.

\subsection{Arrival at the grain surface and reflection
\label{sec:ray-tracing}}

Given $u_c$ as a function of $(\theta_{\mathrm{sph}}, \phi_{\mathrm{sph}}, 
\phi_{\mathrm{in}})$, we next
examine trajectories for $(\theta_{\mathrm{sph}}, \phi_{\mathrm{sph}}, 
\phi_{\mathrm{in}}, \theta_{\mathrm{in}})$,
with $N_1 + 1$ values of $\theta_{\mathrm{in}}$ spaced evenly in 
$\cos \theta_{\mathrm{in}} \in [u_c, 1]$.  For each incoming trajectory, we
tabulate the values of $- \bmath{\hat{s}}$ and 
$(\bmath{\hat{r}} - \mathbfit{r}_{\mathrm{cm}}/r_{\mathrm{sph}}) \bmath{\times} 
\bmath{\hat{s}}$ for use in evaluating $\mathbfit{Q}_{F, \mathrm{arr}}$ and 
$\mathbfit{Q}_{\Gamma, \mathrm{arr}}$ (equations \ref{eq:Q-F-arr} and 
\ref{eq:Q-Gamma-arr}).  

Next, we determine where
the trajectory strikes the grain surface.  Starting with the incoming
particle's position and velocity on the enclosing sphere (as described in
\S \ref{sec:incoming-trajectories}), we advance the particle along its 
trajectory, in steps of length $10^{-3} r_{\mathrm{sph}}$, until the particle
reaches the grain interior.  The final and penultimate steps bracket the
intersection of the trajectory with the grain surface.  The intersection
point is then accurately found by 10 repeated bisections of this bracketing
interval.  This trajectory-tracing algorithm is not employed for the cases
where $\cos \theta_{\mathrm{in}} = u_c$ (since the arrival location on the
grain surface was obtained when $u_c$ was determined) and 
$\cos \theta_{\mathrm{in}} = 1$ [since the trajectory is radial in this case,
it reaches the surface at
$(\theta, \phi) = (\theta_{\mathrm{sph}}, \phi_{\mathrm{sph}})$].  

Finally, we determine the values of $- \bmath{\hat{s}}_f$ and 
$(\bmath{\hat{r}}_f - \mathbfit{r}_{\mathrm{cm}}/r_{\mathrm{sph}}) \bmath{\times} 
\bmath{\hat{s}}_f$ for use in evaluating $\mathbfit{Q}_{F, \mathrm{spec}}$ and 
$\mathbfit{Q}_{\Gamma, \mathrm{spec}}$ (equations \ref{eq:Q-F-spec} and 
\ref{eq:Q-Gamma-spec}).  The surface normal vector $\bmath{\hat{N}}$ at the 
point where the particle arrives at the grain is calculated using equation 
(\ref{eq:n-hat}).  Since
$|\mathbfit{T}_{\theta} \bmath{\times} \mathbfit{T}_{\phi}| = 0$ when 
$\sin \theta = 0$, this case must be treated differently.  Instead, we 
evaluate $\bmath{\hat{N}}$ for a small value of $\sin \theta$ and several 
evenly spaced values of $\phi$ and take the average of these for 
$\bmath{\hat{N}}$ when $\sin \theta = 0$.  The velocity 
$\mathbfit{v}_r$ of the reflected particle is related to the velocity 
$\mathbfit{v}_i$ of the incoming particle by the law of reflection:
\be
\mathbfit{v}_r = \mathbfit{v}_i - 2 (\mathbfit{v}_i \bmath{\cdot} 
\bmath{\hat{N}}) \bmath{\hat{N}} .
\ee
We employ the same procedure as described in the previous paragraph to
follow the trajectory of the reflected particle until it ultimately 
reaches the enclosing sphere (perhaps after multiple reflections on the
grain surface).   

\subsection{Integrals over the reduced speed}

Prior to computing torques, we generate, using {\sc mathematica}, interpolation
tables for the function $I_s(p, s_d, \beta)$ defined in equation (\ref{eq:i-s}) 
with 20,001 values of $\beta$ for the value of 
$s_d$ under consideration and $p=3$ and 4.

\subsection{Characterization of the grain surface}

We divide the grain surface into $N_2 \times N_2$ patches, evenly spaced in
$\cos \theta$ and $\phi$.  
Using the approach described in \S \ref{sec:ray-tracing}, we follow the 
trajectory of a particle departing the surface along the normal vector at
the centre of each patch.  We record whether or not the departing particle
escapes to infinity or strikes the grain elsewhere ($\kappa_{\mathrm{esc}} = 1$
or 0).  If it escapes, then we record the vectors $\bmath{\hat{N}}$, 
$\mathbfit{r}_{\mathrm{surf}}$, and 
$(\mathbfit{r}_{\mathrm{surf}} - \mathbfit{r}_{\mathrm{cm}})/a_{\mathrm{eff}}$ for use 
in evaluating the force and torque associated with outgoing particles.  If the
departing particle strikes the grain elsewhere, then we record the index
values of the patch that it strikes.

\subsection{Torque evaluations}

With the computational results from the preceding sections in hand, it is
now straightforward to evaluate all of the efficiency factors.  
For outgoing scenario (2), we take the departure point for the outgoing
particle to be the centre of the surface patch in which the incoming
particle arrived.  Similarly, when an outgoing particle strikes the grain
surface elsewhere, we assume that the particle immediately departs along
the surface normal in the centre of the patch that was struck.  

We compute torques for $(N_3 + 1, N_3)$ values of $(\theta_{\mathrm{gr}}, 
\phi_{\mathrm{gr}})$ and average over $N_3$ values of $\Phi_2$ (when averaging
over rotation about $\bmath{\hat{a}}_1$).  
For $Q_{\mathrm{arr}}$, $\mathbfit{Q}_{\Gamma, \mathrm{arr}}$, 
$\mathbfit{Q}_{\Gamma, \mathrm{spec}}$, $\mathbfit{Q}_{F, \mathrm{arr}}$,
$\mathbfit{Q}_{F, \mathrm{spec}}$, $\mathbfit{Q}_{\Gamma, \mathrm{spec}}(s_d=0)$,
$\mathbfit{Q}^{\prime}_{\Gamma, \mathrm{spec}}$, $\mathbfit{Q}_{F, \mathrm{spec}}(s_d=0)$,
$\mathbfit{Q}^{\prime}_{F, \mathrm{spec}}$, $\mathbfit{Q}_{\Gamma, \mathrm{out}, (2)}$,
$\mathbfit{Q}_{F, \mathrm{out}, (2)}$, 
$\mathbfit{Q}_{\Gamma, \mathrm{out}, (2)}(s_d = 0)$,
$\mathbfit{Q}^{\prime}_{\Gamma, \mathrm{out}, (2)}$,
$\mathbfit{Q}_{\Gamma, \mathrm{drag, \, out}, (2)}(s_d = 0)$,
and $\mathbfit{Q}^{\prime}_{\Gamma, \mathrm{drag, \, out}, (2)}$, 
integrals are evaluated with $(N_1 + 1, N_1, N_1, N_1 + 1)$ values of
$(\theta_{\mathrm{sph}}, \phi_{\mathrm{sph}}, \phi_{\mathrm{in}}, \theta_{\mathrm{in}})$.  
For $Q_{\mathrm{arr}}(s_d = 0)$, $Q^{\prime}_{\mathrm{arr}}$, 
$\mathbfit{Q}_{\Gamma, \mathrm{arr}}(s_d=0)$, 
$\mathbfit{Q}^{\prime}_{\Gamma, \mathrm{arr}}$,
$\mathbfit{Q}_{F, \mathrm{arr}}(s_d=0)$, and $\mathbfit{Q}^{\prime}_{F, \mathrm{arr}}$,
integrals are evaluated with $(N_1 + 1, N_1, N_1)$ values of
$(\theta_{\mathrm{sph}}, \phi_{\mathrm{sph}}, \phi_{\mathrm{in}})$.
Integrals for $\mathbfit{Q}_{\Gamma, \mathrm{out}, (1)}/Q_{\mathrm{arr}}$, 
$\mathbfit{Q}_{\Gamma, \mathrm{drag, \, out}, (1)}/Q_{\mathrm{arr}}$, and
$\mathbfit{Q}_{F, \mathrm{out}, (1)}/Q_{\mathrm{arr}}$ are evaluated with 
$(4096, 4096)$ values of $(\theta_{\mathrm{surf}}, \phi_{\mathrm{surf}})$.  

For the efficiency factors in the extreme supersonic limit, we first 
evaluate $\mu_{\mathrm{hit}}$ for $(N_1 + 1, N_1, N_3 + 1, N_3)$ values of  
$(\theta, \phi, \theta_{\mathrm{gr}}, \phi_{\mathrm{gr}})$.  We employ the 
trajectory-tracing algorithm described in \S \ref{sec:ray-tracing},
except that we start at location $(\theta, \phi)$ on the surface and move 
outward along $\bmath{\hat{N}}(\theta, \phi)$ to determine whether or not 
$(\theta, \phi)$ is shadowed by another part of the grain.  The integrals
in equations \ref{eq:Q-arr-sup} and 
\ref{eq:Q-gamma-arr-sup}--\ref{eq:Q-gamma-drag-out-sup} are then easily
evaluated.

\subsection{Code verification:  spherical grains}

Consider a uniform, spherical grain that drifts through the gas but does
not spin.  All of the contributions to the torque vanish and analytical 
results for the arrival rate and force
are available as functions of $s_d$.  From \citet{BW65}, 
\be
Q_{\mathrm{arr}} = \pi^{1/2} \exp(-s_d^2) + \pi \left( s_d + \frac{1}{2 s_d}
\right) \mathrm{erf}(s_d)
\ee
where ``erf'' denotes the error function.  Taking asymptotic limits,
\be
Q_{\mathrm{arr}}(s_d = 0) = 2 \pi^{1/2} , 
\ee
\be
Q^{\prime}_{\mathrm{arr}} = 0 , 
\ee
and 
\be
Q_{\mathrm{arr}}(\mathrm{supersonic}) = \pi s_d .
\ee

\citet*{BWA65} found that
$\mathbfit{Q}_{F, \mathrm{spec}} = 0$ and 
\be
\mathbfit{Q}_{F, \mathrm{arr}} = - \pi^{1/2} \left[ \left( s_d + \frac{1}{2 s_d} 
\right) \exp(- s_d^2) + \left( 1 + s_d^2 - \frac{1}{4 s_d^2} \right) \pi^{1/2}
\mathrm{erf}(s_d) \right] \bmath{\hat{s}}_d .
\ee
This reduces to the classic Epstein drag formula in the extreme subsonic 
limit, with $\mathbfit{Q}_{F, \mathrm{arr}}(s_d = 0) = 0$ and 
\be
\mathbfit{Q}^{\prime}_{F, \mathrm{arr}} = - \frac{8}{3} \pi^{1/2} \, 
\bmath{\hat{s}}_d .
\ee
In the extreme supersonic limit, 
$\mathbfit{Q}_{F, \mathrm{arr}} = - \pi s_d^2 \, \bmath{\hat{s}}_d$.  

Our codes reproduce all of these results for a spherical grain, for
which all of the GRS expansion coefficients $a_{lm}$ and $b_{lm}$ vanish.
We tested for numerous combinations of 
$(\theta_{\mathrm{gr}}, \phi_{\mathrm{gr}})$, in the extreme subsonic and 
supersonic limits and with $s_d = 1$.  

\section{Torque Calculations:  Computational Results}
\label{sec:torque-results0}

\subsection{Arrival rate and torques \label{sec:torque-results}}

In this section, we present computational results for grain 1.  

In order to check for convergence of the numerical integrals that appear
in the expressions for the efficiency factors, we first construct a table
of data used in computing the integrands with $N_1 = 2^7 = 128$.  Recall that
there are $N_1$ values of $\phi_{\mathrm{sph}}$ and $\phi_{\mathrm{in}}$ and
$N_1 + 1$ values of $\cos \theta_{\mathrm{sph}}$ and $\cos \theta_{\mathrm{in}}$.
Since $N_1$ is a power of 2, the tabulated data can be used to evaluate
the integrals with $N_1 = 16$, 32, 64, and 128.  We find that 64 is often
sufficient, though 128 is sometimes required.  
For $s_d = 10$, even $N_1 = 128$ is not sufficient for full convergence.  
For efficiency factors associated with outgoing particles, we typically 
adopt $N_2 = 256$ (recall that we divide the surface into $N_2^2$ patches
when examining the trajectories of outgoing particles).  We also ran some
computations with $N_2 = 128$ and 512 to check for convergence in this
parameter.  The number of orientations $(\theta_{\mathrm{gr}}, \phi_{\mathrm{gr}})$
(of the grain body relative to the drift velocity) for which quantities are
computed affects the convergence of the rotationally averaged values.  We
have tried $N_3 = 32$, 64, and 128 (as well as 256 in the case of the 
supersonic limit).  

\begin{figure}
\begin{minipage}{9.1cm}
\includegraphics[width=90mm]{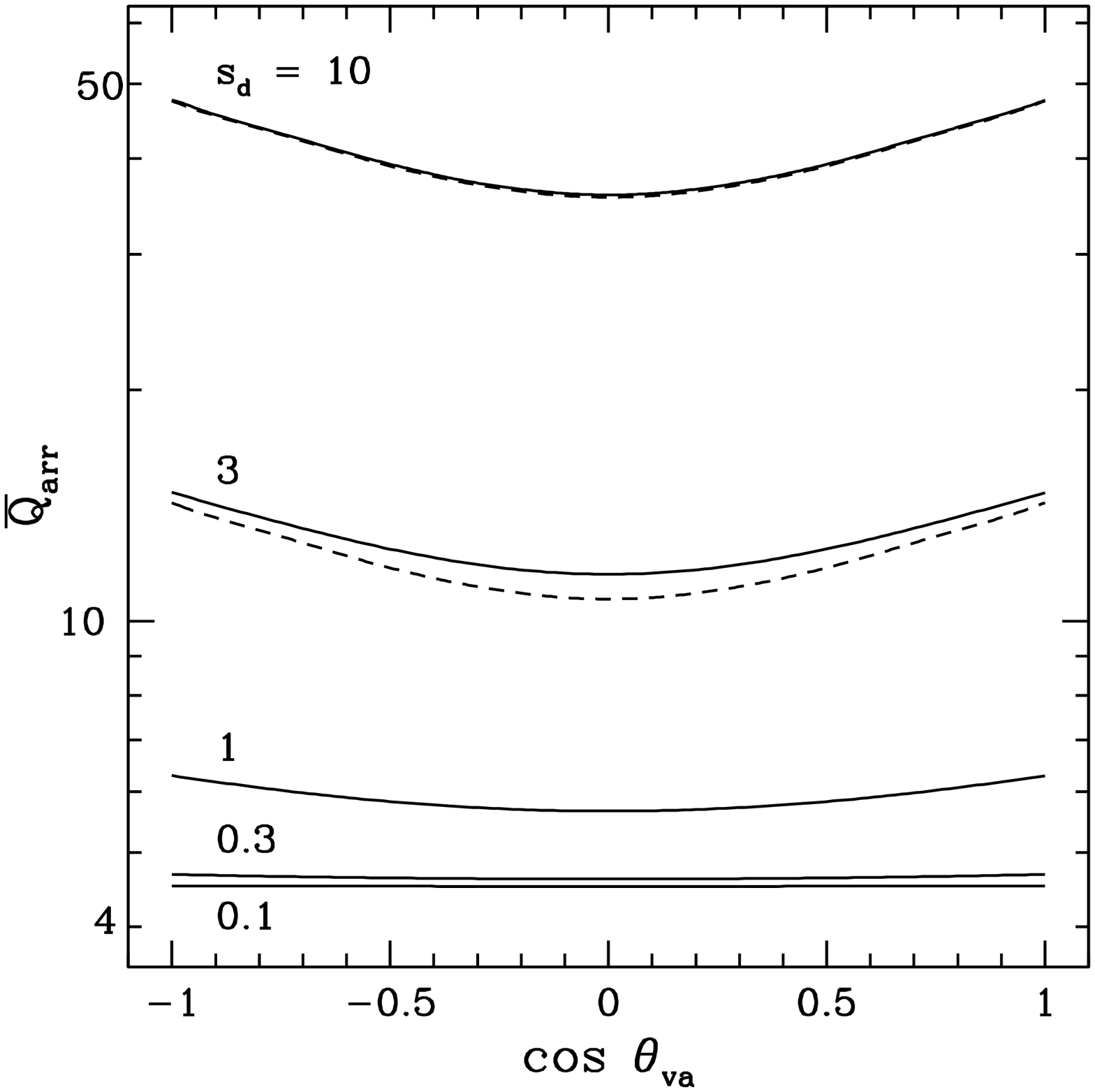}
\end{minipage}
\begin{minipage}{9.1cm}
\includegraphics[width=90mm]{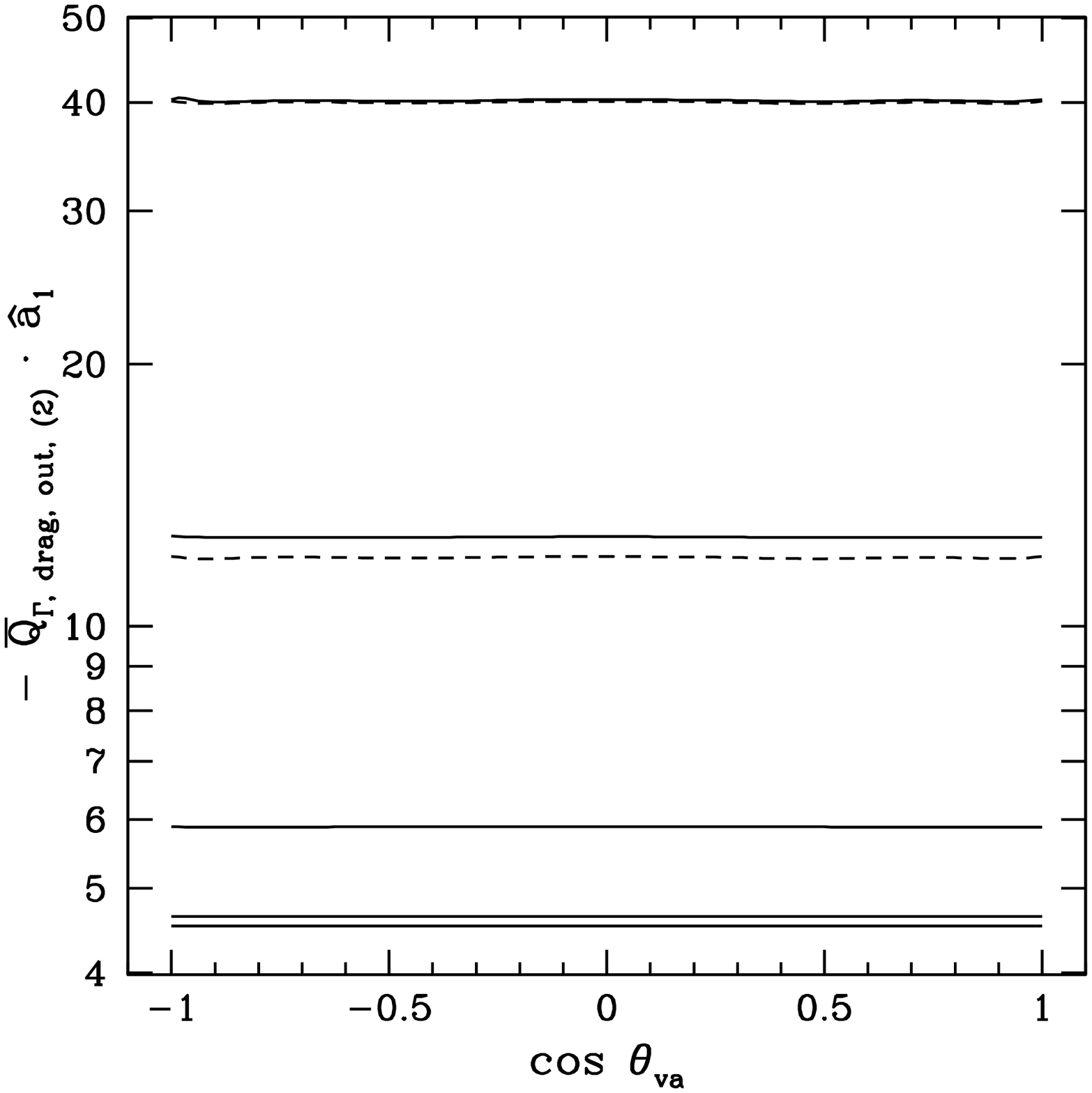}
\end{minipage}
\caption{
Left:
the grain 1 efficiency factor $\overbar{Q}_{\mathrm{arr}}$ for the rate at which 
gas atoms 
arrive at the grain surface, averaged over rotation about $\bmath{\hat{a}}_1$, 
as a function of the angle $\theta_{va}$ between $\bmath{\hat{a}}_1$ and the 
grain velocity for various values of the reduced grain drift speed $s_d$.  
Right:  the drag torque efficiency factor
$\mathbfit{Q}_{\Gamma, \mathrm{drag, \, out}, (2)}$
(component along $\bmath{\hat{a}}_1$) for the same values of $s_d$ (higher
curves are for higher $s_d$).  In both cases,  
dashed curves are the result for the extreme supersonic limit, scaled to 
$s_d = 3$ and 10.  
        }
\label{fig:q_arr}
\end{figure}

Fig. \ref{fig:q_arr} shows $\overbar{Q}_{\mathrm{arr}}$ for various values of $s_d$; 
from equation (\ref{eq:q_arr_sd0}), $Q_{\mathrm{arr}}(s_d = 0) = 4.57$.  The
dashed curves in Fig. \ref{fig:q_arr} are for the extreme supersonic limit,
scaled to $s_d = 3$ and 10.  Recall that, for the subsonic limit, the 
first-order dependence on $s_d$ ($Q^{\prime}_{\mathrm{arr}}$) vanishes.  Since 
$\bmath{\hat{a}}_1$ is the principal axis of greatest moment of inertia, the 
grain presents its largest cross-sectional area to the flowing gas when 
$\bmath{\hat{a}}_1$ lies along the velocity vector.  This explains the 
dependence of $\overbar{Q}_{\mathrm{arr}}$ on $\cos \theta_{va}$, which is most 
pronounced in the supersonic limit.  

\begin{figure}
\begin{minipage}{9.1cm}
\includegraphics[width=90mm]{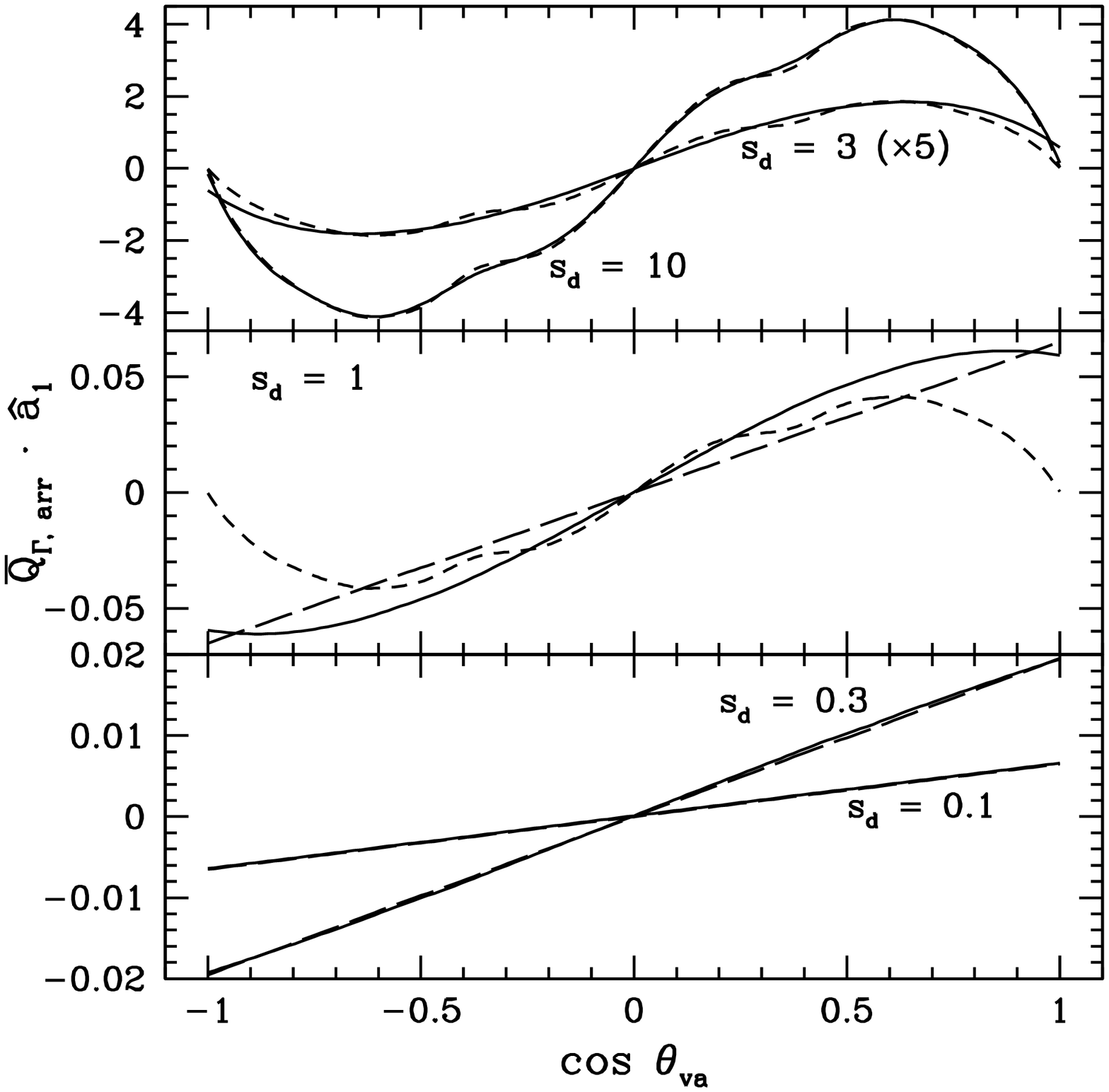}
\end{minipage}
\begin{minipage}{9.1cm}
\includegraphics[width=90mm]{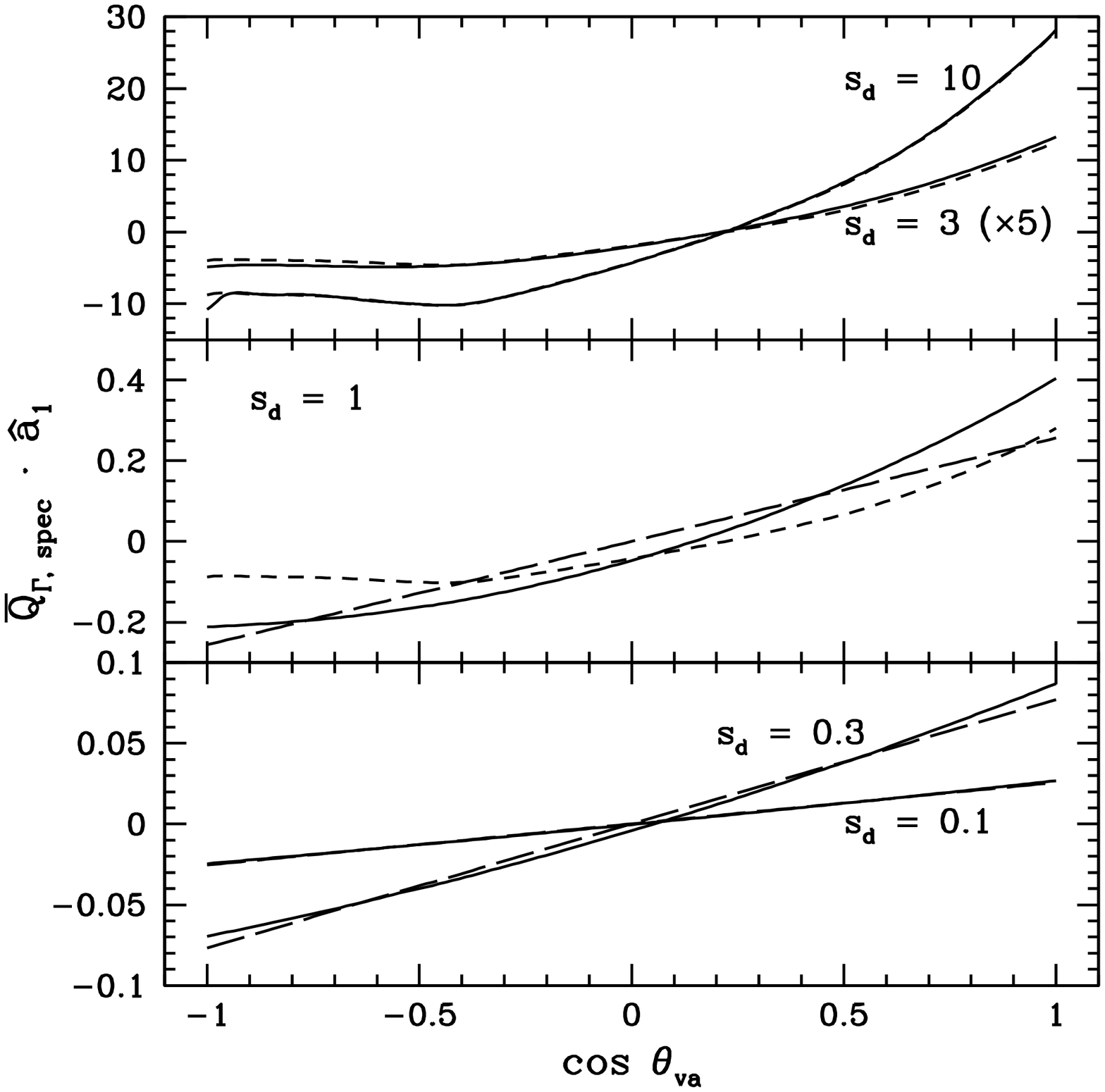}
\end{minipage}
\medskip
\begin{minipage}{9.1cm}
\includegraphics[width=90mm]{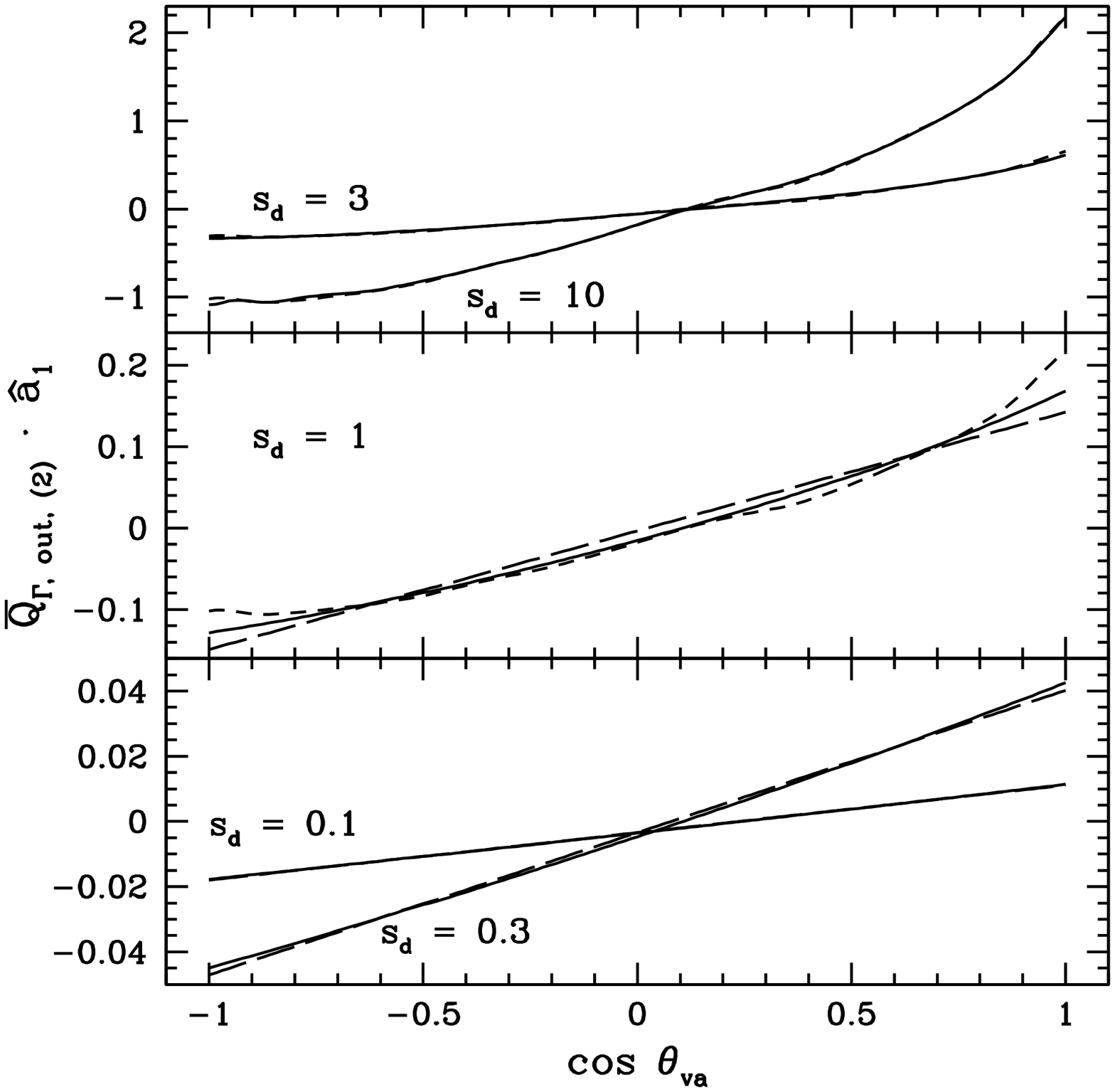}
\end{minipage}
\caption{
Rotationally averaged grain 1 efficiency factor for the 
torque component along $\bmath{\hat{a}}_1$ due to the arrival of gas atoms 
($\overbar{\mathbfit{Q}}_{\Gamma, \mathrm{arr}}$), specular reflection
($\overbar{\mathbfit{Q}}_{\Gamma, \mathrm{spec}}$), and outgoing scenario 2
($\overbar{\mathbfit{Q}}_{\Gamma, \mathrm{out, (2)}}$).
The lower, middle, and upper subpanels are for $s_d =$ (0.1, 0.3), 1.0, and 
(3.0, 10.0) respectively.  Long-dashed (short-dashed) curves are results for 
the extreme supersonic (subsonic) limits, scaled to the corresponding value of
$s_d$.   
        }
\label{fig:q1_arr}
\end{figure}

\begin{figure}
\begin{minipage}{9.1cm}
\includegraphics[width=90mm]{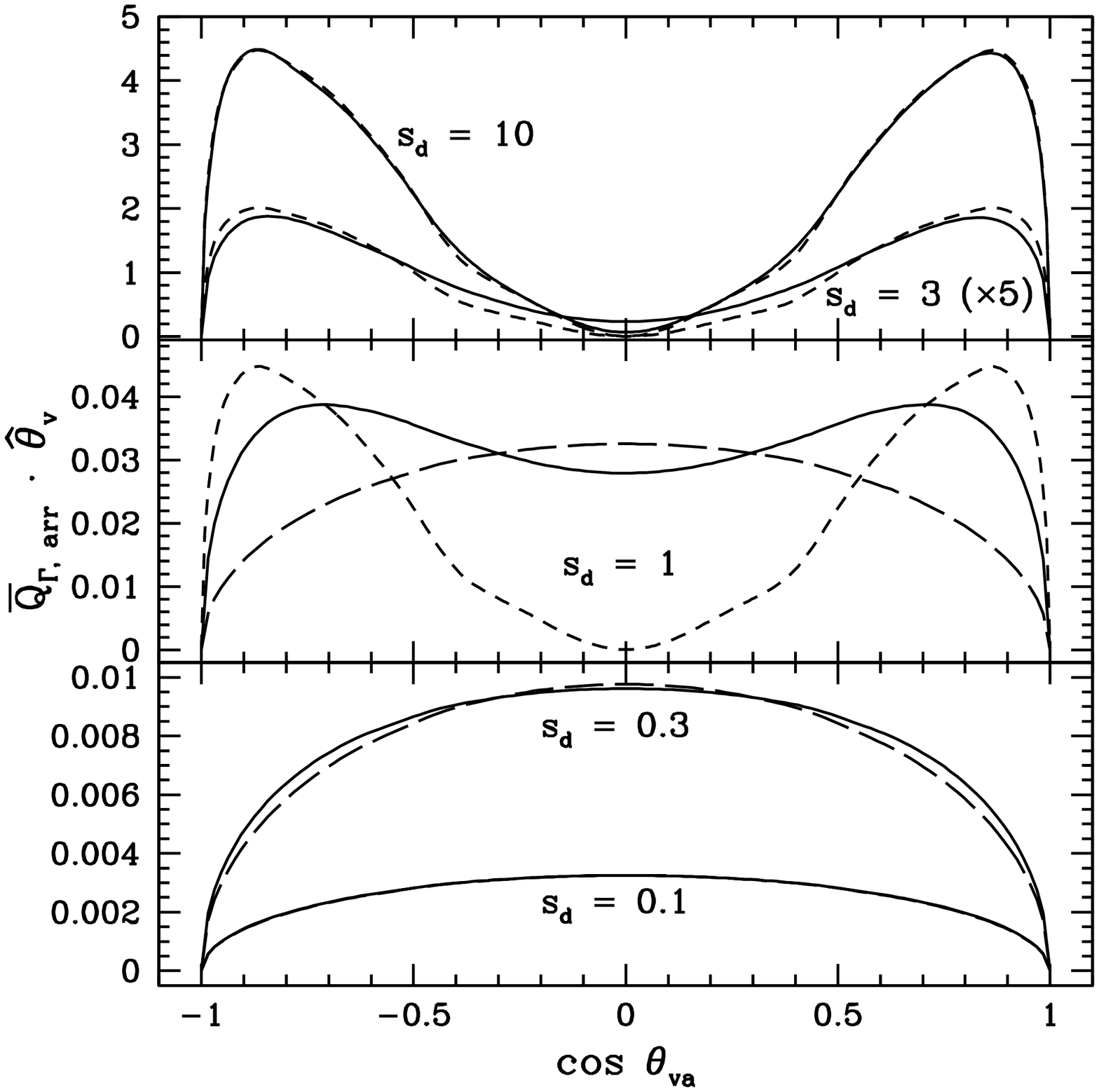}
\end{minipage}
\begin{minipage}{9.1cm}
\includegraphics[width=90mm]{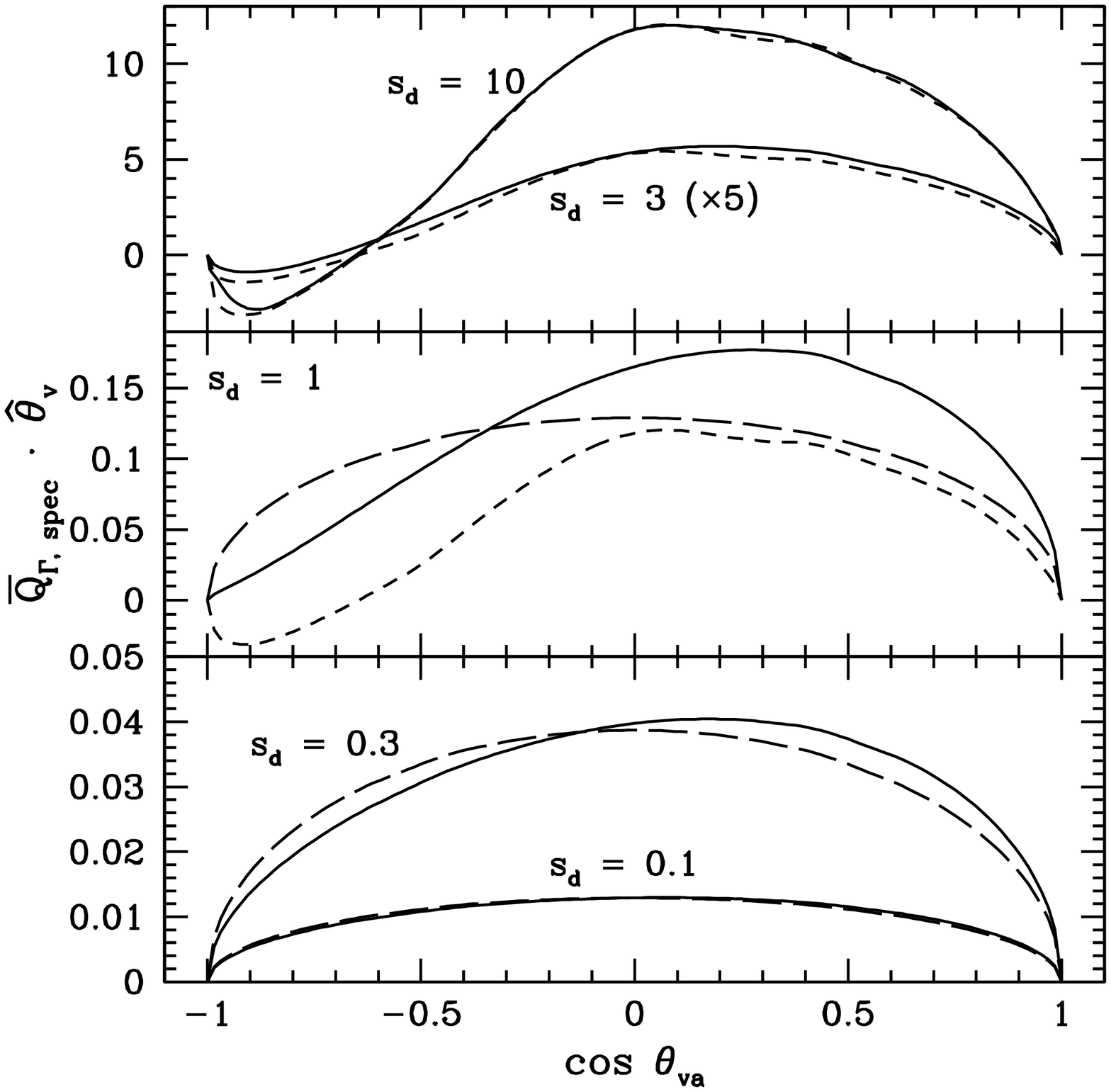}
\end{minipage}
\medskip
\begin{minipage}{9.1cm}
\includegraphics[width=90mm]{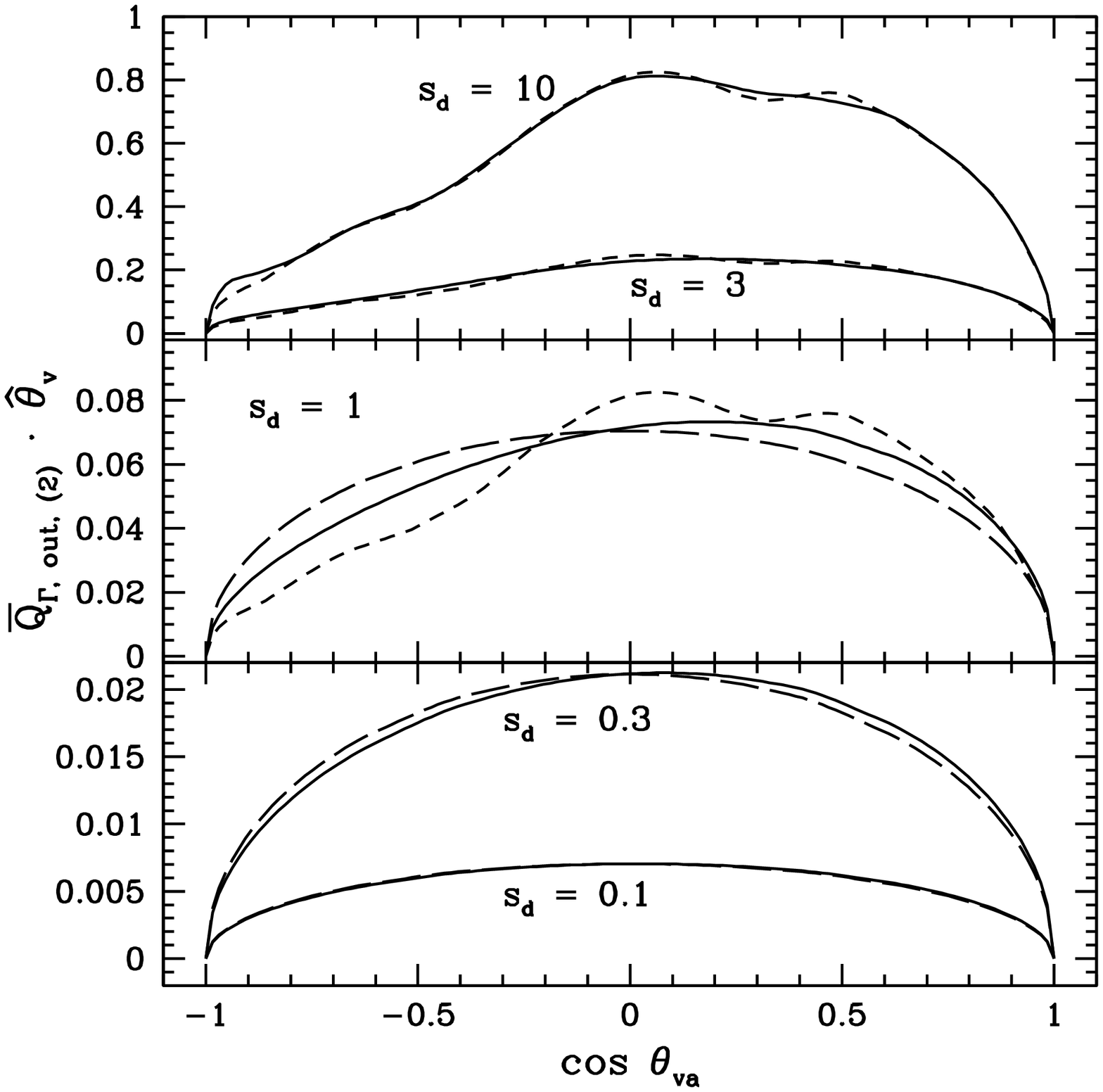}
\end{minipage}
\begin{minipage}{9.1cm}
\includegraphics[width=90mm]{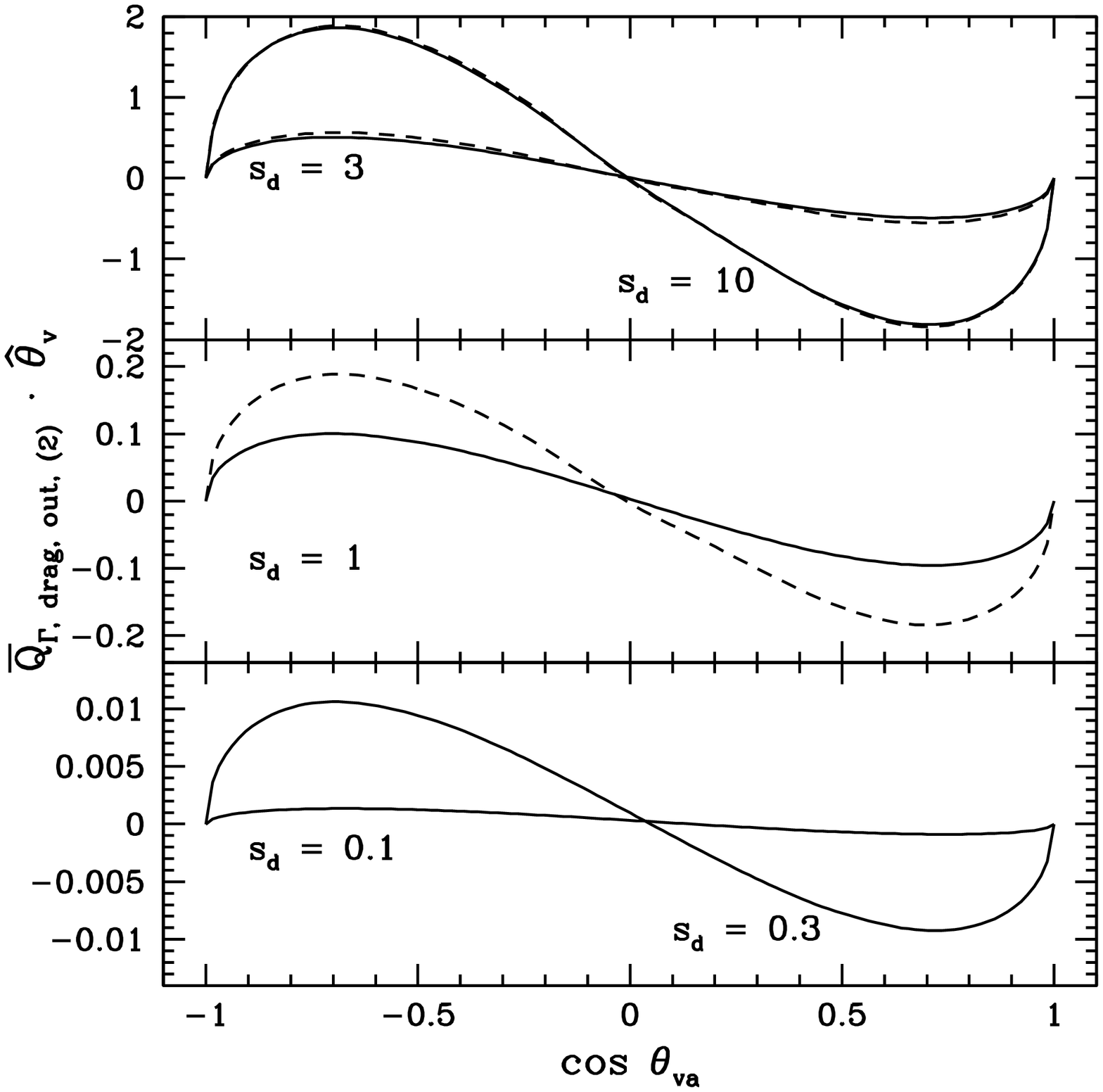}
\end{minipage}
\caption{
Same as Fig. \ref{fig:q1_arr}, except for the component along 
$\bmath{\hat{\theta}}_v$ and including the drag torque efficiency 
$\mathbfit{Q}_{\Gamma, \mathrm{drag, \, out}, (2)}$.
        }
\label{fig:q2_arr}
\end{figure}

\begin{figure}
\begin{minipage}{9.1cm}
\includegraphics[width=90mm]{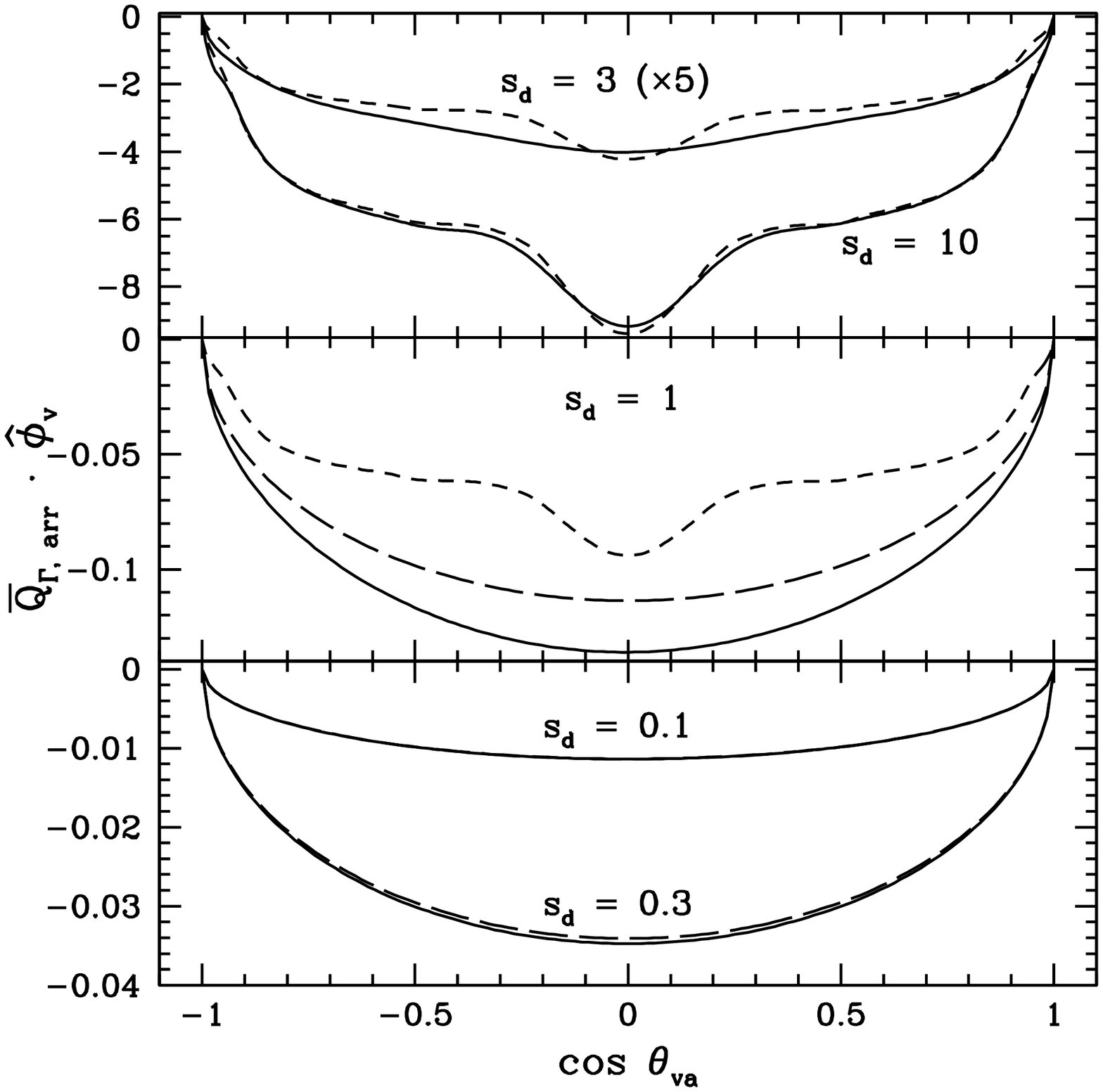}
\end{minipage}
\begin{minipage}{9.1cm}
\includegraphics[width=90mm]{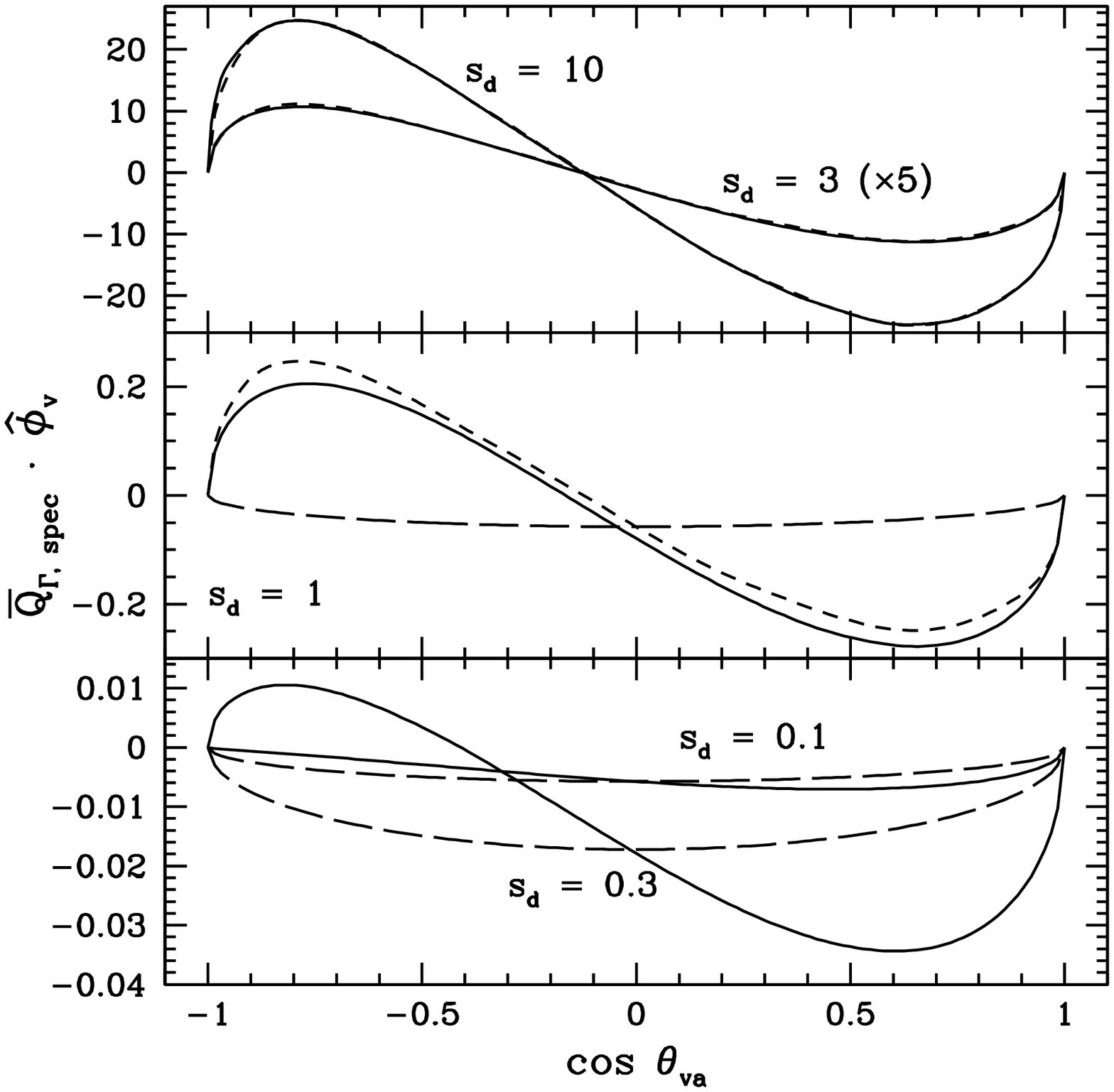}
\end{minipage}
\medskip
\begin{minipage}{9.1cm}
\includegraphics[width=90mm]{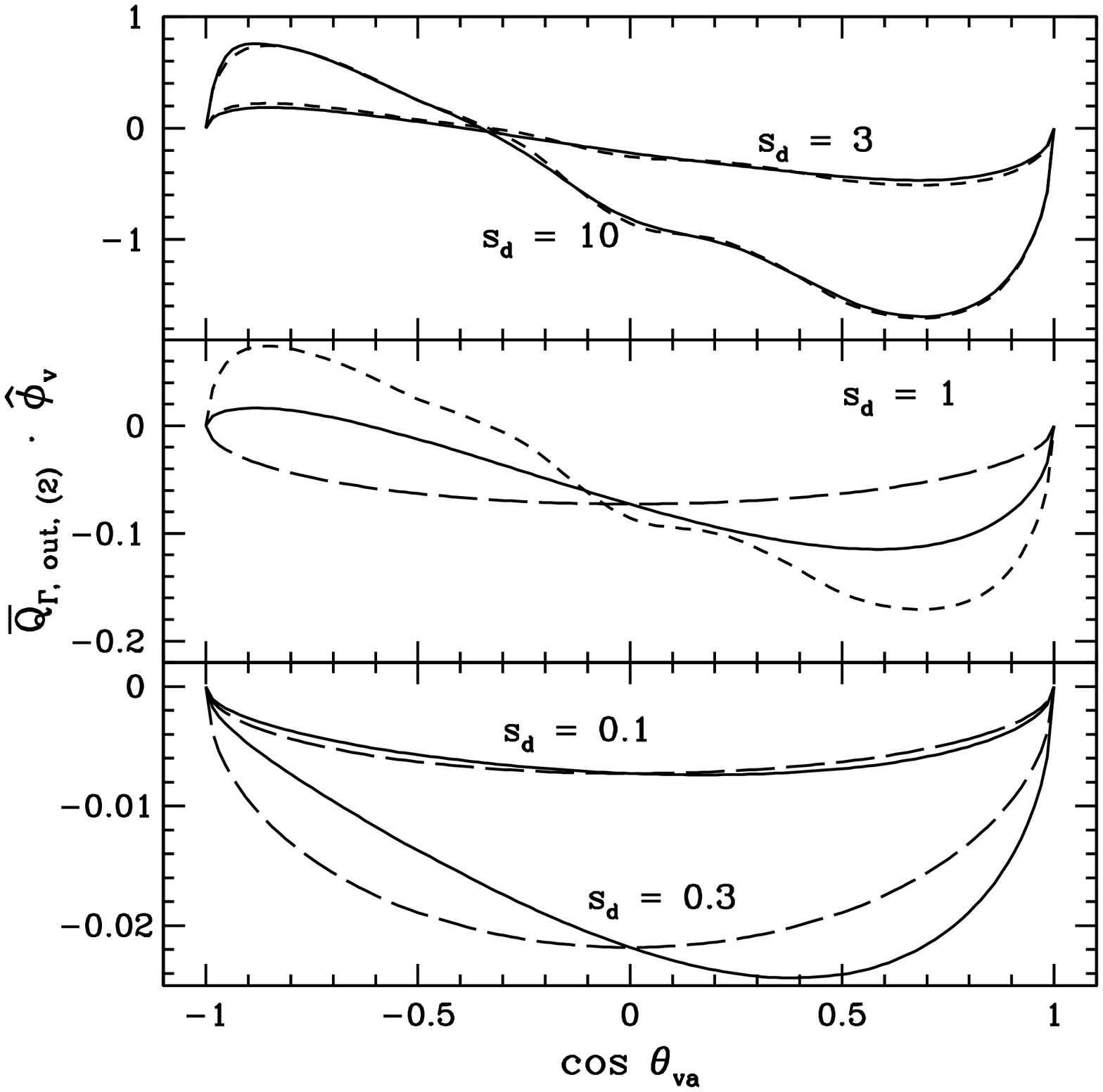}
\end{minipage}
\begin{minipage}{9.1cm}
\includegraphics[width=90mm]{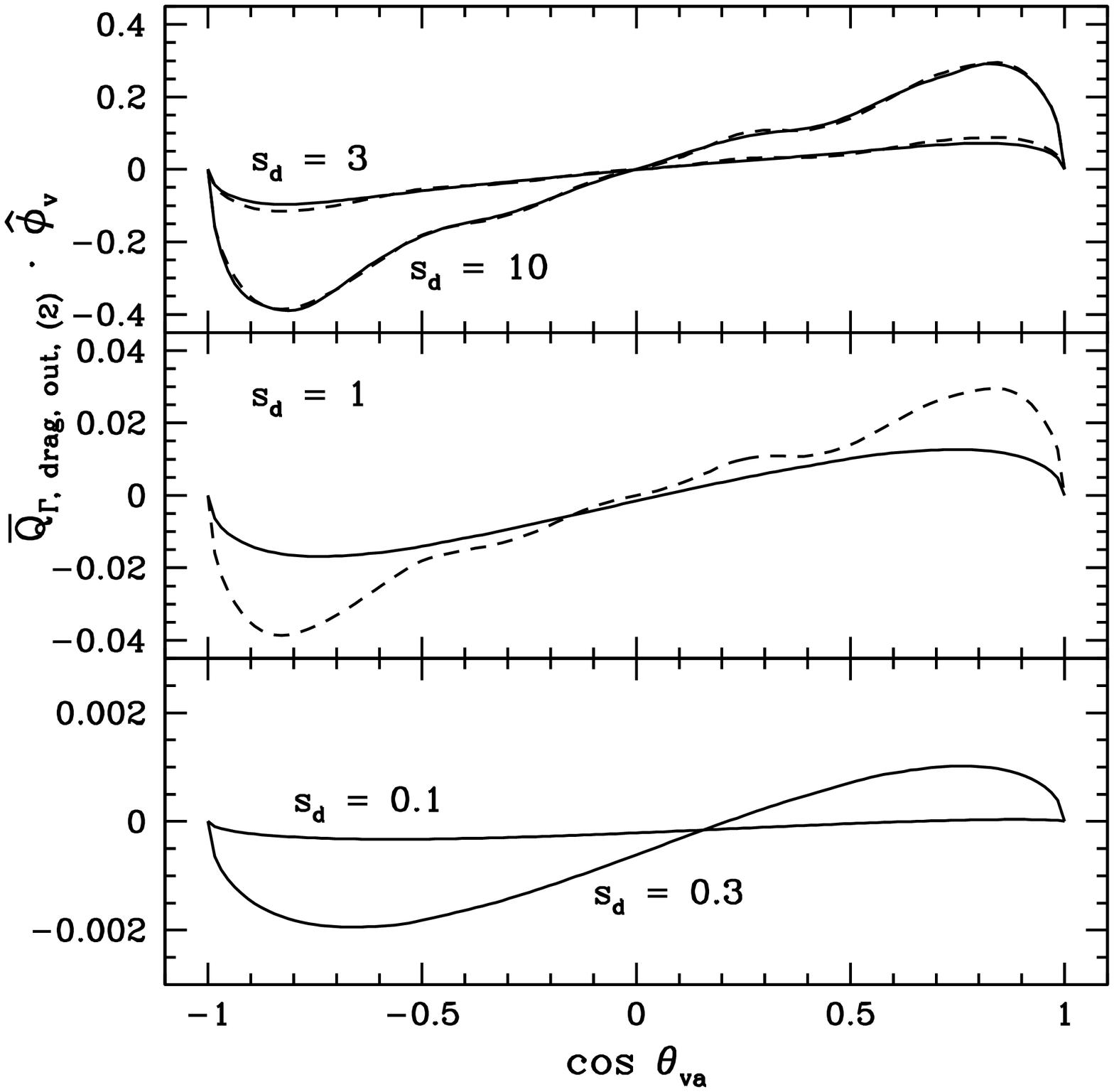}
\end{minipage}
\caption{
Same as Fig. \ref{fig:q2_arr}, except for the component along 
$\bmath{\hat{\phi}}_v$.
        }
\label{fig:q3_arr}
\end{figure}

Figs. \ref{fig:q1_arr}--\ref{fig:q3_arr} show the components of the
rotationally averaged efficiency factor for the torque due to arriving atoms 
($\overbar{\mathbfit{Q}}_{\Gamma, \mathrm{arr}}$), specular reflection
($\overbar{\mathbfit{Q}}_{\Gamma, \mathrm{spec}}$), and outgoing scenario 2
($\overbar{\mathbfit{Q}}_{\Gamma, \mathrm{out, (2)}}$), along $\bmath{\hat{a}}_1$,
$\bmath{\hat{\theta}}_v$, and $\bmath{\hat{\phi}}_v$ (defined in the last 
sentence in \S \ref{sec:suprathermal}).  In the absence of an interstellar 
magnetic field, these components drive rotation about $\bmath{\hat{a}}_1$, 
alignment of $\bmath{\hat{a}}_1$ with respect to the direction 
$\bmath{\hat{s}}_d$ of the grain drift, and precession of $\bmath{\hat{a}}_1$ 
about $\bmath{\hat{s}}_d$.  The solid curves are results for
$s_d = 0.1$, 0.3, 1.0, 3.0, and 10.0.  Results computed in the extreme 
subsonic limit and scaled to $s_d = 0.1$, 0.3, and 1.0 are displayed as 
long-dashed curves.  Similarly, results computed in the extreme 
supersonic limit and scaled to $s_d = 1.0$, 3.0, and 10.0 are displayed as 
short-dashed curves.  

As seen in Figs. \ref{fig:q_arr}--\ref{fig:q3_arr}, results associated with
arriving atoms for 
$s_d = 0.1$ and 0.3 agree very well with those for the subsonic limit and  
results for $s_d = 3$ and 10 agree very well with those for the supersonic 
limit.  Different computer
codes are used for computing results in the cases of a specified value of
$s_d$, the extreme subsonic limit, and the extreme supersonic limit.  As
described in the previous sections, the algorithm for the subsonic 
(supersonic) limit is somewhat (very) different from that for a specified
$s_d$.  Thus, the agreement of the results is confirmation of the validity
of the codes.  

The following features of $\overbar{\mathbfit{Q}}_{\Gamma, \mathrm{arr}}$  
exhibited 
in Figs. \ref{fig:q1_arr} and \ref{fig:q2_arr} are worth noting:  1.  
$\overbar{\mathbfit{Q}}_{\Gamma, \mathrm{arr}} \bmath{\cdot} \bmath{\hat{a}}_1$ is 
an odd function of $\cos \theta_{va}$ and is proportional to $\cos \theta_{va}$
for subsonic drift, 2.  
$\overbar{\mathbfit{Q}}_{\Gamma, \mathrm{arr}} \bmath{\cdot} \bmath{\hat{\theta}}_v$
is an even function of $\cos \theta_{va}$ and is proportional to 
$\sin \theta_{va}$ for subsonic drift, 3.  
$\overbar{\mathbfit{Q}}_{\Gamma, \mathrm{arr}} \bmath{\cdot} \bmath{\hat{a}}_1$ and
$\overbar{\mathbfit{Q}}_{\Gamma, \mathrm{arr}} \bmath{\cdot} \bmath{\hat{\theta}}_v$ 
have the same sign when $\cos \theta_{va} > 0$, 4.  
$\overbar{\mathbfit{Q}}_{\Gamma, \mathrm{arr}} \bmath{\cdot} \bmath{\hat{\theta}}_v
(\cos \theta_{va} = 0) \rightarrow 0$ as $s_d \rightarrow \infty$, 5.  
$\overbar{\mathbfit{Q}}_{\Gamma, \mathrm{arr}} \bmath{\cdot} \bmath{\hat{a}}_1
(\cos \theta_{va} = \pm 1) \rightarrow 0$ as $s_d \rightarrow \infty$, 6.  
$\overbar{\mathbfit{Q}}_{\Gamma, \mathrm{arr}} \bmath{\cdot} \bmath{\hat{\theta}}_v
(\cos \theta_{va} = \pm 1) = 0$.  As shown
in Appendix \ref{sec:arrival-special-results}, 
these properties are satisfied for all grain shapes.  Our
computational results exhibit most of these features for all 13 grains, 
providing further evidence that the code is robust.  There are slight deviations
from the expected form for 
$\overbar{\mathbfit{Q}}_{\Gamma, \mathrm{arr}} \bmath{\cdot} \bmath{\hat{\theta}}_v$
in the subsonic regime for grains 3, 10, and 11, and somewhat larger 
deviations for grains 5 and 9, suggesting that the computations are not fully 
converged in these cases.  In addition, the computational result for 
$\overbar{\mathbfit{Q}}_{\Gamma, \mathrm{arr}} \bmath{\cdot} \bmath{\hat{a}}_1$
is always slightly offset in $\cos \theta_{va}$; i.e.~it passes through zero
at a value of $\cos \theta_{va}$ slightly different from zero.  

In producing the curves in Figs. \ref{fig:q_arr}--\ref{fig:q3_arr}, we
adopted $N_1 = 128$ for $s_d = 0.1$--3.0 and $N_1 = 256$ for $s_d = 10$.
Given the close agreement between the results for $s_d = 10$ with the 
supersonic results scaled to $s_d =10$, we will simply adopt the latter for
grains 2--13.  This greatly reduces the computational time.  

Appendix \ref{sec:subsonic-special-results} notes some features that 
characterize all of the rotationally averaged torque efficiencies in the 
extreme subsonic limit.  We have verified that our results display these
features for all 13 grain shapes.

Fig. \ref{fig:q_arr} also displays the $\bmath{\hat{a}}_1$-component of the 
rotationally averaged drag torque efficiency factor 
$\overbar{\mathbfit{Q}}_{\Gamma, \mathrm{drag, \, out}, (2)}$
computed for outgoing scenario (2).  These results agree well with those 
computed in the extreme supersonic limit and scaled to $s_d = 3$ and 10.  
In the limit of low $s_d$, the results tend towards that found for $s_d = 0$:  
$\mathbfit{Q}_{\Gamma, \mathrm{drag, \, out}, (2)} \bmath{\cdot} \bmath{\hat{a}}_1 
= -4.51$.  We found an extremely weak first-order dependence of 
$\mathbfit{Q}_{\Gamma, \mathrm{drag, \, out}, (2)}$ on $s_d$; 
i.e.~$\mathbfit{Q}^{\prime}_{\Gamma, \mathrm{drag, \, out}, (2)} \bmath{\cdot} 
\bmath{\hat{a}}_1 \ll 1$
(and likewise for the other components).
We computed the second-order term and found that its inclusion substantially 
overestimates $Q_{\Gamma, \mathrm{drag, \, out}, (2)}$ for small $s_d$.  Evidently a 
power-series expansion converges slowly in the low-$s_d$ limit.  Figs.
\ref{fig:q2_arr} and \ref{fig:q3_arr} display $\bmath{\hat{\theta}}_v$- 
and $\bmath{\hat{\phi}}_v$-components of 
$\overbar{\mathbfit{Q}}_{\Gamma, \mathrm{drag, \, out}, (2)}$.  
Curves for the subsonic limit are not displayed because of the poor convergence
behaviour.  


For grains 2--13, plots of the rotationally averaged arrival efficiency 
$\overbar{Q}_{\mathrm{arr}}$ versus $\cos \theta_{va}$ look very similar to that 
for grain 1, but with somewhat smaller
magnitudes when $\alpha =3$ than when $\alpha = 2$.  Plots of the various
torque efficiencies versus $\cos \theta_{va}$ generally show a wide diversity
of shapes, again with the magnitudes often smaller when $\alpha =3$ than
when $\alpha =2$.  The components of the drag torque efficiency along
$\bmath{\hat{a}}_1$ and $\bmath{\hat{\theta}}_v$ are broadly similar, but the 
component along $\bmath{\hat{\phi}}_v$ varies considerably among the grain 
shapes.  

In outgoing scenario (1), the torque and drag efficiencies are both proportional
to the arrival efficiency $Q_{\mathrm{arr}}$.  Since the angle $\theta_{va}$
does not change when the grain rotates around $\bmath{\hat{a}}_1$, 
$\overbar{Q}_{\mathrm{arr}}$ remains constant for this motion.  Thus, the 
components of $\overbar{\mathbfit{Q}}_{\Gamma, \mathrm{out}, (1)}$ and 
$\overbar{\mathbfit{Q}}_{\Gamma, \mathrm{drag, out}, (1)}$ along 
$\bmath{\hat{\theta}}_v$ and $\bmath{\hat{\phi}}_v$ vanish. (We assume that
the time when an atom or molecule departs the grain surface is uncorrelated
with the arrival time of the atom.)  The components of the drag 
efficiency along $\bmath{\hat{a}}_1$ 
are given in Table \ref{tab:Qout2-over-Qarr}. For most of the grain shapes,
$\mathbfit{Q}_{\Gamma, \mathrm{out}, (1)} \bmath{\cdot} \bmath{\hat{a}}_1 / 
Q_{\mathrm{arr}}$ is consistent with zero, having not converged when 
evaluated using $(4096)^2$ patches on the surface.  The exceptions are
grain 4, for which $\mathbfit{Q}_{\Gamma, \mathrm{out}, (1)} \bmath{\cdot} 
\bmath{\hat{a}}_1 / Q_{\mathrm{arr}} = 0.00122$, and grains 2 and 3, for which 
$\mathbfit{Q}_{\Gamma, \mathrm{out}, (1)} \bmath{\cdot} \bmath{\hat{a}}_1 / 
Q_{\mathrm{arr}}$ appears to 
converge to $\sim -3 \times 10^{-5}$ and $\sim 6 \times 10^{-6}$, respectively.

\begin{table}
\caption{Drag efficiency factors in outgoing scenario (1).}
\label{tab:Qout2-over-Qarr}
\begin{tabular}{ll}
\hline
Grain index & 
$\mathbfit{Q}_{\Gamma, \mathrm{drag, out}, (1)} \bmath{\cdot} \bmath{\hat{a}}_1 / 
Q_{\mathrm{arr}}$\\
\hline
1 & -0.979\\
2 & -1.091\\
3 & -1.004\\
4 & -1.052\\
5 & -0.821\\
6 & -0.888\\
7 & -1.054\\
8 & -0.904\\
9 & -0.864\\
10 & -0.819\\
11 & -0.911\\
12 & -0.873\\
13 & -0.907\\
\hline
\end{tabular}
\end{table}

\subsection{Forces}

As noted in \S \ref{sec:force},
$\mathbfit{Q}_{F, \mathrm{arr}}$ and $\mathbfit{Q}_{F, \mathrm{spec}}$ both vanish 
when $s_d=0$.  Thus, the force is entirely drag when $f_{\mathrm{spec}} = 1$.  
The component of the rotationally averaged drag force antiparallel to the 
grain's velocity is
comparable in magnitude to that on a sphere (ranging between about 
75 and 230 per cent of that for a sphere in all cases) but varies with 
$\theta_{va}$, with its maximum value when $\cos \theta_{va} = \pm 1$ and
minimum near $\cos \theta_{va} = 0$.  There
is also a component perpendicular to the grain's velocity which vanishes at
$\cos \theta_{va} = \pm 1$ and near $\cos \theta_{va} = 0$ and can reach values
as high as about 30 per cent of the drag force on a spherical grain.  

For outgoing scenario (2), the force can be non-zero when $s_d=0$.  However,
we have found that this term does not contribute substantially even when 
$s_d=0.1$.  The drag force in this case is qualitatively and quantitatively 
similar to that in the case of specular reflection.

For outgoing scenario (1), $\mathbfit{Q}_{F, \mathrm{out}, (1)}$ is proportional to 
$Q_{\mathrm{arr}}$ and its direction is fixed in grain-body coordinates; only
the component along $\bmath{\hat{a}}_1$ is non-zero when averaged over the grain
rotation.  For most shapes, 
$\mathbfit{Q}_{F, \mathrm{out}, (1)} \bmath{\cdot} \bmath{\hat{a}}_1 / 
Q_{\mathrm{arr}}$ is consistent with zero, having not converged when 
evaluated using $(4096)^2$ patches on the surface.
The exceptions are grains 3 and 4, for which 
$\mathbfit{Q}_{F, \mathrm{out}, (1)} \bmath{\cdot} \bmath{\hat{a}}_1 / Q_{\mathrm{arr}} 
= 0.00128$ and $-0.00344$, respectively, and grain 2, for which 
$\mathbfit{Q}_{F, \mathrm{out}, (1)} \bmath{\cdot} \bmath{\hat{a}}_1 / 
Q_{\mathrm{arr}}$ appears to converge to $\sim 2 \times 10^{-5}$.    
The drag force associated with the arriving
atoms only is similar to that for the above cases, with a somewhat smaller
range of magnitudes.  

\section{Dynamics}
\label{sec:dynamics}

\subsection{Equations of motion}

Consider a coordinate system $(x_B, y_B, z_B)$ with the interstellar 
magnetic field aligned along $\bmath{\hat{z}}_B$.  Take the velocity 
$\mathbfit{s}_d$ of the grain relative to the gas to lie in the $x_B$--$z_B$ 
plane, at angle $\psi_v$ to $\bmath{\hat{z}}_B$.  Assume that the grain
rotates steadily about $\bmath{\hat{a}}_1$, whose orientation is described by 
spherical coordinates $(\xi, \phi_B)$:
\be
\bmath{\hat{a}}_1 = \sin \xi \cos \phi_B \, \bmath{\hat{x}}_B + 
\sin \xi \sin \phi_B \, \bmath{\hat{y}}_B + \cos \xi \, \bmath{\hat{z}}_B . 
\ee
With these definitions, the angle between the grain velocity and 
$\bmath{\hat{a}}_1$ is given by
\be
\cos \theta_{va} = \sin \psi_v \sin \xi \cos \phi_B + \cos \psi_v \cos \xi .
\ee
The transformation between the coordinates 
$(\theta_v, \phi_v)$ introduced at the end of \S \ref{sec:suprathermal} and 
$(\xi, \phi_B)$ is given by
\be
\bmath{\hat{\xi}} \bmath{\cdot} \bmath{\hat{\theta}}_v = \bmath{\hat{\phi}}_B 
\bmath{\cdot} \bmath{\hat{\phi}}_v =
b_1 \equiv \frac{\cos \psi_v \sin \xi - \sin \psi_v \cos \xi \cos \phi_B}{\sin 
\theta_{va}} , 
\ee
\be
- \bmath{\hat{\xi}} \bmath{\cdot} \bmath{\hat{\phi}}_v = \bmath{\hat{\phi}}_B 
\bmath{\cdot} \bmath{\hat{\theta}}_v = 
b_2 \equiv \frac{\sin \psi_v \sin \phi_B}{\sin \theta_{va}} . 
\ee
When $\sin \theta_{va} = 0$, $\bmath{\hat{\theta}}_v = \bmath{\hat{\xi}}$ and
$\bmath{\hat{\phi}}_v = \bmath{\hat{\phi}}_B$ (i.e.~$b_1 = 1$ and $b_2 = 0$).  

In spherical coordinates, the rotationally averaged mechanical torque is given 
by
\be
\overbar{\bmath{\Gamma}}_{\mathrm{mech}} = m n v^2_{\mathrm{th}} a_{\mathrm{eff}}^3 
\left[ J^0_v(\xi, \phi_B) \, \bmath{\hat{\xi}} + G^0_v(\xi, \phi_B) \, 
\bmath{\hat{\phi}}_B + H^0_v(\xi, \phi_B) \, \bmath{\hat{a}}_1 \right] ; 
\ee
from equations \ref{eq:total-torque} and \ref{eq:q-total-torque}, 
\be
J^0_v(\xi, \phi_B) = b_1 \overbar{\mathbfit{Q}}_{\Gamma, \mathrm{mech}} \bmath{\cdot} 
\bmath{\hat{\theta}}_v - b_2
\overbar{\mathbfit{Q}}_{\Gamma, \mathrm{mech}} \bmath{\cdot} \bmath{\hat{\phi}}_v ,
\ee
\be
G^0_v(\xi, \phi_B) = b_2 \overbar{\mathbfit{Q}}_{\Gamma, \mathrm{mech}} \bmath{\cdot} 
\bmath{\hat{\theta}}_v + b_1 \overbar{\mathbfit{Q}}_{\Gamma, \mathrm{mech}} 
\bmath{\cdot} \bmath{\hat{\phi}}_v , 
\ee
\be
H^0_v(\xi, \phi_B) = \overbar{\mathbfit{Q}}_{\Gamma, \mathrm{mech}} \bmath{\cdot} 
\bmath{\hat{a}}_1 .
\ee

Like the mechanical torque, the drag torque can vary as a function of 
$\theta_{va}$ and may have components along $\bmath{\hat{\xi}}$ and 
$\bmath{\hat{\phi}}_B$ as well as along $\bmath{\hat{a}}_1$.  Defining 
\be
Q_{\Gamma, \mathrm{drag}, 0} = - \overbar{\mathbfit{Q}}_{\Gamma, \mathrm{drag}}(\cos 
\theta_{va} = 0) \bmath{\cdot} \bmath{\hat{a}}_1
\ee
and the drag time-scale as
\be
\tau_{\mathrm{drag}} = \frac{I_1}{m n v_{\mathrm{th}} a_{\mathrm{eff}}^4
Q_{\Gamma, \mathrm{drag}, 0}}
= 2.47 \times 10^5 \, \mathrm{yr} \left( \frac{\rho}{3 \, 
\mathrm{g} \, \mathrm{cm}^{-3}} \right) \left( \frac{a_{\mathrm{eff}}}
{0.2 \, \mu \mathrm{m}} \right) \left( \frac{T_{\mathrm{gas}}}{100 \, 
\mathrm{K}} \right)^{-1/2} \left( \frac{m}{m_p} \right)^{-1/2} \left(
\frac{n}{30 \, \mathrm{cm}^{-3}} \right)^{-1} \frac{\alpha_1}
{Q_{\Gamma, \mathrm{drag}, 0}} ,
\ee
the rotationally averaged drag torque can be expressed as
\be
\overbar{\bmath{\Gamma}}_{\mathrm{drag}} = \frac{I_1 \omega}{\tau_{\mathrm{drag}}} \ 
\frac{\overbar{\mathbfit{Q}}_{\Gamma, \mathrm{drag}}}{Q_{\Gamma, \mathrm{drag}, 0}} .
\ee
In analogy with the mechanical torque,
\be
\overbar{\bmath{\Gamma}}_{\mathrm{drag}} = \frac{I_1 \omega}{\tau_{\mathrm{drag}}} 
\left[ - H^0_{\mathrm{drag}}(\xi, \phi_B) \, \bmath{\hat{a}}_1 + J^0_{\mathrm{drag}}
(\xi, \phi_B) \, \bmath{\hat{\xi}} + G^0_{\mathrm{drag}}(\xi, \phi_B) \, 
\bmath{\hat{\phi}}_B \right]
\ee
with
\be
H^0_{\mathrm{drag}}(\xi, \phi_B) = - 
\frac{\overbar{\mathbfit{Q}}_{\Gamma, \mathrm{drag}} 
\bmath{\cdot} \bmath{\hat{a}}_1}{Q_{\Gamma, \mathrm{drag}, 0}} , 
\ee
\be
J^0_{\mathrm{drag}}(\xi, \phi_B) = 
\frac{b_1 (\overbar{\mathbfit{Q}}_{\Gamma, \mathrm{drag}} \bmath{\cdot} 
\bmath{\hat{\theta}}_v) - b_2 (\overbar{\mathbfit{Q}}_{\Gamma, \mathrm{drag}} 
\bmath{\cdot} \bmath{\hat{\phi}}_v)}{Q_{\Gamma, \mathrm{drag}, 0}} , 
\ee
\be
G^0_{\mathrm{drag}}(\xi, \phi_B) = 
\frac{b_2 (\overbar{\mathbfit{Q}}_{\Gamma, \mathrm{drag}} \bmath{\cdot} 
\bmath{\hat{\theta}}_v) + b_1 (\overbar{\mathbfit{Q}}_{\Gamma, \mathrm{drag}} 
\bmath{\cdot} \bmath{\hat{\phi}}_v)}{Q_{\Gamma, \mathrm{drag}, 0}} .
\ee

We will consider five separate cases for the mechanical torque.  In the
first, $f_{\mathrm{spec}} = 1$, i.e.~all of the arriving atoms reflect 
specularly.  In the other cases, $f_{\mathrm{spec}} = 0$; we consider both
outgoing scenarios (1) and (2) with the outgoing particles either H atoms or
H$_2$ molecules.  Since we have not evaluated the drag torque for 
specular reflection, we will simply adopt the drag efficiency for outgoing 
atoms under scenario (2) in this case.  

In order to ascertain the potential of the mechanical torque in aligning
grains, we will examine the rotational dynamics under the action of 
only the mechanical, drag, and magnetic torques.  This final torque, 
due to the interaction of the grain's Barnett magnetic moment with the
interstellar magnetic field, is given by
\be
\bmath{\Gamma}_B = \bmath{\hat{\phi}}_B I_1 \Omega_B \omega \sin \xi 
\ee
where the precession frequency is \citep[see e.g.][]{WD03}
\be
\Omega_B \approx 25 \, \mathrm{yr}^{-1} \left( \frac{a}{0.1 \, \mu \mathrm{m}}
\right)^{-2} \left( \frac{\alpha_1 \rho}{3 \, \mathrm{g} \, \mathrm{cm}^{-3}} 
\right)^{-1} \left( \frac{\chi_0}{3.3 \times 10^{-4}} \right) \left( 
\frac{B}{5 \mu \mathrm{G}} \right) ;
\ee
$\chi_0$ is the static magnetic susceptibility of the grain material.
The following analysis closely follows that in \citet{DW97}
with mechanical torques taking the place of radiative torques.  
The equation of motion,
\be
\frac{\mathrm{d}\mathbfit{J}}{\mathrm{d}t} = I_1 \frac{d}{dt}\left(\omega 
\bmath{\hat{a}}_1 \right) = 
\overbar{\bmath{\Gamma}}_{\mathrm{mech}} + \overbar{\bmath{\Gamma}}_{\mathrm{drag}} 
+ \bmath{\Gamma}_B
\ee
yields three component equations:
\be
\label{eq:domega-dt1}
\frac{\mathrm{d}\omega^{\prime}}{\mathrm{d}t^{\prime}} = M_v H^0_v(\xi, \phi_B) - 
H^0_{\mathrm{drag}}(\xi, \phi_B) \, \omega^{\prime} , 
\ee
\be
\label{eq:dxi-dt1}
\frac{\mathrm{d}\xi}{\mathrm{d}t^{\prime}} = M_v \ \frac{J^0_v(\xi, \phi_B)}
{\omega^{\prime}} + J^0_{\mathrm{drag}}(\xi, \phi_B) , 
\ee
\be
\label{eq:dphi-dt}
\frac{\mathrm{d}\phi_B}{\mathrm{d}t^{\prime}} = \Omega_B \tau_{\mathrm{drag}} + 
M_v \ \frac{G^0_v(\xi, \phi_B)}{\omega^{\prime} \sin \xi} + \frac{G^0_{\mathrm{drag}}
(\xi, \phi_B)}{\sin \xi} , 
\ee
where
\be
t^{\prime} = \frac{t}{\tau_{\mathrm{drag}}} , 
\ee
\be
\omega^{\prime} = \frac{\omega}{\omega_T} , 
\ee
the thermal rotation rate is given by \citep[see e.g.][]{DW97} 
\be
\label{eq:omega-thermal}
\omega_T = \left( \frac{15 kT_{\mathrm{gas}}}{8 \pi \alpha_1 \rho 
a_{\mathrm{eff}}^5} \right)^{1/2} 
= 1.66 \times 10^5 \, \mathrm{s}^{-1} \, \alpha_1^{-1/2} \left( 
\frac{T_{\mathrm{gas}}}{100 \, \mathrm{K}} \right)^{1/2} \left( \frac{\rho}{3 \, 
\mathrm{g} \, \mathrm{cm}^{-3}} \right)^{-1/2} \left( \frac{a_{\mathrm{eff}}}
{0.1 \, \mu \mathrm{m}} \right)^{-2.5} ,
\ee
and 
\be
M_v = \frac{4}{Q_{\Gamma, \mathrm{drag}, 0}} \left( \frac{\pi \alpha_1 \rho
a_{\mathrm{eff}}^3}{15 m} \right)^{1/2}
= 7.75 \times 10^4 \left( \frac{a_{\mathrm{eff}}}{0.1 \, \mu \mathrm{m}}
\right)^{3/2} \left( \frac{\rho}{3 \, \mathrm{g} \, \mathrm{cm}^{-3}}
\right)^{1/2} \left( \frac{m}{m_p} \right)^{-1/2} \frac{\alpha_1^{1/2}}
{Q_{\Gamma, \mathrm{drag}, 0}} .
\ee

Typically, $\Omega_B \tau_{\mathrm{drag}}$ greatly exceeds all of the other 
terms on the right-hand sides in equations 
(\ref{eq:domega-dt1})--(\ref{eq:dphi-dt}).  Thus, we will approximate the
motion in $\phi_B$ as uniform precession and average over this motion in
equations (\ref{eq:domega-dt1}) and (\ref{eq:dxi-dt1}):
\be
\label{eq:domega-dt}
\frac{\mathrm{d}\omega^{\prime}}{\mathrm{d}t^{\prime}} = M_v H_v(\xi) - 
H_{\mathrm{drag}}(\xi) \, \omega^{\prime} , 
\ee
\be
\label{eq:dxi-dt}
\frac{\mathrm{d}\xi}{\mathrm{d}t^{\prime}} = M_v \ \frac{J_v(\xi)}
{\omega^{\prime}} + J_{\mathrm{drag}}(\xi)
\ee
with 
\be
H_v(\xi) = \frac{1}{2 \pi} \int_0^{2\pi} \mathrm{d}\phi_B \, H^0_v(\xi, \phi_B)
\ee
and likewise for $J(\xi)$, $H_{\mathrm{drag}}(\xi)$, and $J_{\mathrm{drag}}(\xi)$.
Of course, when $\xi \ll 1$, the terms in equation (\ref{eq:dphi-dt}) with 
$\sin \xi$ in the denominator cannot be neglected compared with 
$\Omega_B \tau_{\mathrm{drag}}$.  However, in this case the orientation of 
the grain in space hardly depends on $\phi_B$, so the assumption of a 
uniform distribution in $\phi_B$ does not introduce significant error.  The
terms in equations (\ref{eq:dxi-dt1}) and (\ref{eq:dphi-dt}) with 
$\omega^{\prime}$ in the denominator can be comparable to 
$\Omega_B \tau_{\mathrm{drag}}$ for sufficiently small $\omega^{\prime}$.  
However, the analysis already fails for such small $\omega^{\prime}$ since
the assumption of steady rotation about $\bmath{\hat{a}}_1$ is only justified 
in the limit of suprathermal rotation.  

\subsection{Stationary points}

Setting $\mathrm{d}\xi/\mathrm{d}t^{\prime} = 0$ and 
$\mathrm{d}\omega^{\prime}/\mathrm{d}t^{\prime} = 0$ in equations 
(\ref{eq:domega-dt}) and (\ref{eq:dxi-dt}), we find that stationary 
points $(\xi_s, \omega^{\prime}_s)$ occur where 
$\xi_s$ is a zero of the function
\be
\label{eq:Z_v}
Z_v(\xi) = J_v(\xi) H_{\mathrm{drag}}(\xi) + J_{\mathrm{drag}}(\xi) H_v(\xi) .
\ee
Since $J_v$ and $J_{\mathrm{drag}}$ both vanish at $\xi = 0$ and $\pi$, there are 
always stationary points at these $\xi$.  For a given $\xi_s$, 
\be
\omega^{\prime}_s = \frac{M_v H_v(\xi_s)}{H_{\mathrm{drag}}(\xi_s)} .
\ee

A stationary point is characterized by linearizing equations
(\ref{eq:domega-dt}) and (\ref{eq:dxi-dt}) about 
$(\xi_s, \omega^{\prime}_s)$:
\be
\label{eq:linearized1}
\frac{\mathrm{d}\xi}{\mathrm{d}t^{\prime}} \approx A_l (\xi - \xi_s) + B_l 
(\omega^{\prime} - \omega^{\prime}_s) , 
\ee
\be
\frac{\mathrm{d}\omega^{\prime}}{\mathrm{d}t^{\prime}} \approx C_l (\xi - \xi_s) 
+ D_l (\omega^{\prime} - \omega^{\prime}_s)
\ee
with 
\be
A_l = \frac{M_v}{\omega^{\prime}_s} \left. \frac{\mathrm{d}J_v}
{\mathrm{d}\xi} \right|_{\xi = \xi_s} + \left. \frac{\mathrm{d}
J_{\mathrm{drag}}}{\mathrm{d}\xi} \right|_{\xi = \xi_s} , 
\ee
\be
B_l = - \frac{M_v J_v(\xi_s)}{(\omega^{\prime}_s)^2} , 
\ee
\be
C_l = M_v \left. \frac{\mathrm{d}H_v}{\mathrm{d}\xi} \right|_{\xi = \xi_s} -
\omega^{\prime}_s \left. \frac{\mathrm{d}H_{\mathrm{drag}}}{\mathrm{d}\xi} 
\right|_{\xi = \xi_s} , 
\ee
\be
\label{eq:D_l}
D_l = - H_{\mathrm{drag}}(\xi_s) .
\ee
The displacement from the stationary point is proportional to 
$\exp(\lambda_l t^{\prime})$ where
\be
\label{eq:eigenvalue}
\lambda_l = \frac{A_l + D_l \pm [ (A_l + D_l)^2 - 4 (A_l D_l - B_l C_l)]^{1/2}}
{2} .
\ee
Thus, the stationary point is stable (an `attractor') if 
\be
\label{eq:attractor-conditions}
A_l + D_l < 0 \ \ \ \mathrm{and} \ \ \ 
B_l C_l - A_l D_l < 0 ; 
\ee
otherwise it is unstable (a `repeller').  The time-scale for approach to
the stationary point is $- [\mathrm{Re}(\lambda_l)]^{-1}$.  
We expect the alignment to be characterized by the longer relaxation time
and have verified this by
numerically integrating equations (\ref{eq:domega-dt}) and
(\ref{eq:dxi-dt}) for various values of $s_d$ and $\psi_v$. Thus, the 
alignment time is
\be
\label{eq:tau-align}
\tau_{\mathrm{align}} = - \frac{2 \tau_{\mathrm{drag}}}{A_l + D_l + 
[(A_l - D_l)^2 + 4 B_l C_l]^{1/2}} .
\ee

\subsection{Crossover points}

Crossover points, where $\omega^{\prime}$ crosses zero, can only occur at
angles $\xi_c$ where $J_v(\xi_c) = 0$; otherwise there is a singularity
in equation (\ref{eq:dxi-dt}).  Since $J_v(\xi)$ vanishes at $\xi = 0$ 
and $\pi$, crossovers are always found at these angles.  The polarity of a 
crossover is the sign of $\mathrm{d}\omega^{\prime}/\mathrm{d}t^{\prime}$; from 
equation (\ref{eq:domega-dt}), 
\be
\mathrm{polarity} \ = \ \mathrm{sign}[H_v(\xi_c)] .
\ee
A crossover attractor is a crossover for which trajectories with $\xi$
near $\xi_c$ converge to the crossover, whereas the trajectories diverge
from $\xi_c$ for a crossover repeller.  At a crossover repeller, only the
single trajectory with $\xi = \xi_c$ passes through the crossover.  Since
this occurs with infinitesimal probability, physical crossovers only occur at 
crossover attractors.  For $(\xi, \omega^{\prime})$ near a crossover 
$(\xi_c, 0)$, 
\be
\frac{\mathrm{d}\xi}{\mathrm{d}t^{\prime}} \approx J_{\mathrm{drag}}(\xi_c) + 
\left[ \frac{M_v}{\omega^{\prime}} \left. \frac{\mathrm{d}J_v}{\mathrm{d}
\xi} \right|_{\xi = \xi_c} + \left. \frac{\mathrm{d}J_{\mathrm{drag}}}
{\mathrm{d}\xi} \right|_{\xi = \xi_c} \right] (\xi - \xi_c) .
\ee
As a trajectory with $\xi$ near $\xi_c$ approaches $\omega^{\prime} = 0$,
the first term in brackets dominates.  For a crossover with positive
polarity, trajectories converge on the crossover if 
$\mathrm{d}^2 \xi/\mathrm{d}(t^{\prime})^2 < 0$ when $\omega^{\prime} < 0$ and 
$\mathrm{d}^2 \xi/\mathrm{d}(t^{\prime})^2 > 0$ when $\omega^{\prime} > 0$.  
The opposite conditions apply for a crossover of negative polarity.  Thus, a 
crossover is a crossover attractor if
\be
\frac{1}{H_v(\xi_c)} \left. \frac{\mathrm{d}J_v}{\mathrm{d}\xi} 
\right|_{\xi = \xi_c} > 0 .
\ee

\subsection{Results \label{sec:dynamics-results}}

We have examined the grain rotational dynamics for 
$s_d = 0.1$, 0.3, 1.0, 3.0, and 10.0. 
Table \ref{tab:parameters-for-dynamics}
gives the adopted values of the relevant parameters.  The speed of the
outgoing particle is $v_{\mathrm{out}} = (2kT_{\mathrm{dust}}/m_p)^{1/2}$ for H 
atoms (with $T_{\mathrm{dust}}$ the temperature of the grain)
and $v_{\mathrm{out}} = (E_{\mathrm{H}2}/m_p)^{1/2}$ for H$_2$ molecules.  

\begin{table}
\caption{Adopted parameter values for grain rotational dynamics.} 
\label{tab:parameters-for-dynamics}
\begin{tabular}{ll}
\hline
$a_{\mathrm{eff}}$ & $0.2 \, \mu \mathrm{m}$ \\
$\rho$ & $3.0 \, \mathrm{g} \, \mathrm{cm}^{-3}$ \\
$\chi_0$ & $3.3 \times 10^{-4}$ \\
$B$ & $5.0 \, \mu \mathrm{G}$ \\
$T_{\mathrm{dust}}$ & 15 K \\
$T_{\mathrm{gas}}$ & 100 K \\
$E_{\mathrm{H}2}$ & 0.2 eV \\
$n_{\mathrm{H}}$ & $30 \, \mathrm{cm}^{-1}$ \\
\hline
\end{tabular}
\end{table}

\begin{figure}
\includegraphics[width=100mm]{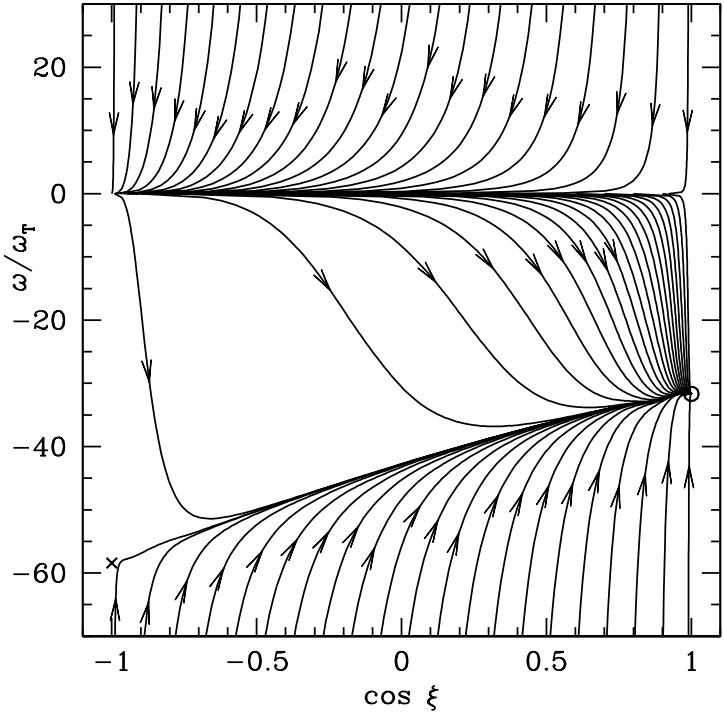}
\caption{
Trajectory map for grain 1, $s_d = 0.1$, $\psi_v = 89^{\circ}$, and atoms as the 
outgoing particles in scenario (2).  
The rotational speed $\omega$ is normalized to the
thermal value $\omega_{T}$ (equation \ref{eq:omega-thermal}); $\xi$ is the
alignment angle.  Attractors (repellers) are indicated by the open 
circles (crosses). 
        }
\label{fig:trajectories-sd0.1.a2r1.out2.at.psi89}
\end{figure}

Fig. \ref{fig:trajectories-sd0.1.a2r1.out2.at.psi89} is a ``trajectory map'', 
which shows how 
$(\cos \xi, \omega^{\prime})$ evolves for grain 1 with $s_d = 0.1$ when the 
outgoing particles are H atoms in scenario (2) 
and $\psi_v = 89^{\circ}$.  This map features an attractor at 
$(\cos \xi, \omega^{\prime}) = (1, -31.7)$, a repeller at $(-1, -58.4)$, 
a crossover attractor at $\cos \xi = -1$ and a crossover repeller at
$\cos \xi = 1$.  \citet{DW97} classified trajectory maps in 
three categories.  This is an example of a noncyclic map, in which all of
the trajectories land on the attractor, and it exhibits perfect alignment with
the magnetic field, since $\xi = 0$ for the attractor.  
The other categories are cyclic maps, which exhibit no attractors, so that
the grain state must cycle between crossovers indefinitely, and 
semicyclic maps, for which the grain state may either land on an attractor
or cycle between crossovers.  
Since our analysis assumes that the grain angular momentum always 
lies along $\bmath{\hat{a}}_1$, we cannot follow the dynamics through the 
crossovers and determine which of these possibilities actually occurs for
semicyclic maps.  

\begin{figure}
\includegraphics[width=160mm]{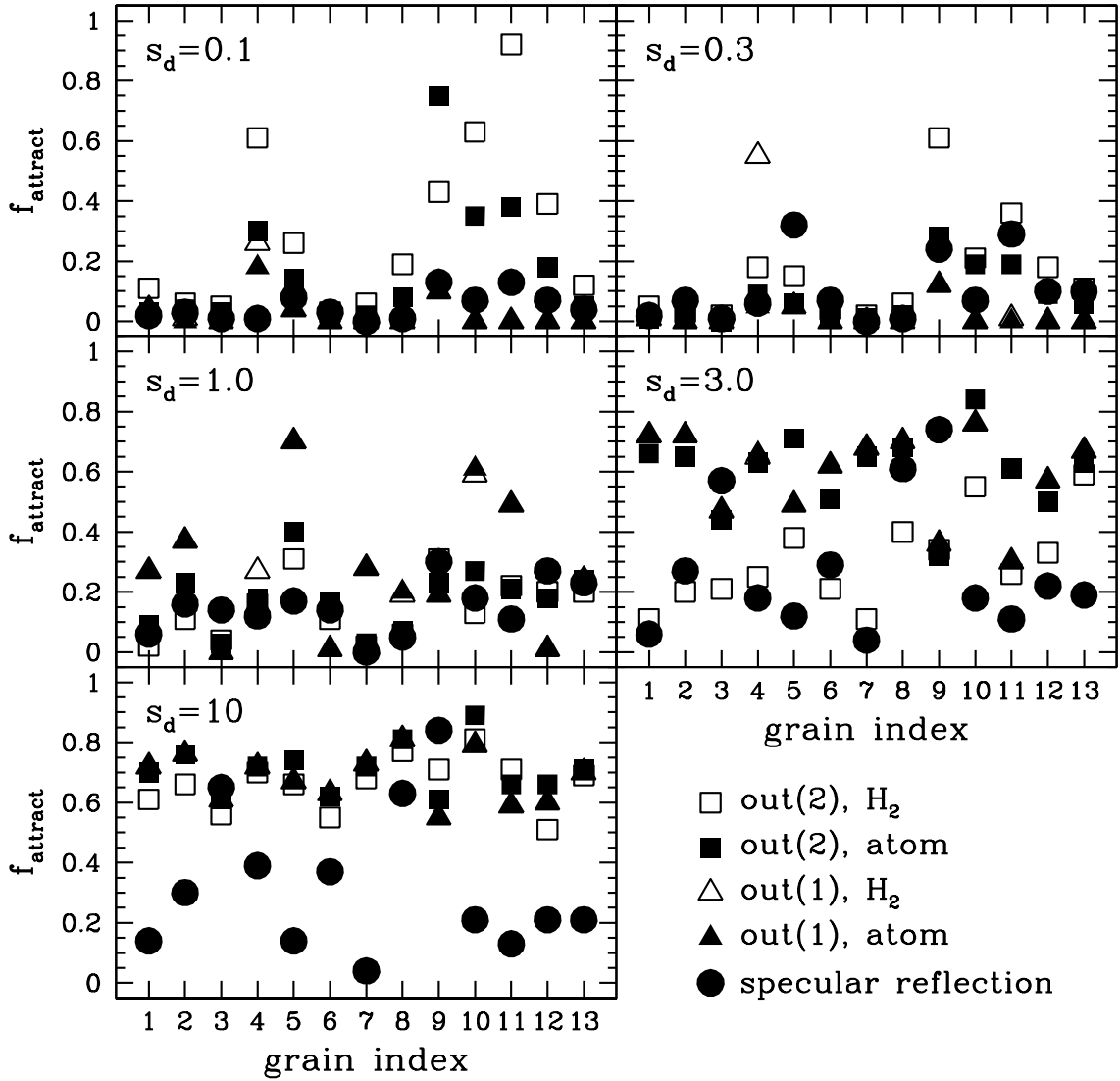}
\caption{
For each of 65 cases (13 grain shapes and five assumptions regarding outgoing
particles) and with five values of $s_d$ as indicated in each panel, 
the fraction $f_{\mathrm{attract}}$ of values of $\psi_v$ 
(uniformly distributed in $\cos \psi_v$) for which the trajectory map 
contains one or more attractors satisfying the alignment-conducive conditions 
described in the text.  
        }
\label{fig:f_attract}
\end{figure}

In future work, we will relax the assumption that the angular momentum always 
lies along $\bmath{\hat{a}}_1$, enabling a firm conclusion regarding the 
effectiveness
of mechanical torques in aligning grains.  Here we attempt to gain some 
insight by examining the incidence of attractors satisfying the following
three conditions that are conducive to alignment:  
(1)  $|\cos \xi| \ge 1/3$, ensuring that the `Rayleigh reduction factor'
characterizing the alignment effectiveness is positive \citep{LD85};
(2)  $\tau_{\mathrm{align}} \le 10^6 \, \mathrm{yr}$, in order to be competitive
with radiative torques \citep{DW97};
(3)  $\omega / \omega_T \ge 3$, so as to avoid disalignment due to 
collisions with gas particles \citep{LH07mech, HL08}.  
For each of the 13 grain shapes, five assumptions regarding the outgoing 
particles, and five values of $s_d$, we consider 100 values of $\psi_v$ between
0 and $\pi/2$, uniformly spaced in $\cos \psi_v$.  Fig. \ref{fig:f_attract}
shows $f_{\mathrm{attract}}$, the fraction of values of $\psi_v$ for which the
trajectory map contains one or more attractors (making it noncyclic or
semicyclic) satisfying the above three conditions.  

With the exception of grain 4, we include only the torque associated with the 
arriving atoms in the case of outgoing scenario (1), since 
$\mathbfit{Q}_{\Gamma, \mathrm{out}, (1)} \bmath{\cdot} \bmath{\hat{a}}_1 / 
Q_{\mathrm{arr}}$ is consistent with zero for most of the shapes.  Thus, except
for grain 4, the open and filled triangles are coincident in Fig. 
\ref{fig:f_attract}.  As noted in 
\S \ref{sec:torque-results}, $\mathbfit{Q}_{\Gamma, \mathrm{out}, (1)} \bmath{\cdot} 
\bmath{\hat{a}}_1 / Q_{\mathrm{arr}}$ appears to converge to a small but non-zero 
value for grains 2 and 3.  Including the associated torque does not alter the 
value of $f_{\mathrm{attract}}$ in any case for grain 3, but does alter its value
for grain 2 by $\pm 0.01$ in some cases.  

The fraction $f_{\mathrm{attract}}$ varies considerably depending on grain shape,
outgoing particle characteristics, and $s_d$.  On the whole, 
$f_{\mathrm{attract}}$ is larger for supersonic drift than for subsonic 
drift, suggesting more effective alignment in the former case.  This may
be partially offset by the result that maps with attractors tend to be 
noncyclic in cases of subsonic drift and semicyclic in cases of 
supersonic drift (except that semicyclic character always dominates in the
case of specular reflection).  Even for subsonic drift, $f_{\mathrm{attract}}$ can 
approach unity for some grain shapes and outgoing scenarios.  Thus, the simple 
analysis assuming that $\mathbfit{J} \parallel \bmath{\hat{a}}_1$
indicates that alignment via mechanical torques may be viable for both
subsonic and supersonic drift; a more detailed study that relaxes this
assumption is needed.  

\subsection{Analysis}

We have found that 
the increase in $f_{\mathrm{attract}}$ with $s_d$ is commonly due to an
increase in the number of attractors as $s_d$ increases rather than an
increase in the fraction of attractors that satisfy the three imposed
conditions.  To gain some insight into this observation, consider how
the overall magnitudes and the shapes of the torques (i.e.~plots of the
rotationally averaged 
efficiency factor components as functions of $\cos \theta_{va}$) vary with $s_d$
and how these affect the incidence of attractors and the associated values
of $\omega/\omega_T$ and $\tau_{\mathrm{align}}$.  

Denote the overall
magnitudes of the mechanical and drag torque efficiencies by 
$\tilde{Q}_{\Gamma, \mathrm{mech}}$ and $\tilde{Q}_{\Gamma, \mathrm{drag}}$, 
respectively.  Both of these tend to increase with $s_d$.  As noted in \S
\ref{sec:supersonic}, in the extreme supersonic limit, the torques 
associated with arriving atoms and specular reflection increase in 
proportion to $s_d^2$ and the outgoing-particle and drag torques increase
in proportion to $s_d$.  In the subsonic limit (\S \ref{sec:subsonic}), 
$\mathbfit{Q}_{\Gamma, \mathrm{arr}} \propto s_d$. The other mechanical torques
can be non-vanishing when $s_d=0$; thus, their overall magnitude does not
necessarily increase monotonically with $s_d$ in the subsonic limit
(though this happens to be the case for grain 1).  As seen in Fig.
\ref{fig:q_arr}
(recall that $\mathbfit{Q}_{\Gamma, \mathrm{drag, out, (1)}} \propto Q_{\mathrm{arr}}$),
$\tilde{Q}_{\Gamma, \mathrm{drag}}$ increases very slowly with $s_d$ in the
subsonic limit and $\tilde{Q}_{\Gamma, \mathrm{drag}} \propto s_d$ in the 
supersonic limit.  

If only the arriving atoms contributed to the mechanical torque, then 
its shape would not depend on $s_d$ in either the extreme subsonic or 
supersonic limits, though it would depend on $s_d$ for intermediate values of
$s_d$.  The shape of the total mechanical torque can vary in all of the regimes
that we examine since it is the sum of two terms (arrival plus specular 
reflection or outgoing scenario 1 or 2) that vary with $s_d$ in different 
ways.  (Of course, for sufficiently large $s_d$, the shape does not depend
on $s_d$, but at $s_d=10$ the magnitude of 
$\overbar{\mathbfit{Q}}_{\Gamma, \mathrm{arr}}$ does not yet overwhelm that of 
$\overbar{\mathbfit{Q}}_{\Gamma, \mathrm{out}}$.  Also, the 
shapes of $\overbar{\mathbfit{Q}}_{\Gamma, \mathrm{spec}}$ and 
$\overbar{\mathbfit{Q}}_{\Gamma, \mathrm{out}}$ 
individually can vary with $s_d$ at low $s_d$ since they do not necessarily 
vanish when $s_d=0$.)  The shape of the drag torque also varies with $s_d$,
but we have found that the dynamics is not greatly modified if the drag
components perpendicular to $\bmath{\hat{a}}_1$ are ignored.  Furthermore, the 
variation of 
$\overbar{\mathbfit{Q}}_{\Gamma, \mathrm{drag, out}, (2)} \cdot \bmath{\hat{a}}_1$ 
with $\theta_{va}$ is mild.  

Now consider how the various quantities that affect the dynamics depend on
$\tilde{Q}_{\Gamma, \mathrm{mech}}$ and $\tilde{Q}_{\Gamma, \mathrm{drag}}$:  
$J_v$ and $H_v$ are both proportional to 
$\tilde{Q}_{\Gamma, \mathrm{mech}}$, $\tau_{\mathrm{drag}}$ and $M_v$ are both
proportional to $\tilde{Q}^{-1}_{\Gamma, \mathrm{drag}}$, and 
$J_{\mathrm{drag}}$ and $H_{\mathrm{drag}}$ are both independent of 
the overall torque magnitudes.  

Thus, the function $Z_v(\xi)$ 
(equation \ref{eq:Z_v}) is proportional to $\tilde{Q}_{\Gamma, \mathrm{mech}}$.
Since stationary points are located at $\xi$ for which $Z_v(\xi) = 0$, 
the incidence of stationary points depends only on the shapes of the
torques, not on their overall magnitudes.  We have found that the 
great majority of the attractors lie at $\xi = 0$ or
$\pi$, where stationary points are always located.
Since the suprathermality
$\omega^{\prime}_s \propto \tilde{Q}_{\Gamma, \mathrm{mech}} / \tilde{Q}_{\Gamma, \mathrm{drag}}$, it tends to increase with $s_d$ and approaches values as large as
$10^5$ in some cases when $s_d = 10$.  

The terms $A_l$ and $D_l$ that appear in the linearized dynamical equations
(equations \ref{eq:linearized1}--\ref{eq:D_l}) are independent of the overall 
torque magnitudes while 
$B_l \propto \tilde{Q}_{\Gamma, \mathrm{drag}} \tilde{Q}^{-1}_{\Gamma, \mathrm{mech}}$
and
$C_l \propto \tilde{Q}_{\Gamma, \mathrm{mech}} \tilde{Q}^{-1}_{\Gamma, \mathrm{drag}}$.
Thus, the conditions for a stationary point to be an attractor
(equation \ref{eq:attractor-conditions}) are independent of the overall torque
magnitude.  The 
increase in the number of attractors with $s_d$ must result from the change 
in the shape of the mechanical torque as $s_d$ increases.  

Recall that $\overbar{\mathbfit{Q}}_{\Gamma, \mathrm{drag}}$ can usually be 
approximated as constant and antiparallel to $\bmath{\hat{a}}_1$ without 
dramatically altering the dynamics.  With this assumption, $B_l = 0$ and
$D_l = -1$.  Thus, the condition for a stationary point to be an attractor
(equation \ref{eq:attractor-conditions})
is $A_l < 0$.  For a given value of $\psi_v$ (except $\sin \psi_v = 0$) and 
the stationary point at $\xi = 0$, $C_l = 0$ and 
\be
A_l =  \frac{Q_{\theta}(\cos \psi_v) \cot \psi_v - Q^{\prime}_{\theta}(\cos \psi_v)
\sin \psi_v}{2 Q_{a1}(\cos \psi_v)}
\ee
where $Q_{\theta}(\cos \theta_{va}) = \overbar{\mathbfit{Q}}_{\Gamma, \mathrm{mech}}
(\cos \theta_{va}) \bmath{\cdot} \bmath{\hat{\theta}}_v$, 
$Q^{\prime}_{\theta}(\cos \theta_{va}) = \mathrm{d}Q_{\theta}(\cos \theta_{va})
/\mathrm{d}(\cos \theta_{va})$, and 
$Q_{a1}(\cos \theta_{va}) = \overbar{\mathbfit{Q}}_{\Gamma, \mathrm{mech}}(\cos 
\theta_{va}) \bmath{\cdot} \bmath{\hat{a}}_1$.
Note that we take $\cos \theta_{va}$ as the argument of the rotationally
averaged efficiency factors here.  If $\sin \psi_v = 0$, then
\be
\label{eq:A_l-psi_v0}
A_l = - \sin \psi_v \, \frac{Q^{\prime}_{\theta}(\cos \psi_v)}{Q_{a1}(\cos \psi_v)} 
= \frac{1}{Q_{a1}(\cos \psi_v)} \left. \frac{\mathrm{d} Q_{\theta}(\cos 
\theta_{va})}{\mathrm{d} \theta_{va}} \right|_{\theta_{va} = \psi_v} .
\ee  
Thus, if $\sin \psi_v \ne 0$ and $Q_{a1}(\cos \psi_v) \ne 0$, then, 
for a given $\psi_v$, 
the condition to have an attractor at $\xi =0$ ($A_l < 0$) 
can be expressed in terms of the shape of the $a_1$- and $\theta_v$-components 
of the rotationally averaged mechanical torque efficiency factor as 
\be
\label{eq:attractor-xi0}
Q_{a1}(\cos \psi_v) \left[ Q^{\prime}_{\theta}(\cos \psi_v) \sin^2 \psi_v - 
Q_{\theta}(\cos \psi_v) \cos \psi_v \right] > 0   
\ee
\citep[c.f. \S 6.3.2 of][]{LH07rad}.
The condition for the stationary point at $\xi = \pi$ to be an attractor is
identical except that $\cos \psi_v$ is replaced with $- \cos \psi_v$.  Of
course, an attractor at $\xi = 0$ or $\pi$ characterized by suprathermal 
rotation will not arise when $Q_{a1}(\cos \psi_v) = 0$.
Evidently the range of values of $\cos \theta_{va}$ for which inequality 
(\ref{eq:attractor-xi0}) is satisfied varies considerably among GRS shapes for 
a given value of $s_d$ and tends to increase with $s_d$ for a given grain 
shape.  

Since the denominator in the expression for $\tau_{\mathrm{align}}$ in equation
(\ref{eq:tau-align}) does not depend on the overall torque magnitudes, the
alignment time varies with $s_d$ in exactly the same way as the drag time, 
namely in proportion to $\tilde{Q}^{-1}_{\Gamma, \mathrm{drag}}$.  Thus, the
distribution of alignment times does not vary substantially as $s_d$ increases
through the subsonic regime but does decrease with $s_d$ in the supersonic
regime.  

Now we will apply the above observations to the dynamics in the case of 
outgoing scenario (1).  Except for grain 4, the torque associated
with the outgoing atoms/molecules is negligible compared with the torque 
associated with the incoming atoms in this scenario.  Thus, 
$Q_{a1}(\cos \theta_{va}) = \overbar{\mathbfit{Q}}_{\Gamma, \mathrm{arr}}(\cos 
\theta_{va}) \bmath{\cdot} \bmath{\hat{a}}_1$ and
$Q_{\theta}(\cos \theta_{va}) = \overbar{\mathbfit{Q}}_{\Gamma, \mathrm{arr}}(\cos 
\theta_{va}) \bmath{\cdot} \bmath{\hat{\theta}}_v$.  The incidence of 
attractors at $\xi = 0$ or $\pi$ as a function of $s_d$ can be explained from 
the grain-shape-independent properties of
$Q_{a1}$ and $Q_{\theta}$ derived in Appendix \ref{sec:arrival-special-results}.
(For the remainder of this discussion, it will be implicit that the attractors
under discussion lie at $\xi = 0$ or $\pi$.)  

First, since $Q_{a1}(0) = 0$, no attractors with suprathermal rotation are
expected for $\cos \psi_v = 0$.  Secondly, when $\cos \psi_v = \pm 1$, the
condition for an attractor is that $A_l$ given by equation 
(\ref{eq:A_l-psi_v0}) must be less than zero.  Since $Q_{a1}$ and 
$Q_{\theta}$ have the same sign when $\cos \theta_{va} > 0$ and 
$Q_{\theta}(\cos \theta_{va} = \pm 1) = 0$, 
$\mathrm{d}Q_{\theta}/\mathrm{d}\theta_{va}$ has the same sign as $Q_{a1}$ for
the stationary point at $\xi=0$.  Thus, $A_l > 0$ and this point is not an
attractor.  A similar analysis shows that the stationary point at $\xi = \pi$
is also not an attractor.  

Aside from the above special cases, the condition for an attractor is 
inequality (\ref{eq:attractor-xi0}).  Suppose $Q_{a1}(\cos \theta_{va}) > 0$
when $X_1 < \cos \theta_{va} < X_2$.  In this case, inequality 
(\ref{eq:attractor-xi0}) is equivalent to 
\be
Q_{\theta}(\cos \psi_v) < \csc \psi_v \, Q_{\theta}(X_0) \ \ , \ \ 
X_1 < \cos \psi_v < X_0 \ \ , 
\ee
\be
Q_{\theta}(\cos \psi_v) > \csc \psi_v \, Q_{\theta}(X_0) \ \ , \ \ 
X_0 < \cos \psi_v < X_2 
\ee 
for some $X_0$ such that $X_1 < X_0 < X_2$.  If $Q_{a1}(\cos \theta_{va}) < 0$,
then the inequality signs relating $Q_{\theta}(\cos \psi_v)$ and 
$\csc \psi_v \, Q_{\theta}(X_0)$ are reversed in the above condition.  

For 9 of the 13 grain shapes, including grain 1, $Q_{\theta}(\cos \theta_{va})$
has the same sign for the entire range of $\cos \theta_{va}$ (-1 to 1).  Thus,
if $Q_{a1}(\cos \theta_{va}) > 0$ when $\cos \theta_{va} > 0$, then 
$Q_{\theta}(\cos \theta_{va}) > 0$ when $\cos \theta_{va} > 0$ and, for a given
$\psi_v$, the condition for an attractor at $\xi = 0$ is 
\be
Q_{\theta}(\cos \psi_v) > \csc \psi_v \, Q_{\theta}(\cos \theta_{va} = 0) .
\ee
If $Q_{a1}(\cos \theta_{va} > 0) < 0$, then 
$Q_{\theta}(\cos \theta_{va} > 0) < 0$ and the condition for an attractor at 
$\xi = 0$ is 
\be
Q_{\theta}(\cos \psi_v) < \csc \psi_v \, Q_{\theta}(\cos \theta_{va} = 0) .
\ee
Thus, the condition for an attractor at $\xi = 0$ (assuming $Q_{\theta}$ has 
the same sign for the entire range of $\cos \theta_{va}$) is
\be
\label{eq:attractor-arr-xi0}
|Q_{\theta}(\cos \psi_v)| > \csc \psi_v \, |Q_{\theta}(\cos \theta_{va} = 0)| .
\ee
The condition for an attractor at $\xi = \pi$ is identical. 
In the extreme subsonic regime, 
$\overbar{\mathbfit{Q}}_{\Gamma, \mathrm{arr}} \bmath{\cdot} \bmath{\hat{\theta}_v}
(\cos \theta_{va}) \propto \sin \theta_{va}$, so the condition for an 
attractor is $\sin \psi_v > 1$.  Thus, for the idealized conditions considered
here for outgoing scenario (1), i.e.~only the torque associated with the 
arriving atoms is significant and only attractors at $\xi = 0$ and $\pi$ are
considered, we expect $f_{\mathrm{attract}} = 0$ in the subsonic regime.  In the 
extreme supersonic limit, 
$\overbar{\mathbfit{Q}}_{\Gamma, \mathrm{arr}} \bmath{\cdot} \bmath{\hat{\theta}_v}
(\cos \theta_{va} = 0) \rightarrow 0$ and condition
(\ref{eq:attractor-arr-xi0}) is satisfied
for an expanding range of values of $\psi_v$.  As a result, 
$f_{\mathrm{attract}}$ increases dramatically as $s_d$ increases.  

As seen in Fig. \ref{fig:f_attract},
our computed $f_{\mathrm{attract}}$ for outgoing scenario (1) does not equal zero 
in the subsonic regime for grains 2, 5, and 9.  
For grain 2, this occurs because the computational result for 
$Q_{a1}(\cos \theta_{va})$ crosses zero at $\cos \theta_{va} \approx -0.018$
rather than at zero.  As a result, $Q_{a1}$ has the wrong sign for a small
range of $\cos \theta_{va}$, yielding spurious attractors at $\xi = \pi$ for 
$\psi_v$ very close to $90^{\circ}$.  Although this slight offset in 
$Q_{a1}(\cos \theta_{va})$ afflicts the computational results for all grains,
it is only large enough to affect $f_{\mathrm{attract}}$ for grain 2.  
For grains 5 and 9, deviations of the shape of the computed 
$Q_{\theta}(\cos \theta_{va})$ from $\sin \theta_{va}$ are responsible for the
spurious attractors.  We have generated versions of Fig.
\ref{fig:f_attract} for the subsonic regime using torques 
computed with $N_1 = 64$ rather than 128.  (Recall that 
there are $N_1$ values of $\phi_{\mathrm{sph}}$ and $\phi_{\mathrm{in}}$ and
$N_1 + 1$ values of $\cos \theta_{\mathrm{sph}}$ and $\cos \theta_{\mathrm{in}}$.)
With $N_1=64$, $f_{\mathrm{attract}}$ for outgoing scenario (1) is somewhat higher 
for grains 2, 5, and 9 and also non-zero for grains 3, 6, 10, and 11.  
Thus, $f_{\mathrm{attract}}$ approaches zero as the numerical resolution 
increases, in agreement with our idealized model.  

In the case of grain 4, 
$\mathbfit{Q}_{\Gamma, \mathrm{out}, (1)} \bmath{\cdot} \bmath{\hat{a}}_1$ is
not negligible.  For $s_d = 0.1$, 
$\overbar{\mathbfit{Q}}_{\Gamma, \mathrm{arr}} \bmath{\cdot} \bmath{\hat{a}_1}
\approx -7 \times 10^{-3} \cos \theta_{va}$, $\mathbfit{Q}_{\Gamma, \mathrm{out}, (1)}
\bmath{\cdot} \bmath{\hat{a}}_1 / Q_{\mathrm{arr}} = 0.00122$, and 
$Q_{\mathrm{arr}} \approx 4.3$.  Evaluating $v_{\mathrm{out}}/v_{\mathrm{th}}$ for
the cases of atomic and molecular outgoing particles using the parameter values
in Table \ref{tab:parameters-for-dynamics}, we find that 
$Q_{a1} \approx Q_0 - 7 \times 10^{-3} \cos \theta_{va}$ with 
$Q_0 \approx 1.9 \times 10^{-3}$ for atoms and 
$Q_0 \approx 1.7 \times 10^{-2}$ for molecules.  Due to the upward shift of
$Q_{a1}$, $Q_{a1}$ and $Q_{\theta}$ have opposite signs when 
$0 < \cos \theta_{va} < 0.28$ for atoms and when $0 < \cos \theta_{va} < 1$
for molecules.  Thus, attractors arise at $\xi = 0$ for $\cos \psi_v$ between
0 and 0.28 (0 and 1) for atomic (molecular) outgoing particles.  Since these
attractors do not all satisfy the conditions on $\omega^{\prime}$ and
$\tau_{\mathrm{align}}$ for effective alignment, $f_{\mathrm{attract}}$ is less than
0.28 and 1 in these cases.  A similar analysis applies when $s_d = 0.3$.  Thus,
it appears that subsonic mechanical torques can yield effective alignment 
even in the case of outgoing scenario (2) if the grain shape is such that
$\mathbfit{Q}_{\Gamma, \mathrm{out}, (1)} \bmath{\cdot} \bmath{\hat{a}}_1$ is
not negligible, but that such shapes are rare.  Of course, we are unable to
draw a strong conclusion on this point since we have only examined 13 shapes.  

Finally, consider the dynamics assuming specular reflection or outgoing
scenario (2) in the subsonic limit.  As shown in Appendix 
\ref{sec:subsonic-special-results}, the rotationally averaged torque
efficiency factors exhibit the same functional dependence on 
$\cos \theta_{va}$ as for the torque associated with arriving atoms, except
that $\overbar{\mathbfit{Q}}_{\Gamma, i} \cdot \bmath{\hat{a}}_1$ 
[$i=$ spec or out(2)] can be non-zero when $s_d=0$.  Thus, just as with the
torque associated with outgoing particles for grain 4 and outgoing scenario
(1), the zero of $Q_{a1}$ shifts away from $\cos \theta_{va} = 0$, opening up
a range of values of $\cos \psi_v$ for which attractors may occur at either
$\xi = 0$ or $\xi = \pi$.  If the simplifying assumptions of this discussion
(e.g.~neglect of complications in the drag torque and the possibility of 
attractors at values of $\cos \xi$ other than $\pm 1$) remain valid, then 
$f_{\mathrm{attract}}$ could reach values as high as
\be
f_{\mathrm{attract, max}} = \max \left[ \frac{|\mathbfit{Q}_{\Gamma, \mathrm{mech}}(s_d=0)
\cdot \bmath{\hat{a}}_1|}{s_d |\overbar{\mathbfit{Q}}^{\prime}_{\Gamma, \mathrm{mech}}
\cdot \bmath{\hat{a}}_1|} , 1 \right]
\ee
where $\mathbfit{Q}_{\Gamma, \mathrm{mech}}$ includes contributions from both
arriving atoms and reflected or outgoing particles 
(see equation \ref{eq:q-total-torque}) and the prime denotes the term linear in
$s_d$ (see \S \ref{sec:subsonic}).  

\begin{figure}
\includegraphics[width=100mm]{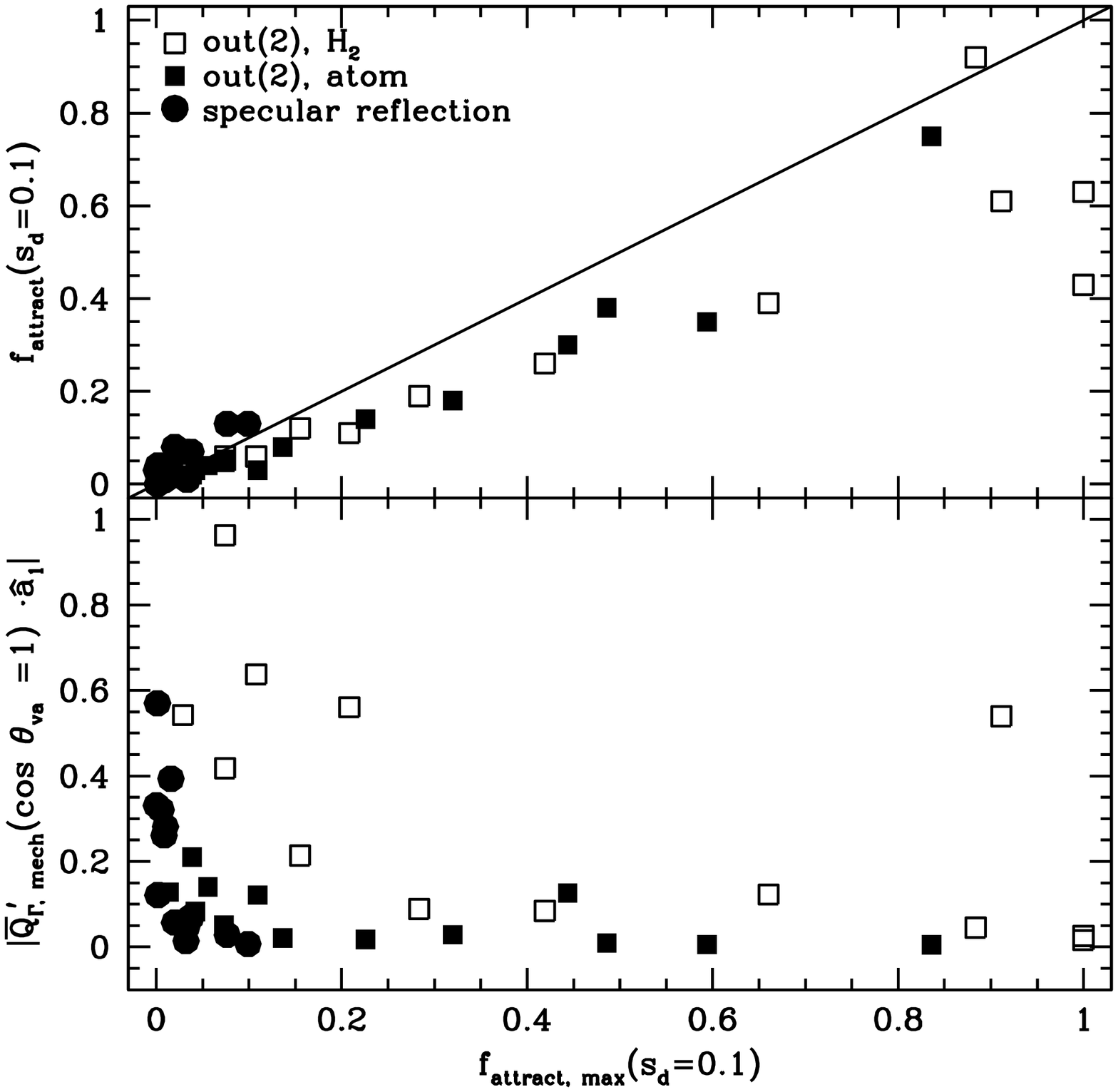}
\caption{
$f_{\mathrm{attract}}$ (upper panel) and 
$|\overbar{\mathbfit{Q}}^{ \prime}_{\Gamma, \mathrm{mech}}(\cos \theta_{va} =1) \cdot 
\bmath{\hat{a}}_1|$ (lower panel)
versus $f_{\mathrm{attract, max}}$ for $s_d=0.1$
        }
\label{fig:f_attract_max}
\end{figure}

The upper panel in Fig. \ref{fig:f_attract_max} shows $f_{\mathrm{attract}}$
versus $f_{\mathrm{attract, max}}$ for $s_d=0.1$ for all 13 grain shapes, 
considering specular reflection and both outgoing scenarios.  In most cases,
$f_{\mathrm{attract}} < f_{\mathrm{attract, max}}$, commonly because
$\tau_{\mathrm{align}} > 10^6 \, \mathrm{yr}$ for the attractors.  In cases where
$f_{\mathrm{attract}} > f_{\mathrm{attract, max}}$, the simplifying assumptions may
be violated or there may be error due to insufficient numerical resolution
in the torque evaluations.  

The lower panel in Fig. \ref{fig:f_attract_max} shows 
$|\overbar{\mathbfit{Q}}^{ \prime}_{\Gamma, \mathrm{mech}}(\cos \theta_{va} =1) \cdot 
\bmath{\hat{a}}_1|$, a measure of the overall magnitude of the mechanical 
torque, again for $s_d=0.1$.  Note that, typically, the cases for which 
$f_{\mathrm{attract}}$ is substantial are characterized by low torque magnitudes.
This is expected from the above analysis:  the mechanical torque in the 
subsonic regime is the sum of a constant term that persists when $s_d=0$
and a term proportional to $s_d$.  The larger the former is in comparison
to the latter, the larger the range of $\cos \psi_v$ for which attractors
can occur.  The one case for which both $f_{\mathrm{attract}}$ and the torque
magnitude are large is for H$_2$ molecules departing in scenario (2) from
grain 4.  As with outgoing scenario (1), this grain happens to experience an
unusually large torque even when not drifting relative to the gas.  

As seen in Fig. \ref{fig:f_attract},
for some of the cases with relatively large $f_{\mathrm{attract}}$ when 
$s_d=0$, $f_{\mathrm{attract}}$ is lower when $s_d=0.3$; the term that persists
when $s_d=0$ is relatively less important when $s_d=0.3$ than when $s_d=0.1$.
As discussed earlier, the increase in $f_{\mathrm{attract}}$ as $s_d$ increases
beyond 1 is due to the change in the shape of the torques; the
torque that persists for zero drift is unimportant in these cases.  

These results suggest that grain shapes which are most susceptible to 
alignment by mechanical torques in the subsonic regime may typically 
experience relatively weak mechanical torques, which therefore are more 
likely to be dominated by other types of torques (e.g.~radiative torques).
In future work, we will evaluate the radiative torques on the 13 grains
considered here and examine the dynamics in full (rather than including only
a subset of the torques) for a range of interstellar environments.

\subsection{Comparison with Lazarian \& Hoang (2007a)}

In their study of radiative torques, Lazarian \& Hoang (2007a, hereafter
LH07a) found that
the ratio $R_{LH} = Q_{e1}^{\mathrm{max}}/Q_{e2}^{\mathrm{max}}$ for a grain 
correlates 
well with the grain's alignment characteristics. Here $Q_{e1}$ is the 
component of the radiative torque efficiency factor along the direction 
$\bmath{S}$ of the radiation field anisotropy and $Q_{e2}$ is the component 
perpendicular to $\bmath{S}$ and in the plane spanned by $\bmath{S}$ and 
$\bmath{\hat{a}}_1$. The superscript `max' indicates the maximum 
absolute value
as a function of the angle between $\bmath{S}$ and $\bmath{\hat{a}}_1$.
Their fig. 24 shows how the incidence of attractors with high angular
momentum (i.e.~the types that we examine here) varies with $\psi$ (the angle 
between the magnetic field direction and $\bmath{S}$) as a function of
$R_{LH}$.  When $1 < R_{LH} < 2$, no or very few high-$J$ attractors are 
expected. As $R_{LH}$ increases above 2, high-$J$ attractors arise near
$\psi = 0$ and extend to larger values of $\psi$ as $R_{LH}$ increases.
Similarly, as $R_{LH}$ decreases below 1, high-$J$ attractors arise near
$\psi = 90^{\circ}$ and extend to lower values of $\psi$ as $R_{LH}$ decreases.

It is of interest to check whether an analogous ratio describes the 
alignment by mechanical torques for the grains examined in this work. 
Thus, we define $R_{LH}^{\mathrm{mech}}$ in the same way as $R_{LH}$, 
considering the total mechanical torque efficiency (equation 
\ref{eq:q-total-torque})
and the components along $\bmath{\hat{z}}_v$ and $\bmath{\hat{x}}_v$ 
in place of $Q_{e1}$ and $Q_{e2}$, respectively (see \S 
\ref{sec:suprathermal}).

Whereas LH07a considered values of $R_{LH}$ from 0.1
to $> 20$, $R_{LH}^{\mathrm{mech}}$ for the cases considered here ranges
from $\approx 0.003$ to $\approx 2.5$ and is less than $\approx 1.4$ for
nearly all cases. Thus, we do not have the opportunity to compare the
alignment behaviour for large values of the ratio. 

Consider first outgoing scenario (1), for which only the torque
associated with the arriving atoms is relevant (except for grain 4).  
For a given grain shape, $R_{LH}^{\mathrm{mech}}$ decreases as 
$s_d$ increases, since $\bmath{\hat{z}}_v$ is parallel to 
$\bmath{\hat{s}}_d$. When $s_d=0.1$, 
$R_{LH}^{\mathrm{mech}} \approx 1.2$--1.3 and there is a very low 
incidence of attractors (except for grain 4), consistent with the results
in LH07a for alignment by radiative torques. As
$s_d$ increases (and $R_{LH}^{\mathrm{mech}}$ decreases), the incidence
of attractors increases and are concentrated towards $\psi_v = 90^{\circ}$, 
again as expected from fig. 24 in LH07a. However, the range of values of
$\psi_v$ for which attractors arise varies considerably among the grain
shapes and is not correlated with $R_{LH}^{\mathrm{mech}}$. 

For the other scenarios, the results diverge even further from those in 
LH07a.  For example, for outgoing scenario (2) with outgoing H atoms, 
$R_{LH}^{\mathrm{mech}} \approx 1.3$--1.4 for all grain shapes when 
$s_d = 0.1$. Whereas fig. 24 in LH07a indicates no high-$J$ attractors (or 
perhaps only when $\psi_v \approx 90^{\circ}$) for this value of the 
ratio, we find attractors over a range of values of $\psi_v$. The extent
of this range varies considerably with grain shape, with the maximum 
$\psi_v$ at $90^{\circ}$ and the minimum between $25^{\circ}$ and 
$85^{\circ}$. For some other cases, the various grain shapes exhibit a 
larger range of values of $R_{LH}^{\mathrm{mech}}$. For these values of the
ratio, it is expected from LH07a that the range of $\psi_v$ for which 
high-$J$ attractors arise should increase as $R_{LH}^{\mathrm{mech}}$ 
decreases. We do not find such a correlation. 

Thus, it appears that the ratio $R_{LH}^{\mathrm{mech}}$ is not generally a
reliable guide to the character of the alignment driven by mechanical 
torques. We will revisit this question in our upcoming work relaxing the
assumption that the grain angular momentum always lies along 
$\bmath{\hat{a}}_1$.

\section{Conclusions} \label{sec:conclusions}

In this study, we have developed theoretical and computational tools for 
evaluating the mechanical torques experienced by irregularly shaped,  
drifting grains.  We have examined various assumptions about how the
colliding gas particles depart the grain (specular reflection, 
departure from an arbitrary location on the grain versus the location 
at which the incoming particle arrived, departure in atomic versus 
molecular form).  We developed computer codes for all of these scenarios.
Arbitrary values of the drift speed can be accommodated, as well as the
extreme subsonic and supersonic limits.  The codes were verified by 
comparing with known results (e.g.~for spherical grains), by comparing the
results for fairly high (low) values of $s_d$ (the drift speed divided by the
gas thermal speed) with the results for the supersonic (subsonic) limit, 
and by verifying that features of the torques common to all grain shapes
were exhibited.  

After evaluating the torques for 13 different grain shapes, we examined
the rotational dynamics assuming steady rotation about the principal axis
of greatest moment of inertia, $\bmath{\hat{a}}_1$.  We introduced the
quantity $f_{\mathrm{attract}}$ to characterize the efficiency of alignment by
mechanical torques (\S \ref{sec:dynamics-results}).  For subsonic drift,
$f_{\mathrm{attract}}$ varies considerably with grain shape and, for some 
shapes, with the assumptions regarding the departure of atoms/molecules
from the grain.  The efficiency of subsonic alignment is primarily determined
by the magnitude of the torque on a non-drifting grain relative to the 
torque that increases in proportion to the drift speed.   
(More precisely, it is the component of the torque along $\bmath{\hat{a}}_1$
that matters.)  Thus, efficient alignment by mechanical torques in the 
subsonic regime may require that the torques be relatively weak, in which 
case they may be dominated by other types of torques.  

As the drift speed increases from the subsonic to the supersonic regime,
$f_{\mathrm{attract}}$ tends to increase, suggesting efficient alignment for
all grains and most departure scenarios.  Efficient alignment can result
even for outgoing scenario (1), in which the location of a departing 
atom/molecule on the grain surface is not correlated with the location of 
arrival \citep[c.f. \S 11.7 of][]{LH07rad}.  The increase in 
$f_{\mathrm{attract}}$ with $s_d$ results from changes in the shape of the
torques rather than from an increase in the torque magnitudes.

Future work will examine the dynamics without 
assuming rotation about $\bmath{\hat{a}}_1$ and will consider the case that
the outgoing molecules depart from special sites on the grain surface
\citep{Purcell79}.  We will also examine alignment by radiative torques for
the 13 grains in this study.  By examining the full dynamics under a range of
interstellar conditions we hope to clarify the relative importance of the
various candidate aligning processes and the environments in which they are
operative.

\bibliographystyle{mnras}
\bibliography{mybib} 

\appendix

\section{Calculation Details} 
\label{sec:calc-details}

\subsection{Torque Due to Incoming Atoms \label{sec:calc-details-incoming}}

Employing equations (\ref{eq:r-hat}) and (\ref{eq:s-hat}), the cross products
in the expression for $\mathbfit{Q}_{\Gamma, \mathrm{arr}}$ 
(equation \ref{eq:Q-Gamma-arr}) are
\be
\bmath{\hat{r}} \bmath{\times} \bmath{\hat{s}} =
\sin \theta_{\mathrm{in}} \left[(\cos \theta \cos \phi \sin
\phi_{\mathrm{in}} + \sin \phi \cos \phi_{\mathrm{in}}) \bmath{\hat{x}} + (\cos 
\theta \sin \phi \sin \phi_{\mathrm{in}} - \cos \phi \cos \phi_{\mathrm{in}}) 
\bmath{\hat{y}} - \sin \theta \sin \phi_{\mathrm{in}} \, \bmath{\hat{z}} \right]
\ee
and
\be
\frac{\mathbfit{r}_{\mathrm{cm}}}{r_{\mathrm{sph}}} \bmath{\times} \bmath{\hat{s}} = 
\frac{1}{r_{\mathrm{sph}}} \left[ (y_{\mathrm{cm}} s_{0z} - z_{\mathrm{cm}} s_{0y}) 
\bmath{\hat{x}} + (z_{\mathrm{cm}} s_{0x} - x_{\mathrm{cm}} s_{0z}) \bmath{\hat{y}}
+ (x_{\mathrm{cm}} s_{0y} - y_{\mathrm{cm}} s_{0x}) \bmath{\hat{z}} \right]
\ee
with
\be
s_{0x} \equiv \bmath{\hat{s}} \bmath{\cdot} \bmath{\hat{x}} = - \cos 
\theta \cos \phi \sin \theta_{\mathrm{in}} \cos \phi_{\mathrm{in}} + \sin \phi \sin
\theta_{\mathrm{in}} \sin \phi_{\mathrm{in}} - \sin \theta \cos \phi \cos
\theta_{\mathrm{in}} , 
\ee
\be
s_{0y} \equiv \bmath{\hat{s}} \bmath{\cdot} \bmath{\hat{y}} =
- \cos \theta \sin \phi \sin
\theta_{\mathrm{in}} \cos \phi_{\mathrm{in}} - \cos \phi \sin \theta_{\mathrm{in}}
\sin \phi_{\mathrm{in}} - \sin \theta \sin \phi \cos \theta_{\mathrm{in}} , 
\ee
\be
s_{0z} \equiv \hat{s} \bmath{\cdot} 
\bmath{\hat{z}} = \sin \theta \sin \theta_{\mathrm{in}}
\cos \phi_{\mathrm{in}} - \cos \theta \cos \theta_{\mathrm{in}} .
\ee
Thus, 
\be
\left( \bmath{\hat{r}} - \frac{\mathbfit{r}_{\mathrm{cm}}}{r_{\mathrm{sph}}} \right) 
\bmath{\times} \bmath{\hat{s}} = A_x \bmath{\hat{x}} + A_y \bmath{\hat{y}} + 
A_z \bmath{\hat{z}}
\ee
with
\be
A_x = \sin \theta_{\mathrm{in}} (\cos \theta \cos \phi \sin
\phi_{\mathrm{in}} + \sin \phi \cos \phi_{\mathrm{in}}) -
\frac{y_{\mathrm{cm}} s_{0z} - z_{\mathrm{cm}} s_{0y}}{r_{\mathrm{sph}}}
\ee
\be
A_y = \sin \theta_{\mathrm{in}} (\cos \theta \sin \phi \sin
\phi_{\mathrm{in}} - \cos \phi \cos \phi_{\mathrm{in}}) -
\frac{z_{\mathrm{cm}} s_{0x} - x_{\mathrm{cm}} s_{0z}}{r_{\mathrm{sph}}}
\ee
\be
A_z = - \sin \theta_{\mathrm{in}} \sin \theta \sin \phi_{\mathrm{in}} -
\frac{x_{\mathrm{cm}} s_{0y} - y_{\mathrm{cm}} s_{0x}}{r_{\mathrm{sph}}} .
\ee

\subsection{Torque Due to Outgoing Particles \label{sec:calc-details-outgoing}}

In equation (\ref{eq:Q-Gamma-out-1}),

\be
\eta_S \, \frac{\mathbfit{r}_{\mathrm{surf}}}{a_{\mathrm{eff}}} \bmath{\times} 
\bmath{\hat{N}} =
\frac{r_{\mathrm{surf}}^3}{a_{\mathrm{eff}}} \, \nu_N
\left\{ \left[ - w_2(\theta, \phi)
\sin \theta \sin \phi + w_3(\theta, \phi) \cot \theta \cos \phi \right]
\bmath{\hat{x}} 
+ \left[ w_2(\theta, \phi) \sin \theta \cos \phi + w_3(\theta, \phi)
\cot \theta \sin \phi \right] \bmath{\hat{y}} - w_3(\theta, \phi) 
\bmath{\hat{z}} \right\} .
\ee

\subsection{Extreme Subsonic Limit \label{sec:calc-details-subsonic}}

Equations (\ref{eq:q-gamma-arr-sd0}) and (\ref{eq:q-gamma-prime-arr})
simplify to 
\be
\label{eq:q-gamma-arr-sd0-2}
\mathbfit{Q}_{\Gamma, \mathrm{arr}}(s_d = 0) = \frac{1}{8 \pi} \left( 
\frac{r_{\mathrm{sph}}}{a_{\mathrm{eff}}} \right)^3 \int_{-1}^1 \mathrm{d}(\cos 
\theta) \int_0^{2 \pi} \mathrm{d}\phi \int_0^{2 \pi} \mathrm{d}\phi_{\mathrm{in}} 
\left[ \mathbfit{A}_1 (1 - u_c^3) + \mathbfit{A}_2 (1 - u_c^2)^{3/2} \right]
\ee
and 
\begin{eqnarray}
\mathbfit{Q}^{\prime}_{\Gamma, \mathrm{arr}} & = & \pi^{-3/2} \left( 
\frac{r_{\mathrm{sph}}}{a_{\mathrm{eff}}} \right)^3 \int_{-1}^1 \mathrm{d}(\cos 
\theta) \int_0^{2 \pi} \mathrm{d}\phi \int_0^{2 \pi} \mathrm{d}\phi_{\mathrm{in}} 
\left\{ \frac{1}{2} \beta_1 \mathbfit{A}_1
(1 - u_c^4) + \frac{1}{2} \beta_2 \mathbfit{A}_2 (1 - u_c^2)^2 + \right. 
\nonumber \\
& & \left. \frac{1}{4} \left( \beta_2 \mathbfit{A}_1 + \beta_1 
\mathbfit{A}_2 \right)
\left[ \cos^{-1} u_c + u_c (1 - 2 u_c^2) \sqrt{1 - u_c^2} \right] \right\}
\end{eqnarray}
where
\be
A_x = A_{1x} \cos \theta_{\mathrm{in}} + A_{2x} \sin \theta_{\mathrm{in}} , 
\ee
\be
s_{0x} = s_{0x, 1} \cos \theta_{\mathrm{in}} + s_{0x, 2} \sin \theta_{\mathrm{in}}
\ee
(and likewise for $y$ and $z$),
\be
s_{0x, 1} = - \sin \theta \cos \phi , 
\ee
\be
s_{0x, 2} = - \cos \theta \cos \phi \cos \phi_{\mathrm{in}} + \sin \phi \sin  
\phi_{\mathrm{in}} , 
\ee
\be
s_{0y, 1} = - \sin \theta \sin \phi , 
\ee
\be
s_{0y, 2} = - \cos \theta \sin \phi \cos \phi_{\mathrm{in}} - \cos \phi \sin 
\phi_{\mathrm{in}} , 
\ee
\be
s_{0z, 1} = - \cos \theta , 
\ee
\be
s_{0z, 2} = \sin \theta \cos \phi_{\mathrm{in}} , 
\ee
\be
A_{1x} = - \frac{y_{\mathrm{cm}} s_{0z, 1} - z_{\mathrm{cm}} s_{0y, 1}}{r_{\mathrm{sph}}} ,
\ee
\be
A_{2x} = \cos \theta \cos \phi \sin \phi_{\mathrm{in}} + \sin \phi \cos
\phi_{\mathrm{in}} - \frac{y_{\mathrm{cm}} s_{0z, 2} - z_{\mathrm{cm}} s_{0y, 2}}
{r_{\mathrm{sph}}} ,
\ee
\be
A_{1y} = - \frac{z_{\mathrm{cm}} s_{0x, 1} - x_{\mathrm{cm}} s_{0z, 1}}{r_{\mathrm{sph}}} ,
\ee
\be
A_{2y} = \cos \theta \sin \phi \sin \phi_{\mathrm{in}} - \cos \phi \cos
\phi_{\mathrm{in}} - \frac{z_{\mathrm{cm}} s_{0x, 2} - x_{\mathrm{cm}} s_{0z, 2}}
{r_{\mathrm{sph}}} ,
\ee
\be
A_{1z} = - \frac{x_{\mathrm{cm}} s_{0y, 1} - y_{\mathrm{cm}} s_{0x, 1}}{r_{\mathrm{sph}}} ,
\ee
\be
A_{2z} = - \sin \theta \sin \phi_{\mathrm{in}}
- \frac{x_{\mathrm{cm}} s_{0y, 2} - y_{\mathrm{cm}} s_{0x, 2}}{r_{\mathrm{sph}}} .
\ee

\subsection{Drag Force in the Extreme Subsonic Limit 
\label{sec:calc-details-force-subsonic}}

Equations (\ref{eq:Q-F-arr-sd0}) and (\ref{eq:Q-F-arr-prime}) simplify to
\be
\mathbfit{Q}_{F, \mathrm{arr}}(s_d = 0) = - \frac{1}{8 \pi} \left( 
\frac{r_{\mathrm{sph}}}{a_{\mathrm{eff}}} \right)^2 \int_{-1}^1 \mathrm{d}(\cos 
\theta) \int_0^{2\pi} \mathrm{d}\phi \int_0^{2\pi} \mathrm{d}\phi_{\mathrm{in}} 
\left[ (1-u_c^3) \bmath{\hat{r}} + (1-u_c^2)^{3/2} \mathbfit{M} \right]
\ee
and
\begin{eqnarray}
\mathbfit{Q}^{\prime}_{F, \mathrm{arr}} & = &
- \frac{\pi^{-3/2}}{2} \left( \frac{r_{\mathrm{sph}}}
{a_{\mathrm{eff}}} \right)^2 \int_{-1}^1 \mathrm{d}(\cos \theta) \int_0^{2\pi} 
\mathrm{d}\phi \int_0^{2\pi} \mathrm{d}\phi_{\mathrm{in}} \biggl\{ \beta_1 (1 - 
u_c^4) \bmath{\hat{r}}  + \beta_2 (1-u_c^2)^2 \mathbfit{M} +  \nonumber \\
& & \frac{1}{2} \left[ u_c (1-2 u_c^2) 
(1- u_c^2)^{1/2} + \cos^{-1} u_c \right] (\beta_2 \bmath{\hat{r}} + \beta_1 
\mathbfit{M}) \biggr\} 
\end{eqnarray}
with 
\be
\mathbfit{M} = (\cos \phi_{\mathrm{in}} \cos \theta \cos \phi - 
\sin \phi_{\mathrm{in}} \sin \phi) \bmath{\hat{x}}  + 
(\cos \phi_{\mathrm{in}} \cos \theta
\sin \phi + \sin \phi_{\mathrm{in}} \cos \phi) \bmath{\hat{y}} -  
\cos \phi_{\mathrm{in}} \sin \theta \, \bmath{\hat{z}} .
\ee

\section{Special Results Associated with Arriving Atoms} 
\label{sec:arrival-special-results}

Here we derive some general features of the arrival efficiency and arrival
torque efficiency noted in \S \S \ref{sec:subsonic} and 
\ref{sec:torque-results}.  First, note that when $\cos \theta_{va} = 0$,
$\bmath{\hat{\theta}}_v = - \bmath{\hat{s}}_d$.  Since all of the incoming
gas atoms have velocities along $- \bmath{\hat{s}}_d$ in the limit 
$s_d \rightarrow \infty$, 
$\overbar{\mathbfit{Q}}_{\Gamma, \mathrm{arr}} \bmath{\cdot} \bmath{\hat{\theta}}_v
(\cos \theta_{va} = 0) \rightarrow 0$ as $s_d \rightarrow \infty$.
Similarly, since $\bmath{\hat{a}}_1$ is parallel or antiparallel to 
$\bmath{\hat{s}}_d$ when $\cos \theta_{va} = \pm 1$, 
$\overbar{\mathbfit{Q}}_{\Gamma, \mathrm{arr}} \bmath{\cdot} \bmath{\hat{a}}_1
(\cos \theta_{va} = \pm 1) \rightarrow 0$ as $s_d \rightarrow \infty$.

The remaining results are more readily apparent if we adopt an approach that
dispenses with the enclosing sphere.  For the remainder of this appendix, we
will take the origin at the grain's centre of mass.  For now, redefine 
the grain-body axes $(\bmath{\hat{x}}, \bmath{\hat{y}}, \bmath{\hat{z}})$
such that $\theta_{\mathrm{gr}} = 0$ (i.e.~the grain is
moving along the $\bmath{\hat{z}}$ direction).  As usual, 
$\bmath{\hat{z}}$ and $\bmath{\hat{x}}$ are the reference axes for the 
polar angle $\theta$ and azimuthal angle $\phi$, respectively.
In this case, 
$|\mathbfit{s} + \mathbfit{s}_d|^2 = s^2 + s_d^2 -2 s s_d \cos \theta$.  

The rate at which gas atoms 
coming from within solid angle element $d\Omega$ about the
direction $(\theta, \phi)$, with reduced speeds between $s$ and $s+ds$, 
strike an area element oriented perpendicular to the gas flow and with area
$dA_{\perp}$ is
\be
dR_{\mathrm{arr}} = n v_{\mathrm{th}} \, d\Omega \, \pi^{-3/2} \exp(-|\mathbfit{s} + 
\mathbfit{s}_d|^2) \, s^3 \, ds \, dA_{\perp} .
\ee
The cross-sectional area $A_{\perp}$ that the grain presents to gas atoms
coming from direction $(\theta, \phi)$ can be expressed as
\be
A_{\perp}(\theta, \phi) = \frac{1}{2} \int_0^{2\pi} d\chi \, [b(\theta, \phi, 
\chi)]^2
\ee
where $b(\theta, \phi, \chi)$ is the largest impact parameter for which a 
gas atom collides with the grain as a function of a rotation angle $\chi$ about
the direction $(\theta, \phi)$.  Since $A_{\perp}$ is the same for two 
directions on opposite sides of the sky, 
$A_{\perp}(\pi - \theta, \phi + \pi) =  A_{\perp}(\theta, \phi)$ and
$b(\pi - \theta, \phi + \pi, \chi) =  b(\theta, \phi, \chi)$.  

In the extreme subsonic limit ($s_d \ll 1$) and with the grain-body axes
oriented as noted above, 
$\exp(-|\mathbfit{s} + \mathbfit{s}_d|^2) \approx \exp(-s^2) \, (1+ 2 s s_d
\cos \theta)$.  Thus,
\be
R_{\mathrm{arr}} \approx n v_{\mathrm{th}} \pi^{-3/2} \int_0^{\infty} ds \, s^3 
\exp \left( -s^2 \right) \int_0^{2\pi} d\phi \int_0^1 d(\cos \theta) \left\{
2 + 2 s s_d \left[ \cos \theta + \cos (\pi - \theta) \right] \right\} 
A_{\perp}(\theta, \phi) .
\ee
Since $\cos \theta + \cos (\pi - \theta) = 0$, 
$Q^{\prime}_{\mathrm{arr}}$ (i.e.~the term proportional to $s_d$) vanishes.  

For the remainder of this appendix, choose grain-body
axes such that $(\bmath{\hat{x}}, \bmath{\hat{y}}, \bmath{\hat{z}})$ lie
along $(\bmath{\hat{a}}_2, \bmath{\hat{a}}_3, \bmath{\hat{a}}_1)$.
Taking $\chi = 0$ along $\bmath{\hat{\theta}}$ and increasing from there towards
$\bmath{\hat{\phi}}$ when $0 \le \theta \le \pi/2$, the angular momentum of
an atom arriving with impact parameter $r_{\perp}$ is
\be
\label{eq:Delta-J-arr-app}
\Delta \mathbfit{J}_{\mathrm{arr}}(\theta, \phi, \chi, r_{\perp}, s) 
= m v_{\mathrm{th}} s r_{\perp} \left[ - (\cos
\chi \sin \phi + \sin \chi \cos \theta \cos \phi) \, \bmath{\hat{x}} + 
(\cos \chi \cos \phi - \sin \chi \cos \theta \sin \phi) \, \bmath{\hat{y}} + 
\sin \chi \sin \theta \, \bmath{\hat{z}} \right] 
\ee
for $0 \le \theta \le \pi/2$.  
With the area element $dA_{\perp} = r_{\perp} dr_{\perp} d\chi$, the efficiency 
factor for the arrival torque is
\be
\mathbfit{Q}_{\Gamma, \mathrm{arr}} = \pi^{-3/2} \int_0^{2\pi} d\phi \int_{-1}^1
d(\cos \theta) \int_0^{2\pi} d\chi \int_0^{b(\theta, \phi, \chi)} dr^{\prime}_{\perp}
r^{\prime}_{\perp} \int_0^{\infty} ds \, s^3\exp \left( -|\mathbfit{s} + 
\mathbfit{s}_d|^2 \right) \frac{\Delta \mathbfit{J}_{\mathrm{arr}}(\theta, \phi, 
\chi, r_{\perp}, s)}{m v_{\mathrm{th}} a_{\mathrm{eff}}}
\label{eq:Q-Gamma-arr-app2}
\ee
where $r^{\prime}_{\perp} = r_{\perp}/a_{\mathrm{eff}}$.  
Given direction $(\theta, \phi)$ and the direction 
$(\theta_{\mathrm{gr}}, \phi_{\mathrm{gr}})$ of the grain's velocity, 
\be
|\mathbfit{s} + \mathbfit{s}_d|^2 = s_d^2 + s^2 - 2 s_d s \left[ \cos
\theta_{\mathrm{gr}} \cos \theta + \sin \theta_{\mathrm{gr}} \sin \theta 
\cos (\phi - \phi_{\mathrm{gr}}) \right] .
\ee
Thus, for direction $(\pi - \theta, \phi + \pi)$, 
\be
|\mathbfit{s} + \mathbfit{s}_d|^2 = s_d^2 + s^2 + 2 s_d s \left[ \cos
\theta_{\mathrm{gr}} \cos \theta + \sin \theta_{\mathrm{gr}} \sin \theta 
\cos (\phi - \phi_{\mathrm{gr}}) \right] .
\ee
Since $\Delta \mathbfit{J}_{\mathrm{arr}}(\pi - \theta, \phi + \pi, \chi, 
r_{\perp}, s) = - \Delta \mathbfit{J}_{\mathrm{arr}}(\theta, \phi, \chi, r_{\perp}, 
s)$, combining the terms for direction $(\theta, \phi)$ and 
$(\pi - \theta, \phi + \pi)$ in the expression for 
$\mathbfit{Q}_{\Gamma, \mathrm{arr}}$ yields
\begin{eqnarray}
\mathbfit{Q}_{\Gamma, \mathrm{arr}}(\theta_{\mathrm{gr}}, \phi_{\mathrm{gr}}) 
& = & \pi^{-3/2} \int_0^{2\pi} d\phi \int_0^1
d(\cos \theta) \int_0^{2\pi} d\chi \int_0^{b(\theta, \phi, \chi)} dr^{\prime}_{\perp}
r^{\prime}_{\perp} \int_0^{\infty} ds \, s^3 \exp \left[ - \left( s_d^2 + s^2
\right) 
\right] \frac{\Delta \mathbfit{J}_{\mathrm{arr}}(\theta, \phi, \chi, r_{\perp}, s)}
{m v_{\mathrm{th}} a_{\mathrm{eff}}} \nonumber \\
& & \times
\left( \exp \left\{ 2 s_d s \left[ \cos \theta_{\mathrm{gr}} \cos \theta + 
\sin \theta_{\mathrm{gr}} \sin \theta \cos (\phi - \phi_{\mathrm{gr}}) \right]
\right\} - \exp \left\{ -2 s_d s \left[ \cos \theta_{\mathrm{gr}} \cos \theta + 
\sin \theta_{\mathrm{gr}} \sin \theta \cos (\phi - \phi_{\mathrm{gr}}) \right]
\right\} \right) .
\label{eq:Q-Gamma-arr-app}
\end{eqnarray}
Clearly, $\mathbfit{Q}_{\Gamma, \mathrm{arr}} = 0$ when $s_d = 0$.  
From equations (\ref{eq:rot-avg-transf1})--(\ref{eq:rot-avg-transf3})
with $(\bmath{\hat{x}}, \bmath{\hat{y}}, \bmath{\hat{z}}) = 
(\bmath{\hat{a}}_2, \bmath{\hat{a}}_3, \bmath{\hat{a}}_1)$,
$\theta_{\mathrm{gr}} = \theta_{va}$ and $\phi_{\mathrm{gr}} = \pi/2 - \Phi_2$.
Since $\Phi_2$ is the grain's rotation angle about $\bmath{\hat{a}}_1$, 
\be
\overbar{\mathbfit{Q}}_{\Gamma, \mathrm{arr}}(\theta_{va}) = \frac{1}{2\pi}
\int_0^{2\pi} d\phi_{\mathrm{gr}} \, \mathbfit{Q}_{\Gamma, \mathrm{arr}}(
\theta_{\mathrm{gr}} = \theta_{va}, \phi_{\mathrm{gr}}) .
\ee
Thus,
\begin{eqnarray}
\overbar{\mathbfit{Q}}_{\Gamma, \mathrm{arr}}(\theta_{va}) \bmath{\cdot}
\bmath{\hat{a}}_1 & = & \frac{\pi^{-5/2}}{2} \int_0^{2\pi} d\phi \int_0^1
d(\cos \theta) \sin \theta \int_0^{2\pi} d\chi \sin \chi 
\int_0^{b(\theta, \phi, \chi)} dr^{\prime}_{\perp} (r^{\prime}_{\perp})^2 \int_0^{\infty} ds 
\, s^4 \exp \left[ - \left( s_d^2 + s^2 \right) \right] \int_0^{2\pi} 
d\phi_{\mathrm{gr}} \nonumber \\
& & \times \left( \exp \left\{ 2 s_d s \left[ \cos \theta_{va} \cos \theta + 
\sin \theta_{va} \sin \theta \cos (\phi - \phi_{\mathrm{gr}}) \right]
\right\} - \exp \left\{ -2 s_d s \left[ \cos \theta_{va} \cos \theta + 
\sin \theta_{va} \sin \theta \cos (\phi - \phi_{\mathrm{gr}}) \right]
\right\} \right) .
\end{eqnarray}
Replacing the integration variable $\phi_{\mathrm{gr}}$ with 
$\phi_{\mathrm{gr}} + \pi$ in the second exponential, 
\begin{eqnarray}
\overbar{\mathbfit{Q}}_{\Gamma, \mathrm{arr}}(\theta_{va}) \bmath{\cdot}
\bmath{\hat{a}}_1 & = & \pi^{-5/2} \int_0^{2\pi} d\phi \int_0^1
d(\cos \theta) \sin \theta \int_0^{2\pi} d\chi \sin \chi 
\int_0^{b(\theta, \phi, \chi)} dr^{\prime}_{\perp} (r^{\prime}_{\perp})^2 \int_0^{\infty} ds 
\, s^4 \exp \left[ - \left( s_d^2 + s^2 \right) \right] \int_0^{2\pi} 
d\phi_{\mathrm{gr}} \nonumber \\
& & \times 
\exp \left[ 2 s_d s \sin \theta_{va} \sin \theta \cos (\phi - \phi_{\mathrm{gr}}) 
\right] \sinh \left( 2 s_d s \cos \theta_{va} \cos \theta \right) .
\label{eq:bar-Q-a1-app}
\end{eqnarray}
Thus, $\overbar{\mathbfit{Q}}_{\Gamma, \mathrm{arr}}(\theta_{va}) \bmath{\cdot}
\bmath{\hat{a}}_1$ is an odd 
function of $\cos \theta_{va}$.  Retaining only the first-order term in 
$\sinh (2 s_d s \cos \theta_{va} \cos \theta)$ as $s_d \rightarrow 0$, 
$\overbar{\mathbfit{Q}}_{\Gamma, \mathrm{arr}}(\theta_{va}) \bmath{\cdot}
\bmath{\hat{a}}_1 \propto \cos \theta_{va}$ in the extreme subsonic limit.  

Since $\bmath{\hat{\theta}}_v = \bmath{\hat{x}}_v \cos \theta_{va} - 
\bmath{\hat{z}}_v \sin \theta_{va}$, equations (\ref{eq:dot1}), (\ref{eq:dot3}),
and (\ref{eq:Q_i-xv}) yield
\be
\label{eq:Q-theta-v-app}
\mathbfit{Q}_{\Gamma, \mathrm{arr}} \bmath{\cdot} \bmath{\hat{\theta}}_v = 
- \left( \mathbfit{Q}_{\Gamma, \mathrm{arr}} \bmath{\cdot} \bmath{\hat{x}} \,
\sin \Phi_2 + \mathbfit{Q}_{\Gamma, \mathrm{arr}} \bmath{\cdot} \bmath{\hat{y}}
\, \cos \Phi_2 \right) = 
- \left( \mathbfit{Q}_{\Gamma, \mathrm{arr}} \bmath{\cdot} \bmath{\hat{x}} \,
\cos \phi_{\mathrm{gr}} + \mathbfit{Q}_{\Gamma, \mathrm{arr}} \bmath{\cdot} 
\bmath{\hat{y}} \, \sin \phi_{\mathrm{gr}} \right) .
\ee
From equations (\ref{eq:Delta-J-arr-app}) and (\ref{eq:Q-theta-v-app}), 
\be
\label{eq:Delta-J-arr-app2}
\frac{\Delta \mathbfit{J}_{\mathrm{arr}}(\theta, \phi, \chi, r_{\perp}, s)} 
{m v_{\mathrm{th}} s r_{\perp}} \bmath{\cdot} \bmath{\hat{\theta}}_v = \cos \chi 
\sin (\phi - \phi_{\mathrm{gr}}) + \sin \chi \cos \theta \cos (\phi - 
\phi_{\mathrm{gr}}) .
\ee
From equations (\ref{eq:Q-Gamma-arr-app}) and (\ref{eq:Delta-J-arr-app2}),
$\overbar{\mathbfit{Q}}_{\Gamma, \mathrm{arr}} \bmath{\cdot} \bmath{\hat{\theta}}_v$
contains terms of the form $\sin (\phi - \phi_{\mathrm{gr}}) \exp[2 s_d s \sin 
\theta_{va} \sin \theta \cos (\phi - \phi_{\mathrm{gr}})]$ and 
$\cos (\phi - \phi_{\mathrm{gr}}) \exp[2 s_d s \sin \theta_{va} \sin \theta \cos 
(\phi - \phi_{\mathrm{gr}})]$.  When integrated over $\phi_{\mathrm{gr}}$ 
(0 to $2\pi$), the former yields zero and the latter yields 
$2 \pi I_1(2 s_d s \sin \theta_{va} \sin \theta)$, where $I_1(u)$ denotes
the first-order modified Bessel function of the first kind.  Since $I_1(u)$
is an odd function of $u$, 
\begin{eqnarray}
\overbar{\mathbfit{Q}}_{\Gamma, \mathrm{arr}}(\theta_{va}) \bmath{\cdot} 
\bmath{\hat{\theta}}_v & = & 
2 \pi^{-3/2} \int_0^{2\pi} d\phi \int_0^1 d(\cos \theta)
\cos \theta \int_0^{2\pi} d\chi \, \sin \chi \int_0^{b(\theta, \phi, \chi)}
dr^{\prime}_{\perp} (r^{\prime}_{\perp})^2 \int_0^{\infty} ds \, s^4 \exp \left[- \left( 
s_d^2 + s^2 \right) \right] \nonumber \\ 
& & \times I_1 (2 s_d s \sin \theta_{va} \sin \theta)
\cosh (2 s_d s \cos \theta_{va} \cos \theta) .
\label{eq:bar-Q-theta-v-app}
\end{eqnarray}
Thus, 
$\overbar{\mathbfit{Q}}_{\Gamma, \mathrm{arr}}(\theta_{va}) \bmath{\cdot} 
\bmath{\hat{\theta}}_v$ is an even function of $\cos \theta_{va}$.  
Retaining only the lowest-order terms in 
$\cosh (2 s_d s \cos \theta_{va} \cos \theta)$ and 
$I_1(2 s_d s \sin \theta_{va} \sin \theta)$ as $s_d \rightarrow 0$, 
$\overbar{\mathbfit{Q}}_{\Gamma, \mathrm{arr}}(\theta_{va}) \bmath{\cdot} 
\bmath{\hat{\theta}}_v \propto \sin \theta_{va}$ in the extreme subsonic limit.  
Since $I_1(0) = 0$, 
$\overbar{\mathbfit{Q}}_{\Gamma, \mathrm{arr}}(\theta_{va}) \bmath{\cdot} 
\bmath{\hat{\theta}}_v (\cos \theta_{va} = \pm 1) =0$.
The results derived here for the extreme subsonic limit can also be obtained
from equation (\ref{eq:q-gamma-prime-arr}).  Comparing equations
(\ref{eq:bar-Q-a1-app}) and (\ref{eq:bar-Q-theta-v-app}), we see that
$\overbar{\mathbfit{Q}}_{\Gamma, \mathrm{arr}}(\theta_{va}) \bmath{\cdot} 
\bmath{\hat{a}}_1$ and
$\overbar{\mathbfit{Q}}_{\Gamma, \mathrm{arr}}(\theta_{va}) \bmath{\cdot} 
\bmath{\hat{\theta}}_v$ have the same sign when $\cos \theta_{va} > 0$.

\section{Special Results for the Extreme Subsonic Limit} 
\label{sec:subsonic-special-results}

Adopting the same approach used in the derivation of 
equation (\ref{eq:Q-Gamma-arr-app2}), the rotationally averaged torque
$\overbar{\bmath{\Gamma}}(\theta_{va})$
associated with any process (except outgoing scenario 1) 
in the extreme subsonic limit is given by 
\be
\overbar{\bmath{\Gamma}}_i(\theta_{va}) = \frac{\pi^{-5/2} n v_{\mathrm{th}} 
a_{\mathrm{eff}}^2}{2} \int_0^{2\pi} d\phi_{\mathrm{gr}} \int_0^{2\pi} d\phi \int_{-1}^1 
d(\cos \theta) \int_0^{2\pi} d\chi \int_0^{b(\theta, \phi, \chi)}
dr^{\prime}_{\perp} r^{\prime}_{\perp} \left\{ k_1 + k_2 s_d \left[ \cos 
\theta_{va} \cos \theta + \sin \theta_{va} \sin \theta \cos
(\phi - \phi_{\mathrm{gr}}) \right] \right\} \Delta \mathbfit{J}_i
\label{eq:torque-app-sub}
\ee
where $\Delta \mathbfit{J}_i$ is the angular momentum transferred during an 
event and the subscript $i$ denotes the type of event (arrival of an atom,
specular reflection, departure of a molecule or an atom following sticking).  
From equations (\ref{eq:is3}) and (\ref{eq:is4}), 
$(k_1, k_2) = (3 \sqrt{\pi}/8, 2)$ for $i =$ arr, spec and 
$(k_1, k_2) = (1/2, 3 \sqrt{\pi}/4)$ for $i =$ out(2).  
With grain-body axes chosen such that 
$(\bmath{\hat{x}}, \bmath{\hat{y}}, \bmath{\hat{z}})$ lie
along $(\bmath{\hat{a}}_2, \bmath{\hat{a}}_3, \bmath{\hat{a}}_1)$,
$\Delta \mathbfit{J} \cdot \bmath{\hat{a}}_1 = \Delta J_z$ and 
$\Delta \mathbfit{J} \cdot \bmath{\hat{\theta}}_v = - ( \Delta J_x \cos
\phi_{\mathrm{gr}} + \Delta J_y \sin \phi_{\mathrm{gr}})$ 
(equation \ref{eq:Q-theta-v-app}).  
Since $\Delta \mathbfit{J}_i$ is independent of $\theta_{va}$ and 
$\phi_{\mathrm{gr}}$, equation (\ref{eq:torque-app-sub}) reveals that 
(1)  $\overbar{\mathbfit{Q}}_{\Gamma, i} \cdot \bmath{\hat{\theta}}_v = 0$ when
$s_d = 0$, (2)  $\overbar{\mathbfit{Q}}_{\Gamma, i} \cdot \bmath{\hat{a}}_1$ 
can be non-zero when $s_d = 0$ (though, as shown in Appendix 
\ref{sec:arrival-special-results}, this does not hold for the specific case
of the torque associated with arriving atoms), 
(3)  $\overbar{\mathbfit{Q}}_{\Gamma, i}^{\prime} \cdot \bmath{\hat{a}}_1 \propto 
\cos \theta_{va}$, 
(4)  $\overbar{\mathbfit{Q}}_{\Gamma, i}^{\prime} \cdot \bmath{\hat{\theta}}_v 
\propto \sin \theta_{va}$.

\bsp	
\label{lastpage}
\end{document}